\documentclass[12pt,a4paper]{article}

\usepackage{amsmath,amssymb,graphicx}
\usepackage{bbm}

\usepackage[bulletsep]{collref}

\makeatletter \@addtoreset{equation}{section} \makeatother
\renewcommand{\theequation}{\thesection.\arabic{equation}}

\makeatletter
\let\old@startsection=\@startsection
\let\oldl@section=\l@section
\renewcommand{\@startsection}[6]{\old@startsection{#1}{#2}{#3}{#4}{#5}{#6\mathversion{bold}}}
\renewcommand{\l@section}[2]{\oldl@section{\mathversion{bold}#1}{#2}}
\makeatother

\makeatletter
\let\old@makecaption=\@makecaption
\def\@makecaption{\small\old@makecaption}
\makeatother

\renewcommand{\theequation}{\thesection.\arabic{equation}}

\def\x{{\tt x}}

\begin{document}

\thispagestyle{empty}
\setlength\textwidth{450pt}

\def\thefootnote{\fnsymbol{footnote}}\begin{flushright}
ITEP-TH-59/09\\
LPTENS-09/36 \\
UUITP-25/09\\
\end{flushright}\vskip 0.5cm\begin{center}
\Large{\bf  Integrability and the $AdS_3/CFT_2$ correspondence}
\end{center}\vskip 0.8cm
\begin{center}{\large A. Babichenko$^{1,2}$, B. Stefa\'nski, jr.$^{3}$,
K. Zarembo$^{4,5}$\footnote{Also at ITEP, Moscow, Russia}}
\vskip 0.2cm{\it $^1$ Department of Particle Physics, Weizmann Institute
of Science \\ Rehovot 76100, Israel}
\vskip 0.2cm{\it $^2$ Department of Applied Mathematics,
Holon Institute of Technology\\ Holon 58102, Israel}
\vskip 0.2cm{\it $^3$ Centre for Mathematical Science, City University London,
\\ Northampton Square, London EC1V 0HB, UK}
\vskip 0.2cm{\it $^4$ CNRS – Laboratoire de Physique Th´eorique, ´ Ecole Normale Sup´erieure \\
24 rue Lhomond, 75231 Paris, France}
\vskip 0.2cm{\it $^5$ Department of Physics and Astronomy, Uppsala University \\
SE-751 08 Uppsala, Sweden}

\end{center}
\vskip 1.0cm
\begin{abstract}\noindent
We investigate the $AdS_3/$CFT${}_2$ correspondence for theories
with 16 supercharges using the integrability approach. We construct
Green-Schwarz actions for Type IIB strings on $AdS_3\times S^3\times
M_4$ where $M_4=T^4$ or $S^3\times S^1$ using the coset approach.
These actions are based on a $\mathbb{Z}_{4}$ automorphism of the
super-coset $D(2,1;\alpha)\times D(2,1;\alpha )/SO(1,2)\times
SO(3)\times SO(3)$. The equations of motion admit a representation
in terms of a Lax connection, showing that the system is classically
integrable. We present the finite gap equations for these actions.
When $\alpha=0\,,\,1/2\,,\,1$ we propose a set of quantum Bethe
equations valid at all values of the coupling. The $AdS_3/$CFT${}_2$
duals contain novel massless modes whose role remains to be
explored.
\end{abstract}

\vfill

\setcounter{footnote}{0}
\def\thefootnote{\arabic{footnote}}
\newpage

\renewcommand{\theequation}{\thesection.\arabic{equation}}

\section{Introduction}
\label{sec0}

Several $AdS$/CFT systems possess integrable structures and are
solvable non-per\-tur\-ba\-tive\-ly by Bethe ansatz techniques.
Integrability tools such as  Bethe ansatz, exact S-matrices,
bootstrap and finite-gap integration proved useful in finding the
exact spectrum of the $AdS_5/$CFT${}_4$ and $AdS_4/$CFT${}_3$
systems at the planar/free-string level. We believe that this is not
the end of the story and that other integrable $AdS$/CFT systems
should exist. In this paper we extend the integrability approach to
the $AdS_3/$CFT${}_2$ dual pairs with sixteen supercharges. The
$AdS_3$ backgrounds typically arise from the D1-D5 system in type
IIB string theory \cite{Strominger:1996sh} and are dual to
two-dimensional conformal field theories \cite{Maldacena:1998re}.

The $AdS_3/$CFT${}_2$ duality may appear simpler than its
higher-dimensional counterparts, partly because the conformal
symmetry in two dimensions is larger and more constraining, partly
because problems with string quantization on Ramond-Ramond
backgrounds inevitable in higher dimensions can be circumvented in
$d=3$. Some $AdS_3\times X$ backgrounds can be supported by pure
NSNS flux, which leads to enormous simplification of the worldsheet
CFT. The worldsheet sigma-model then admits an NSR description, has
extended chiral symmetry, and as a consequence is solvable by
representation theory of chiral algebras
\cite{Giveon:1998ns,Elitzur:1998mm,Maldacena:2000hw,Maldacena:2000kv,Maldacena:2001km}.
On the contrary, the RR $AdS_3$ sigma-model \cite{Berkovits:1999im}
does not have useful holomorphicity properties and the usual CFT
methods do not work. In this respect, the $AdS_3$ backgrounds with
the RR flux are as complicated as their higher-dimensional
counterparts such as $AdS_5\times S^5$ or $AdS_4\times CP^3$, which
however are exactly solvable due to their integrability. We would
like to argue that the appropriate method to attack the problem of
string quantization on the RR  $AdS_3$ backgrounds is also the Bethe
ansatz. Clear evidence for integrability in $AdS_3/$CFT${}_2$ comes
from the geometric construction of the Green-Schwarz string action
on $AdS_3\times S^3$
\cite{Rahmfeld:1998zn,Park:1998un,Metsaev:2000mv,Chen:2005uj,Adam:2007ws},
and from the symmetries of the giant magnons on this background
\cite{David:2008yk}.

Within the context of the $AdS$/CFT correspondence, integrability
manifests itself in a number of ways. In this paper we will be
mostly concerned with the string (AdS) side of the duality, where
the construction is particularly simple. The classical integrability
of the string sigma-model follows from the Lax representation of the
equations of motion \cite{Bena:2003wd}. The  monodromy matrix of the
Lax connection then generates an infinite set of local or non-local
commuting conserved charges. Given the monodromy matrix, one can use
the finite-gap methods to solve the equations of motion in terms of
a much simpler set of integral
equations~\cite{Kazakov:2004qf,Beisert:2005bm}. These same equations
arise in the semiclassical limit of the quantum Bethe ansatz
\cite{Beisert:2005fw} that diagonalizes the exact worldsheet
S-matrix \cite{Staudacher:2004tk,Beisert:2005tm}. The experience
with $AdS_5/$CFT${}_4$ \cite{Beisert:2005fw} and $AdS_4/$CFT${}_3$
\cite{Gromov:2008bz,Gromov:2008qe} systems shows that the quantum
Bethe equations are very constrained by symmetries
\cite{Beisert:2005fw,Janik:2006dc} and can be almost uniquely
reconstructed from their semiclassical counterparts
\cite{Arutyunov:2004vx}. It is not clear to us if the Y-system or
the TBA equations
\cite{Gromov:2009tv,Bombardelli:2009ns,Gromov:2009bc,Arutyunov:2009ur}
can be reconstructed from the classical data. The semiclassical
limit is encoded in the Y-system, albeit in a rather non-trivial way
\cite{Gromov:2009tq}.

The key property of the $AdS_5\times S^5$ and $AdS_4\times CP^3$
backgrounds that guarantees their integrability is the geometric
construction of the Green-Schwarz action in terms of the coset
superspace
\cite{Metsaev:1998it,Arutyunov:2008if,Stefanski:2008ik,Gomis:2008jt}\footnote{In
the $AdS_4\times CP^3$ case the coset arises after partially fixing
the kappa-symmetry \cite{Arutyunov:2008if,Gomis:2008jt}.}. If the
coset admits a $\mathbbm{Z}_4$ grading \cite{Berkovits:1999zq},
integrability follows automatically, since the equations of motion
and Maurer-Cartan equations of any $\mathbbm{Z}_4$ coset can be
written as a flatness conditions for a Lax connection
\cite{Bena:2003wd}. It has been known for a long time that the
Green-Schwarz sigma-model on the six-dimensional $AdS_3\times S^3$
background is the $PSU(1,1|2)\times PSU(1,1|2)/SU(1,1)\times SU(2)$
supercoset
\cite{Rahmfeld:1998zn,Park:1998un,Metsaev:2000mv}\footnote{An
alternative construction of the GS action on $AdS_3\times S^3$ is
given in \cite{Pesando:1998wm}.} which, as one can check, possesses
a $\mathbbm{Z}_4$ grading and is therefore integrable
\cite{Chen:2005uj}\footnote{The pure spinor  superstring in this
background is also integrable \cite{Adam:2007ws}.}.

We will consider Type IIB strings on $AdS_3\times S^3\times M_4$
where $M_4=T^4$ or\footnote{When $M_4=K3$ the theory also preserves
16 supersymmetries. In orbifold limits of $K3$ our results
generalise in a straightforward way.} $S^3\times S^1$. The case of $
S^3\times S^1$ in a certain sense is more general although
definitely more complicated. The $AdS_3\times S^3\times S^3\times
S^1$ geometry is a supergravity solution with sixteen supercharges
\cite{Cowdall:1998bu,Boonstra:1998yu,Gauntlett:1998kc,deBoer:1999rh,Gukov:2004ym}
and can be supported by either NSNS or RR three-form flux. In the
NSNS case the duality is relatively well understood, because one can
use the NSR formalism and more or less standard CFT methods to
quantize string in this background
\cite{Elitzur:1998mm,deBoer:1999rh,Giveon:2003ku,Gukov:2004ym}. On
the contrary, the $AdS_3/$CFT${}_2$ duality for  $AdS_3\times
S^3\times S^3\times S^1$ supported by the RR flux perhaps is the
most obscure case among all known $AdS$/CFT pairs. The $AdS$/CFT
correspondence for $AdS_3\times S^3\times S^3\times S^1$  was
discussed in \cite{deBoer:1999rh} and at length in
\cite{Gukov:2004ym}. The string in this RR background cannot be
quantized by any known method. As far as the dual CFT is concerned,
very little is known about it, apart from its rather intricate
symmetries. The CFT is probably a resolution of a permutation
orbifold \cite{deBoer:1999rh}, and displays a number of unusual
features, in particular the non-linear BPS bound
\cite{deBoer:1999rh}. All this makes identification of the spectra
and the moduli spaces on the two sides of the duality problematic
even at the supergravity level \cite{Gukov:2004ym}.

The radii of the two three-spheres ($R_\pm$) and the AdS radius
($l$) are not independent in  $AdS_3\times S^3\times S^3\times S^1$.
The supergravity equations of motion require them to satisfy the
triangle equality:
\begin{equation}\label{supgraradii}
 \frac{1}{R_+^2}+\frac{1}{R_-^2}=\frac{1}{l^2}\,.
\end{equation}
The same triangle equality arises in the invariant bilinear form of
the exceptional Lie superalgebra $\mathfrak{d}(2,1;\alpha )$
\cite{Kac:1977qb,Frappat:1996pb}, eq.~(\ref{str}), and not just by
chance -- the symmetry of the corresponding $AdS$/CFT pair is the
large $N=4$ superconformal algebra \cite{Boonstra:1998yu} whose
rigid part is  $\mathfrak{d}(2,1;\alpha )$. Indeed, the
super-isometries of the $AdS_3\times S^3\times S^3$ background form
two copies of $\mathfrak{d}(2,1;\alpha )$ \cite{Gauntlett:1998kc}.
The parameter $\alpha$, potentially any complex number
\cite{Kac:1977qb,Frappat:1996pb}, is related to the relative size of
the two spheres and takes values between $0$ and $1$ when a suitable
reality condition is imposed on the super-algebra. In view of the
triangle equality (\ref{supgraradii}) we shall use the trigonometric
parameterization:
\begin{equation}\label{radiiphi}
 \alpha =\frac{l^2}{R_+^2}\equiv \cos^2\phi,\qquad
 \frac{l^2}{R_-^2}\equiv \sin^2\phi.
\end{equation}
There are two simplifying limits worth mentioning. One is $\phi =0$.
The radii of AdS and of one of the spheres then become equal, while
the other sphere blows up to an infinite size. By re-compactifying
on $T^3$, we get the $AdS_3\times S^3\times T^4$ background. This
limiting case corresponds to the $\alpha \rightarrow 1$ degeneration
of the $\mathfrak{d}(2,1;\alpha )$ algebra, which up to some abelian
factors contracts to $\mathfrak{psu}(1,1|2)$. The symmetry algebra
of $AdS_3\times S^3$ is indeed $PSU(1,1|2)\times PSU(1,1|2)$. The
other special point is $\phi =\pi /4$, when the two spheres have
equal sizes, $\sqrt{2}$ times smaller than the radius of AdS. The
exceptional superalgebra $\mathfrak{d}(2,1;\alpha )$ (with $\alpha
=1/2$) then coincides with the classical $\mathfrak{osp}(4|2)$
superalgebra from the $\mathfrak{d}(n,m)$ series.

The plan of the paper is the following. In section~\ref{sec1} we
recall the standard classical integrable structure of a general
supercoset with $\mathbb{Z}_{4}$ grading. We then specify the
general construction to the $AdS_3\times S^3\times S^3\times S^1$
case. In section~\ref{sec3} we show that the supercoset action in
the flat space limit reproduces the flat space GS action in a
particular $\kappa$-gauge. In section~\ref{sec4}, we demonstrate
that, up to terms quadratic in fermions,  our supercoset sigma model
is indeed a realization of worldsheet theory of GS string on the
$AdS_{3}\times S^{3}\times S^{3}\times S^{1}$ background with
completely fixed $\kappa$-symmetry identical to the one used in
section~\ref{sec3}. In section~\ref{sec5} we investigate the BMN
limit of the coset sigma model in light cone gauge and compare it to
the supergravity analysis in~\cite{Lu:2002kw,Gukov:2004ym}. We make
some preliminary steps towards going beyond the strict BMN limit. In
section~\ref{sec6} we point out that the type IIB GS string on the
purely RR $AdS_{3}\times S^{3}\times T_{4}$ background can be
treated as a limiting case of the $AdS_{3}\times S^{3}\times
S^{3}\times S^{1}$ super coset action. In section~\ref{sec7} we
discuss the general classical integrability scheme for
$\mathbb{Z}_{4}$ symmetric (super)cosets, and derive the finite gap
equations entirely in terms of the group-theory data. We then derive
classical Bethe equations for $AdS_3\times S^3\times S^3\times S^1$
and discuss their BMN limit. We propose quantum Bethe equations for
two special points, $\phi =\pi /4$ and $\phi =0$, the latter case
corresponding to $AdS_3\times S^3\times T^4$. In section~\ref{sec8}
we conclude with a preliminary discussion of the massless modes,
which is a novel feature of the $AdS_3/$CFT${}_2$ duality compared
to $AdS_5/$CFT${}_4$ or $AdS_4/$CFT${}_3$. In the appendices we
collect the commutation relations of the superalgebra
$\mathfrak{d}(2,1;\alpha)$, the gamma matrix conventions, the
background-field expansion of the general $\mathbbm{Z}_4$ coset
action, and some higher order terms of the near-BMN expansion in
$AdS_3\times S^3\times S^3\times S^1$.

\section{$\mathbbm{Z}_4$ cosets}\label{z4cosets}
\label{sec1}

The space-time supersymmetric action of the superstring in flat
space \cite{Green:1983wt} can be interpreted as a coset sigma-model
\cite{Henneaux:1984mh,McArthur:1999dy}. The coset construction
readily generalizes to curved space, and in particular allows one to
build the Green-Schwarz action for a number of AdS-type backgrounds.
The basic example is the Metsaev-Tseytlin action in $AdS_5\times
S^5$ \cite{Metsaev:1998it} (see \cite{Arutyunov:2009} for a recent
review).  As realized in \cite{Berkovits:1999zq}, an important
feature of the AdS-type cosets is the $\mathbbm{Z}_4$
symmetry\footnote{The manifestly $\mathbbm{Z}_4$ invariant form of
the Metsaev-Tseytlin action is given in \cite{Roiban:2000yy}.}, that
in particular allows one to construct the Wess-Zumino term necessary
for the consistency of the Green-Schwarz action. Integrability
arises as a bonus symmetry in all $\mathbbm{Z}_4$ cosets. Indeed,
the derivation of the Lax pair for the $AdS_5\times S^5$ sigma-model
\cite{Bena:2003wd} does not really depend on the specifics of the
background and relies solely on the existence of the $\mathbbm{Z}_4$
structure.

\subsection{The action and equations of motion}

We start by reviewing the general construction of the sigma-model
action for $\mathbbm{Z}_4$ cosets. A coset $G/H_0$ possesses a
$\mathbbm{Z}_4$ symmetry, if the superalgebra $\mathfrak{g}$ admits
a $\mathbbm{Z}_4$ decomposition:
\begin{equation}
 \mathfrak{g}=\mathfrak{h}_0\oplus\mathfrak{h}_1\oplus\mathfrak{h}_2\oplus\mathfrak{h}_3,
\end{equation}
consistent with the (anti-)commutation relations:
$[\mathfrak{h}_n,\mathfrak{h}_m\}\subset
\mathfrak{h}_{(n+m)\!\!\mod\! 4}$. Equivalently the $\mathbbm{Z}_4$
symmetry is associated with  an order-four automorphism of the Lie
superalgebra $\mathfrak{g}$, which is a linear map $\Omega:
\mathfrak{g}\rightarrow \mathfrak{g}$ that satisfies $[\Omega
(X),\Omega (Y)\}=\Omega ([X,Y\})$ and $\Omega ^4={\rm id}$. The
subspace $\mathfrak{h}_n$ then is defined as a subset of generators
whose $\mathbbm{Z}_4$ charge is $n$, in the basis in which $\Omega $
is diagonal:
\begin{equation}\label{z4action}
 \Omega (\mathfrak{h}_n)=\,{\rm e}\,^{\pi in/2}\mathfrak{h}_n.
\end{equation}
The denominator of the coset is the Lie group of the invariant
subalgebra $\mathfrak{h}_0$ in the $\mathbbm{Z}_4$ decomposition.
The fermion number $F$ is the $\mathbbm{Z}_4$ charge $\!\!\mod 2$,
so that $\mathfrak{h}_0\oplus\mathfrak{h}_2$ is the bosonic
subalgebra of $\mathfrak{g}$, and $\mathfrak{h}_1$, $\mathfrak{h}_3$
consist of Grassmann-odd generators.

The worldsheet embedding in $G/H$ is parameterized by a coset
representative $g(x)\in G$, subject to gauge transformations
$g(x)\rightarrow g(x)h(x)$ with $h(x)\in H_0$. The global $G$-valued
transformations act on $g(x)$ from the left: $g(x)\rightarrow
g'g(x)$. The action and the equation of motion can be written in
terms of the left-invariant current
\begin{equation}\label{cur}
 J_{\bf a} =g^{-1}\partial _{\bf a} g=J_{{\bf a}\,0 }+J_{{\bf a}\,1 }+J_{{\bf a}\,2
 }+J_{{\bf a}\,3 }.
\end{equation}
The $\mathfrak{h}_0$ component of the current transforms as a
connection under the gauge transformations: $J_{\,{\bf a}\,0
}\rightarrow h^{-1}J_{{\bf a}\,0 }h+h^{-1}\partial _{\bf a} h$. The
other three components transform as matter fields in the adjoint:
$J_{{\bf a}\, 1,2,3}\rightarrow h^{-1}J_{{\bf a} \,1,2,3}h$.

The action of the sigma model is\footnote{We use $(+-)$ conventions
for the worldsheet metric, but mostly-plus conventions for the
target-space. The $\varepsilon $-tensor is defined such that
$\varepsilon ^{01}=1$.}
\begin{equation}\label{action}
 S=\int_{}^{}d^2x\,\mathop{\mathrm{Str}}
 \left(\sqrt{h}h^{{\bf a} {\bf b} }J_{{\bf a}\,2 }J_{{\bf b} \,2}+\varepsilon ^{{\bf a} {\bf b} }
 J_{{\bf a}\,1 }J_{{\bf b}\,3 }\right).
\end{equation}
Here $\mathop{\mathrm{Str}}(\cdot \,\cdot )$ denotes the $G$ and
$\mathbbm{Z}_4$ invariant bilinear form on $\mathfrak{g}$. This
action is obviously gauge invariant and $\mathbbm{Z}_4$-symmetric.

The equations of motion for this action and the Bianchi identities
for the currents (the Maurer-Cartan equations) read:
\begin{eqnarray}
 2D_\mathbf{a}\left(\sqrt{-h}h^{\mathbf{a}\mathbf{b}}J_{\mathbf{b}\,2}\right)
 -\varepsilon ^{\mathbf{a}\mathbf{b}}[J_{\mathbf{a}\,1},J_{\mathbf{b}\,1}]
 +\varepsilon ^{\mathbf{a}\mathbf{b}}[J_{\mathbf{a}\,3},J_{\mathbf{b}\,3}]
 &=&0
 \nonumber \\
 \left(\sqrt{-h}h^{\mathbf{a}\mathbf{b}}+\varepsilon
 ^{\mathbf{a}\mathbf{b}}\right)
 [J_{\mathbf{a}\,2},J_{\mathbf{b}\,1}]&=&0
 \nonumber \\
 \left(\sqrt{-h}h^{\mathbf{a}\mathbf{b}}-\varepsilon
 ^{\mathbf{a}\mathbf{b}}\right)
 [J_{\mathbf{a}\,2},J_{\mathbf{b}\,3}]&=&0
 \nonumber \\
 \varepsilon ^{\mathbf{a}\mathbf{b}}\left(
 2D_\mathbf{a}J_{\mathbf{b}\,2}
 +[J_{\mathbf{a}\,1},J_{\mathbf{b}\,1}]
 +[J_{\mathbf{a}\,3},J_{\mathbf{b}\,3}]
 \right)&=&0
 \nonumber \\
 \varepsilon ^{\mathbf{a}\mathbf{b}}\left(D_\mathbf{a}J_{\mathbf{b}\,1}
 +[J_{\mathbf{a}\,2},J_{\mathbf{b}\,3}]\right)&=&0
 \nonumber \\
 \varepsilon^{\mathbf{a}\mathbf{b}}\left(D_\mathbf{a}J_{\mathbf{b}\,3}
 +[J_{\mathbf{a}\,2},J_{\mathbf{b}\,1}]\right)&=&0
 \nonumber \\
 F_{\mathbf{a}\mathbf{b}}+[J_{\mathbf{a}\,2},J_{\mathbf{b}\,2}]
 +[J_{\mathbf{a}\,1},J_{\mathbf{b}\,3}]+[J_{\mathbf{a}\,3},J_{\mathbf{b}\,1}]
 &=&0,
\end{eqnarray}
where $D_\mathbf{a}=\partial _\mathbf{a}+[J_{\mathbf{a}\,0},\cdot]$
and $F_{\mathbf{a}\mathbf{b}}=\partial
_\mathbf{a}J_{\mathbf{b}\,0}-\partial
_\mathbf{b}J_{\mathbf{a}\,0}+[J_{\mathbf{a}\,0},J_{\mathbf{b}\,0}]$.
These equations admit a Lax representation, they are equivalent to
the flatness condition for the connection \cite{Bena:2003wd}
\begin{equation}\label{lax}
 L_\mathbf{a}=J_{\mathbf{a}\,0}+\frac{{\tt x}^2+1}{{\tt x}^2-1}\,J_{\mathbf{a}\,2}
 -\frac{2{\tt x}}{{\tt x}^2-1}\,\,\frac{1}{\sqrt{-h}}\,h_{\mathbf{a}\mathbf{b}}\varepsilon
 ^{\mathbf{b}\mathbf{c}}J_{\mathbf{c}\,2}+\sqrt{\frac{{\tt x}+1}{{\tt x}-1}}\,J_{\mathbf{a}\,1}
 +\sqrt{\frac{{\tt x}-1}{{\tt x}+1}}
 J_{\mathbf{a}\,3}.
\end{equation}
The spectral parameter $\x$ is an arbitrary complex number $\x\neq
\pm 1$. Provided that the currents obey the equations of motion, the
Lax connection satisfies
\begin{equation}\label{flatness}
 \partial _\mathbf{a}L_\mathbf{b}-\partial
 _\mathbf{b}L_\mathbf{a}+[L_\mathbf{a},L_\mathbf{b}]=0.
\end{equation}
And conversely, if the connection $L_\mathbf{a}$ is flat for any
${\tt x}$, the currents satisfy the equations of motion.

The Wilson loop of the Lax connection defines an infinite set of
conserved charges, which include the global Noether charges of the
left group multiplication. These Noether charges are expressed in
term of the gauge-invariant right current that can be obtained from
the left currents $J_{\mathbf{a}\,n}$ by conjugation with the coset
representative $g$:
\begin{equation}\label{globch}
 k^\mathbf{a}=g\left(\sqrt{-h}h^{\mathbf{a}\mathbf{b}}J_{\mathbf{b}\,2}
 -\frac{1}{2}\,\varepsilon ^{\mathbf{a}\mathbf{b}}J_{\mathbf{b}\,1}
 +\frac{1}{2}\,\varepsilon
 ^{\mathbf{a}\mathbf{b}}J_{\mathbf{b}\,3}\right)g^{-1}.
\end{equation}
This current is conserved:
\begin{equation}
 \partial _\mathbf{a}k^\mathbf{a}=0,
\end{equation}
as a consequence of the equations of motion.

\subsection{$AdS_3\times S^3\times S^3\times S^1$ supercoset}
\label{sec2}

The conformal algebra in two dimensions is a two-fold tensor
product, with two factors acting independently on the left and right
movers. The cosets appropriate for the $AdS_3/$CFT${}_2$
correspondence are thus of the form $H\times H/H_0$. If $H$ is a
superalgebra, such a coset will naturally have a $\mathbbm{Z}_4$
structure. Indeed, one can define a $\mathbbm{Z}_4$ automorphism on
$\mathfrak{g}=\mathfrak{h}\oplus\mathfrak{h}$ by combining the
fermion parity with the permutation of the two factors:
\begin{equation}\label{omegad}
 \Omega =
 \begin{pmatrix}
   0  & \mathop{\mathrm{id}}  \\
    (-1)^F & 0  \\
 \end{pmatrix}.
\end{equation}
This map satisfies all necessary requirements: it preserves the
(anti)-commutation relations of $\mathfrak{h}\oplus\mathfrak{h}$ and
squares to $(-1)^F$ such that its forth power is the identity:
$\Omega ^4=\mathop{\mathrm{id}}$.  The $\mathbbm{Z}_4$ grading
associated with the automorphism (\ref{omegad}) is
\begin{eqnarray}\label{z4d21a}
 \mathfrak{h}_0&=&\left\{(X,X)|X\in\mathfrak{h}_{\rm bos}\right\}
 \nonumber \\
 \mathfrak{h}_1&=&\left\{(X,iX)|X\in\mathfrak{h}_{\rm ferm}\right\}
 \nonumber \\
 \mathfrak{h}_2&=&\left\{(X,-X)|X\in\mathfrak{h}_{\rm bos}\right\}
 \nonumber \\
 \mathfrak{h}_3&=&\left\{(X,-iX)|X\in\mathfrak{h}_{\rm ferm}\right\}
\end{eqnarray}
In particular, the invariant subspace is the diagonal bosonic
subalgebra.

Thus for any superalgebra $H$ one can construct a $\mathbbm{Z}_4$
invariant coset sigma-model with the global  $H\times H$ symmetry.
The denominator of the coset is the diagonal bosonic subgroup. The
bosonic part of the action is the sigma-model with the target space
$H_{\rm bos}\times H_{\rm bos}/H_{\rm diag}$ isomorphic
to\footnote{In the gauge orbit $\{(g_Lh,g_Rh)|h\in H_{\rm bos}\}$,
we can pick a representative by taking $h=g_L^{-1}$, or in other
words impose $g_L(x)=1$ as a gauge condition. What remains is the
group manifold $\{(1,g)\}$.} $H_{\rm bos}$. Thus constructed
sigma-model will be automatically integrable.

The construction is completely general and works for any supergroup.
By taking $H=PSU(1,1|2)$ we recover the action of
\cite{Rahmfeld:1998zn,Park:1998un,Metsaev:2000mv} for the
Green-Schwarz string on the $AdS_3\times S^3$ background
\cite{Chen:2005uj}. We can also pick $H=D(2,1;\alpha )$ whose
bosonic subgroup for $0<\alpha <1$ is $SU(1,1)\times SU(2)\times
SU(2)$. When restricted to the bosonic fields, the action reduces to
a sigma-model whose target space is the group manifold of
$SU(1,1)\times SU(2)\times SU(2)$, namely $AdS_3\times S^3\times
S^3$. The $D(2,1;\alpha )\times D(2,1;\alpha )/SU(1,1)\times
SU(2)\times SU(2)$ coset thus describes a Green-Schwarz-type
sigma-model on $AdS_3\times S^3\times S^3$. The equations of motion
following from the coset action admit the Lax representation and
consequently the model is completely integrable.

It is not immediately clear if this model is capable of describing
the Green-Schwarz superstring on $AdS_3\times S^3\times S^3\times
S^1$, because of the missing $S^1$ factor. This factor has to be
added by hand. Similar situation occurs in the hybrid sigma-model on
$AdS_3\times S^3\times T^4$ \cite{Berkovits:1999im}, where the $T^4$
factor is completely orthogonal to the non-linear part of the
action. The hybrid formalism assumes the conformal gauge from the
very beginning \cite{Berkovits:1999im}, and adding an independent
CFT is not a problem provided the total central charge vanishes. On
the contrary, in the Green-Schwarz action {\it all} bosons are
coupled to {\it all} fermions through the kinetic term $\bar{\theta
}^I\partial _\mathbf{a}X^M\Gamma_M\partial _\mathbf{b}\theta ^J $.
The desired decoupling of a bosonic direction is essentially
equivalent to setting one of the Dirac matrices to zero. This does
not sound right at all. In addition, there are 32 fermions in the
Green-Schwarz action compared to sixteen (2 $\times $ the number of
supercharges in $\mathfrak{d}(2,1;\alpha )$) fermion degrees of
freedom in the $D(2,1;\alpha )\times D(2,1;\alpha )/SU(1,1)\times
SU(2)\times SU(2)$ coset. However, half of the 32 Green-Schwarz
fermions are unphysical because of kappa-symmetry. In the next
two sections we will demonstrate that by taking a special gauge
choice of the kappa-symmetry it is possible to decouple the $S^1$
factor in the $AdS_3\times S^3\times S^3\times S^1$ supergeomerty,
and that the resulting kappa-fixed GS action coincides with the
coset model plus one free boson.

The fact that the $S^1$ appears as an extra factor in the action
without any couplings to the other fields can be anticipated from
the structure of the Killing spinors \cite{Gauntlett:1998kc}. In
particular, the momentum Killing vector for $S^1$ does not appear on
the right-hand-side of the anti-commutator of the Killing spinors
for this background and also commutes with all the supercharges
\cite{Gauntlett:1998kc}\footnote{We are grateful to Jerome Gauntlett
for a detailed explanation of these and related results of
\cite{Gauntlett:1998kc}.}.

\section{Flat space limit}
\label{sec3}

In this section we find the flat space limit of the
action~(\ref{action}) for the $D(2,1;\alpha)\times D(2,1;\alpha
)/SU(1,1)\times SU(2)\times SU(2)$ coset, and show that it coincides
with the usual flat space Green-Schwarz action~\cite{Green:1983wt}
in a particular kappa-gauge.

To obtain the flat space limit of the $\mathfrak{d}(2,1;\alpha )^2$ algebra
we should rescale the generators as follows
\begin{eqnarray}
\mathfrak{h}_0&\rightarrow \mathfrak{h}_0\,,\qquad\qquad\qquad
\mathfrak{h}_2&\rightarrow R\mathfrak{h}_2 \\
\mathfrak{h}_1&\rightarrow \sqrt{R}\mathfrak{h}_1\,,\qquad\qquad
\mathfrak{h}_3&\rightarrow \sqrt{R}\mathfrak{h}_3\,,
\end{eqnarray}
and take the $R\rightarrow\infty$ limit. In this limit the
generators of $\mathfrak{h}_0$ become the angular momenta of
$SO(1,2)\times SO(3)\times SO(3)\subset SO(1,9)$, while the elements
of $\mathfrak{h}_2$ become flat space-time momenta; it is easy to
check that, in the $R\rightarrow\infty$ limit, the elements of
$\mathfrak{h}_2$ commute with themselves and the supercharges - as
is the case for flat-space momenta. In the flat space limit the
anti-commutators (\ref{commutation_relations}) become
\begin{eqnarray}
\left\{ Q^I_{ a \alpha {\dot \alpha}}  \,,\, Q^J_{b \beta {\dot
\beta}}\right\} &=& \delta^{IJ} \left[i(\varepsilon\gamma^\mu )_{ab
}\varepsilon _{\alpha\beta}\varepsilon
 _{\dot{\alpha}\dot{\beta}}P_\mu
 +\cos^2\phi \,\varepsilon_{ab}(\varepsilon\gamma^n )_{\alpha\beta}\varepsilon
 _{\dot{\alpha}\dot{\beta}}P_n
 \right. \nonumber \\ &&\left.
 +\sin^2\phi \,\varepsilon_{ab }\varepsilon _{\alpha\beta}(\varepsilon\gamma^{\dot{n}}
 )_{\dot{\alpha }\dot{\beta }}P_{\dot{n}}\right]\,,\label{fsacomm}
\end{eqnarray}
where $I=1,2$, $Q^I=Q^L-i(-1)^I Q^R\in \mathfrak{h}_{2I-1}$ and
$P_\mu=S^L_\mu-S^R_\mu$, $P_n=L^L_n-L^R_n$, $P_{{\dot n}}=R^L_{{\dot
n}}-R^R_{{\dot n}}$, and the indices $L$ and $R$ distinguish the two
copies of $\mathfrak{d}(2,1;\alpha )$.

Recall that the IIB flat space supersymmetry algebra is
\begin{equation}
\left\{ q^I_{{\hat a}} \,,\, q^J_{{\hat b}}\right\} = \delta^{IJ}
\left[C\Gamma^M(1+\Gamma)\right]_{{\hat a}{\hat b}}P_M\,, \label{qq}
\end{equation}
where $I\,,\,J=1\,,\,2$ counts the amount of 10d supersymmetry,
${\hat a}\,,\,{\hat b}=1,\dots,32$ are spinor indices of $SO(1,9)$,
$M=0,\dots,9$ is the 10d vector index, $C$ is the charge conjugation
matrix, $\Gamma^M$ are $32\times 32$ Dirac matrices of $SO(1,9)$ and
$\Gamma$ is the 10d chirality matrix\footnote{In
Appendix~\ref{lesnouvelles} we define a basis for the gamma matrices
that will be particularly useful to the symmetries of the problem.
}.

We would like to identify a sub-algebra of this flat space-time
supersymmetry algebra which has the same form as the flat space
limit of $\mathfrak{d}(2,1;\alpha )^2$. In particular, we want  to
find a subset of 16 fermionic generators which
satisfy~(\ref{fsacomm}).
To this end, we define the projection operators
\begin{equation}\label{kappaprojector}
K^\pm(\phi)\equiv \frac{1}{2}(1\pm \cos \phi \,\Gamma^{012345}
\pm\sin \phi\,\Gamma^{012678})\,.
\end{equation}
Below, we will show that the sixteen supercharges $K^+ q^I_\alpha$ satisfy~(\ref{fsacomm}).
This will allow us to show that the flat space limit of the action~(\ref{action})
for $\mathfrak{g}=\mathfrak{d}(2,1;\alpha )^2$ coincides with
the usual flat space Green-Schwarz action~\cite{Green:1983wt} in a particular kappa-gauge.

First, let us note that
\begin{equation}\label{}
 K^\pm K^\pm=K^\pm,\qquad K^\pm K^\mp=0,
\end{equation}
as required for a projector. Further, $K^\pm$ commute with the 10d
chirality matrix $\Gamma \equiv \Gamma^{0123456789}$ and satisfy
\begin{equation}
 K^{\pm t}C=CK^\mp
\end{equation}
where ${}^t$ indicates the transpose. As a result, projecting with
respect to $K^\pm$ is compatible with both the 10d Majorana and Weyl
conditions. With our choice of gamma matrices, the anticommutator of
the supercharges $q^I$ projected by $K^+$ is
\begin{eqnarray}
\left\{ K^+(\phi)q^I \,,\, K^+(\phi)q^J\right\}
&=& \delta^{IJ} m(\phi)\otimes \left(i \varepsilon \gamma
^\mu\otimes \varepsilon \otimes \varepsilon \,P_\mu+\cos\phi \,
\varepsilon \otimes \varepsilon \gamma ^{n}\otimes \varepsilon \,P_{{n}}\right.\nonumber \\
&&\left. \qquad\qquad\qquad\qquad+\sin \phi\, \varepsilon \otimes
\varepsilon \otimes \varepsilon \gamma ^{\dot{n}}P_{\dot{n}}\right)
\,,\label{projectedsucharges}
\end{eqnarray}
where $m(\phi)$ is a degenerate $4\times 4$ matrix
\begin{equation}
 m(\phi)=
 \begin{pmatrix}
   0  & 1  \\
    0 & 0  \\
 \end{pmatrix}
 \otimes
 \begin{pmatrix}
   1+\cos\phi   & -\sin\phi   \\
    -\sin\phi  & 1-\cos\phi   \\
 \end{pmatrix}.
\end{equation}
The projected supercharges $K^+(\phi)q^I$ have the same commutation
relations as the flat space limit of the $\mathfrak{d}(2,1;\alpha
)^2$ supercharges (the equations~(\ref{projectedsucharges})
and~(\ref{fsacomm}) differ only by normalization of momenta). As a
result the flat-space limit of the $\mathbbm{Z}_4$ coset sigma-model
action will match with the flat space Green-Schwarz action in the
(fully fixed) kappa gauge
\begin{equation}
K^-(\phi )\theta ^I=\theta ^I. \label{kappafix}
\end{equation}
One can explicitly check that the form of the actions is indeed
equivalent; in doing this it is important to start with the 3d form
of the Wess-Zumino term - just as one does for $AdS_5\times
S^5$~\cite{Metsaev:1998it}.

\section{Coset model vs. Green-Schwarz string}
\label{sec4}

In this section we compare the $D(2,1;\alpha )\times D(2,1;\alpha
)/SU(1,1)\times SU(2)\times SU(2)$ supercoset with the Green-Schwarz
action on $AdS_3\times S^3\times S^3\times S^1$, expanded to the
second order in fermions\footnote{We would like to thank A.~Tseytlin
for suggesting this calculation to us.}. As discussed before, the
agreement of the two is not at all obvious. For successful
comparison one has to completely fix the kappa-symmetry in the
Green-Schwarz action in such a way that the $S^1$ factor decouples
from fermions. In this respect, the background is similar to
$AdS_4\times CP^3$, for which the coset description requires
partially fixed kappa-symmetry \cite{Arutyunov:2008if,Gomis:2008jt}.
A novel feature is the complete decoupling of a bosonic direction.
It is known from the experience with $AdS_4\times CP^3$ that the
coset kappa-symmetry gauge may become singular on certain string
configurations \cite{Gomis:2008jt,Cagnazzo:2009zh}. It would thus be
interesting to find the full Green-Schwarz action on $AdS_3\times
S^3\times S^3\times S^1$, perhaps along the lines of the type IIA
$AdS_4\times CP^3$ case \cite{Gomis:2008jt}. For technical reasons
we will restrict ourselves to the quadratic part of the action,
which is known in a closed form for any type IIB supergravity
background \cite{Cvetic:1999zs}.

The Green-Schwarz fermions couple to the metric and to the
three-form RR flux as \cite{Cvetic:1999zs}:
\begin{equation}\label{GSaction}
 \mathcal{L}_{\rm GS}=\left(\sqrt{-h}h^{\mathbf{a}\mathbf{b}}\delta ^{IJ}
 -\varepsilon ^{\mathbf{a}\mathbf{b}}\sigma_3 ^{IJ}\right)
 \bar{\theta }^IE\!\!\!/_\mathbf{a }
 \left(D_\mathbf{b}\delta ^{JK}+\frac{1}{48}\,F\!\!\!/E\!\!\!/_\mathbf{b}\sigma _1^{JK}
 \right)\theta ^K.
\end{equation}
Here $E^A_\mathbf{a}$ is the worldsheet projection of the vierbein:
\begin{equation}
 E^A_\mathbf{a}=\partial _\mathbf{a}X^ME_M^A.
\end{equation}
For any tangent-space tensor we define
$$
 \mathcal{F}\!\!\!/=\mathcal{F}_{A_1\ldots A_n}\Gamma ^{A_1\ldots
 A_n}.
$$
The metric coupling in the covariant derivative is the standard spin
connection:
\begin{equation}
 D_\mathbf{b}=\partial _\mathbf{b}+\frac{1}{4}\,\Omega\!\!\!/ _\mathbf{b},
 \qquad \Omega _\mathbf{b}^{AB}=\partial _\mathbf{b}X^M\Omega
_M^{AB}
\end{equation}
The fermions are Majorana-Weyl spinors of the same chirality :
\begin{equation}
 \Gamma \theta ^I=\theta ^I,\qquad \bar{\theta }^I=\theta
 ^{It}C,\qquad I=1,2.
\end{equation}
The explicit form of the Dirac matrices, as given in the
appendix~\ref{lesnouvelles}, will be important in our calculation.

In the units where the radius of $AdS_3$ is set to one, the metric
and the RR three-form are
\begin{eqnarray}\label{sugrametric}
 ds^2&=& ds^2(AdS_3)+\frac{1}{\cos^2 \phi }\,ds^2(S_+^3)+\frac{1}{\sin^2 \phi
 }\,ds^2(S_-^3)+dU^2, \\
 \label{sugraflux}
 F&=&Vol(AdS_3)+\frac{1}{\cos^2 \phi } \,Vol(S_+^3)+\frac{1}{\sin^2\phi }\,Vol(S_-^3),
\end{eqnarray}
where $ds^2(\mathcal{M})$ and $Vol(\mathcal{M})$ are the standard
metricae and volume forms on $AdS_3$ and $S^3$, and $U$ is the
periodic coordinate on $S^1$.

To find an explicit form of the metric, the spin connection and the
volume form, we can use the fact that $AdS_3\times S^3\times
S^3\times S^1$ is a group manifold  of $SL(2,\mathbbm{R})\times
SU(2)\times SU(2)\times U(1)$. The metric and the volume form can
then be expressed through the Maurer-Cartan forms:
\begin{equation}
 \omega =g^{-1}dg=\omega ^it_i,
\end{equation}
where $t_i$ are the Lie-algebra generators of
$\mathfrak{sl}(2,\mathbbm{R})$, $\mathfrak{su}_L(2)$ or
$\mathfrak{su}_R(2)$, assumed to be canonically normalized: to $\eta
_{ij}=\mathop{\mathrm{diag}}(-++)$ for
$\mathfrak{sl}(2,\mathbbm{R})$ and $\eta _{ij}=\delta _{ij}$ for
$\mathfrak{su}(2)$. The metric and the volume form on a group
manifold are given by
\begin{equation}\label{ds2}
 ds^2=\eta _{ij}\omega ^i\wedge\star \omega ^j,\qquad Vol=f_{ijk}\omega ^j\wedge
 \omega ^j\wedge \omega ^k,
\end{equation}
where $f^i_{\hphantom{i}jk}$ are the structure constants. This
suggests the following local frame on $AdS_3\times S^3\times
S^3\times S^1$:
\begin{equation}\label{locfr}
 E^A_M=\left\{
 \begin{array}{ll}
 \omega ^\mu _M, &M =0,1,2\\
 \frac{1}{\cos\phi }\,\omega ^n_M, &M=3,4,5\\
 \frac{1}{\sin\phi }\,\omega ^{\dot{n}}_M, &M=6,7,8\\
 \delta ^A_9, &M=9.
 \end{array}
 \right.
\end{equation}
The factors of $\cos\phi $ and $\sin\phi $ take into account that
$AdS_3$, $S^3_+$ and $S^3_-$  have different radii which satisfy the
triangle equality (\ref{supgraradii}). For the tangent-space indices
we use the Lie-algebra notations from appendix~\ref{salgebra}.

The spin connection on a group manifold is determined by the
Maurer-Cartan equations:
\begin{equation}
 \Omega ^m_{M\,n}=-\frac{1}{2}\,f^m_{\hphantom{m}nl}\,\omega ^l_M,
\end{equation}
In our case,
\begin{equation}
 \Omega _{M\,AB}= \left\{
 \begin{array}{ll}
 -\frac{1}{2}\,\epsilon _{\mu \nu \lambda }\,\omega ^\lambda _M,
 &M=0,1,2 \nonumber \\
 -\frac{1}{2}\,\epsilon _{mnp}\,\omega ^p_M, &M=3,4,5\nonumber \\
 -\frac{1}{2}\,\epsilon _{\dot{m}\dot{n}\dot{p}}\,\omega ^{\dot{p}}_M,
 &M=6,7,8 \nonumber \\
 0, &M=9.
 \end{array}
 \right.
\end{equation}
Finally, the three-form flux (\ref{sugraflux}) has the following
tangent-space components:
\begin{equation}\label{flux}
 F_{\mu \nu \lambda }=\varepsilon _{\mu \nu \lambda },
 \qquad F_{mnp}=\cos \phi \,\varepsilon _{mnp},
 \qquad F_{\dot{m}\dot{n}\dot{p}}=\sin \phi \,\varepsilon _{\dot{m}\dot{n}\dot{p}}.
\end{equation}
Contracting the RR form with the Dirac matrices we find:
\begin{equation}
 F\!\!\!/=6(\Gamma ^{012}+\cos \phi \,\Gamma ^{345}+\sin \phi \,\Gamma
 ^{678}).
\end{equation}
This matrix is nilpotent: $F\!\!\!/^2=0$, and is proportional to the
projector introduced in the previous section,
eq.~(\ref{kappaprojector}):
\begin{equation}
 F\!\!\!/=12\Gamma ^{012}K^+.
\end{equation}
This elucidates the geometric origin of the gauge-fixing condition
(\ref{kappafix}).

Assuming that the fermions obey the gauge-fixing condition
(\ref{kappafix}), and taking into account the gamma-matrix
identities $K^{\pm t}C=CK^\mp$, $K^\pm\Gamma ^{012}=\Gamma
^{012}K^\mp$ and $D_\mathbf{a}K^\pm=K^\pm D_\mathbf{a}$, we can
bring the Green-Schwarz Lagrangian (\ref{GSaction}) to the following
form:
\begin{equation}\label{GSaction1}
 \mathcal{L}_{\rm GS}=\left(\sqrt{-h}h^{\mathbf{a}\mathbf{b}}\delta ^{IJ}
 -\varepsilon ^{\mathbf{a}\mathbf{b}}\sigma_3 ^{IJ}\right)
 \bar{\theta }^I K^+ E\!\!\!/_\mathbf{a} K^-
 \left(D_\mathbf{b}\delta ^{JK}+\frac{1}{4}\,\Gamma ^{012}K^+
 E\!\!\!/_\mathbf{b}K^-\sigma _1^{JK}
 \right)\theta ^K.
\end{equation}
Quite remarkably, this Lagrangian does not depend on the $S^1$
coordinate $U$. Before the gauge fixing, the fermions coupled to $U$
via the $\partial _\mathbf{a}U\Gamma ^9$ term in
$E\!\!\!/_\mathbf{a}$, but in $K^-E\!\!\!/_\mathbf{a}K^+$ the
$\partial _\mathbf{a}U\Gamma ^9$ term is projected out by virtue of
an easily verifiable identity $K^+\Gamma ^9K^-=0$.  The $S^1$ factor
indeed decouples from fermions after the kappa-symmetry fixing
condition is imposed. It remains to show that the $AdS_3\times
S^3\times S^3$ couplings in the Green-Schwarz action are the same as
in the supercoset model.

Imposing the chiral and the kappa-symmetry conditions
$$
\frac{1-\Gamma }{2}\,\theta^I =0,\qquad K^+\theta^I =0,
$$
and using the explicit form of the projectors,
eqs.~(\ref{chiralprojector}), (\ref{kapexp}), we find that the
worldsheet fermions acquire the following form:
\begin{equation}\label{}
 \theta ^I=\left|+\right\rangle_0\otimes\left|-\right\rangle_\phi
 \otimes\theta^{I\, a\alpha \dot{\alpha }},
\end{equation}
where the spinors $\left|\pm \right\rangle_\phi $ are defined in
(\ref{spinors}). The remaining tri-spinor indices of $\theta^{I\,
a\alpha \dot{\alpha }}$ are acted upon by the three triplets of gamma
matrices $\gamma ^\mu $, $\gamma ^{m}$, $\gamma ^{\dot{m}}$, defined
in (\ref{3ddirac}). Plugging the gauge-fixed fermions into the
Lagrangian (\ref{GSaction1}), we can get rid of the first two
spinors in the tensor product with the help of (\ref{spinident}).
Explicitly,
\begin{eqnarray}\label{defV}
 K^+ E\!\!\!/_\mathbf{b}K^-&\rightarrow&
 -i\omega ^\mu_\mathbf{b} \gamma _\mu +\omega ^{m}_\mathbf{b}\gamma _{m}
 +\omega ^{\dot{m}}_\mathbf{b}\gamma _{\dot{m}}\equiv iV_\mathbf{b} \\
 K^-\Gamma ^{012}K^+&\rightarrow &i\\
 \label{spincfixed}
 \Omega\!\!\!/ _\mathbf{b}&\rightarrow& -\omega ^\mu_\mathbf{b} \gamma
 _\mu
 -i\omega ^{m}_\mathbf{b}\gamma _{m}-i\omega ^{\dot{m}}_\mathbf{b}\gamma _{\dot{m}}
 =V_\mathbf{b},
\end{eqnarray}
and the Green-Schwarz Lagrangian becomes
\begin{equation}\label{tion}
 \mathcal{L}_{\rm GS}=i\left(\sqrt{-h}h^{\mathbf{a}\mathbf{b}}\delta ^{IJ}
 -\varepsilon ^{\mathbf{a}\mathbf{b}}\sigma_3 ^{IJ}\right)
 \bar{\theta }^I V_\mathbf{a}
 \left[\left(\partial _\mathbf{b}+\frac{1}{4}\,V_\mathbf{b}\right)\delta ^{JK}
 -\frac{1}{4}\,V_\mathbf{b}
 \sigma _1^{JK}
 \right]\theta ^K.
\end{equation}
According to (\ref{chargeconjmatr}),
\begin{equation}\label{}
 \bar{\theta }^I_{a\alpha \dot{\alpha }}=\theta ^{I\,b\beta \dot{\beta }}\epsilon
 _{ba}\epsilon _{\beta \alpha }\epsilon _{\dot{\beta }\dot{\alpha }}\,\,
 {\vphantom{\left|+\right\rangle}}_0\!\left\langle +\right|i\sigma ^2\left|-\right\rangle_0
 {\vphantom{\left|-\right\rangle}}_\phi\!\left\langle -\right|
 i\sigma ^2\left|+ \right\rangle_\phi
 =-\theta ^{I\,b\beta \dot{\beta }}\epsilon
 _{ba}\epsilon _{\beta \alpha }\epsilon _{\dot{\beta }\dot{\alpha }}\,.
\end{equation}

This Lagrangian should be compared to the quadratic terms in the
supercoset action. The expansion of the coset Lagrangian to the
second order in fluctuations in an arbitrary bosonic background is
given in appendix~\ref{backexp}. We can specify the general
construction to the case of the $D(2,1;\alpha )\times D(2,1;\alpha
)/SL(2,\mathbbm{R})\times SU(2)\times SU(2)$ coset. The background
bosonic field in (\ref{bdecomposition}) in this case is
parameterized by an element of $H_{\rm bos}\times H_{\rm bos}/H_{\rm
diag}$: $g_B=(g_L,g_R)\sim (g_Lh,g_Rh)$. Fixing the coset gauge by
setting $g_L=1$ we find:
\begin{equation}\label{}
 g^{-1}_B\partial _\mathbf{a}g_B=\omega ^i_\mathbf{a}T_i^R
 =\frac{1}{2}\,\omega ^i_\mathbf{a}\left(T_i^L+T_i^R\right)
 -\frac{1}{2}\,\omega ^i_\mathbf{a}\left(T_i^L-T_i^R\right),
\end{equation}
and thus for the background currents (\ref{bcurrent}) we get:
\begin{eqnarray}\label{}
 K_\mathbf{a}&=&-\frac{1}{2}\,\omega
 ^i_\mathbf{a}\left(T_i^L-T_i^R\right)\nonumber \\
 A_\mathbf{a}&=&\frac{1}{2}\,\omega
 ^i_\mathbf{a}\left(T_i^L+T_i^R\right).
\end{eqnarray}
The fermion part of the fluctuation field in (\ref{bdecomposition})
can be parameterized as
\begin{equation}\label{}
 \mathbbm{X}_{1,3}=2\theta ^{2,1\,\alpha a\dot{a}}\left(Q^L_{\alpha a\dot{a}}\pm
 iQ^R_{\alpha a\dot{a}}\right).
\end{equation}
{}Using the commutation relations of the $\mathfrak{d}(2,1;\alpha )$
algebra listed in appendix~\ref{salgebra}, we obtain:
\begin{equation}\label{OKay}
 [A_\mathbf{a},\theta ^I]=\frac{1}{4}\,V_\mathbf{a}\theta ^I,\qquad
 [K_\mathbf{a},\theta ^I]=-\frac{1}{4}\,V_\mathbf{a}\sigma_1 ^{IJ}\theta
 ^J,
\end{equation}
where $V_\mathbf{a}$ is the same combination of the background
currents and Dirac matrices as in (\ref{defV}), (\ref{spincfixed}).
Plugging $K_\mathbf{a}$ and $A_\mathbf{a}$, as given in
(\ref{OKay}), into the supercoset Lagrangian (\ref{quadaction}), we
find that it exactly agrees with the kappa-fixed Green-Schwarz
action (\ref{tion}).

\section{The BMN limit}
\label{sec5}

In this section we will quantize the coset sigma-model in the
light-cone gauge perturbatively in the sigma-model coupling by
expanding the Lagrangian near a light-cone geodesic. Technically,
this is the same background-field expansion from
appendix~\ref{backexp} as we used in the previous section to compare
the coset sigma-model with the Green-Schwarz string action. The
quadratic action describes the BMN limit \cite{Berenstein:2002jq} of
a point-like string moving along the light-cone geodesic
\cite{Gubser:2002tv} $t=x^0=\varphi$, where $t$ is the global AdS
time, $x^0$ is the worldsheet time direction, and $\varphi $ is an
angle on  $S^3\times S^3\times S^1$.  The quadratic terms can be
read off the general formula (\ref{quadaction}), with the background
taken in the form
\begin{equation}\label{cosetpar}
 g_B=\,{\rm e}\,^{\Xi }\,,\qquad
 \Xi =tD+\varphi J,
\end{equation}
where $D$ is the dilatation generator of $\mathfrak{d}(2,1;\alpha
)^2$ and $J$ is the angular momentum. Higher orders in fluctuations
generate the near-BMN expansion
\cite{Parnachev:2002kk,Callan:2003xr,Callan:2004uv}.

The light-cone frame in $AdS_3\times S^3\times S^3\times  S^1$ is
obtained by combining the global AdS time with an angle on
$S^3\times S^3\times S^1$. The time is conjugate to the dilatation
generator\footnote{The generators of $\mathfrak{d}(2,1;\alpha )$ are
defined in appendix~\ref{salgebra} and the superscripts $L$ and $R$
denote the generators of the left and right $\mathfrak{d}(2,1;\alpha
)$ algebras.}
\begin{equation}\label{dilgen}
 D=S_0^L-S_0^R
\end{equation}
For the angular momentum we have more freedom. Potentially one can
pick an arbitrary element of $\mathfrak{su}(2)\times
\mathfrak{su}(2)\times \mathfrak{u}(1)$:
\begin{equation}\label{jey}
 J=C(L_5^L-L_5^R)+C'(R_8^L-R_8^R)+C''P,
\end{equation}
where $P$ is the generator of the extra $U(1)$ factor. However,
there are two conditions to satisfy: $J$ has to be appropriately
normalized in order for $t\pm\varphi $ to be light-cone directions,
and the light-cone gauge should preserve some supersymmetry. We
shall see that these two conditions uniquely determine all three
coefficients in (\ref{jey}). If we relax the supersymmetry
condition, there are more solutions which we discuss later in
section~\ref{otherlightg}.

In the light-cone gauge $D+J$ is fixed, and determines the internal
length of the string, and $D-J$ becomes the light-cone Hamiltonian:
\begin{equation}\label{}
 H=D-J
\end{equation}
Both should be null elements of the underlying Lie algebra. This
guarantees that the string moves along the light-cone as is
required by the Virasoro constraints for the background
(\ref{cosetpar}). From the explicit form of the metric on
$\mathfrak{d}(2,1;\alpha )$ in appendix~\ref{salgebra}, we find that
$C$, $C'$, $C''$ must satisfy
\begin{equation}\label{zeronorm}
 -\frac{1}{2}+\frac{C^2}{2\cos^2\phi }+\frac{C'{}^2}{2\sin^2\phi }+C''{}^2=0.
\end{equation}
If we also want to preserve supersymmetry, the Hamiltonian $H=D-J$
should commute with some of the supercharges. The eigenvalues of
$\mathop{\mathrm{ad}}H$ on the odd generators of the superalgebra
are proportional to $\pm 1\mp C\mp C'$, where the signs take all
eight possible values. We thus require that
\begin{equation}\label{}
 C'=1-C.
\end{equation}
With this choice, the zero norm condition (\ref{zeronorm}) becomes
\begin{equation}\label{}
 \frac{(C-\cos^2\phi  )^2}{2\cos^2\phi \sin^2\phi }+C''{}^2=0.
\end{equation}
The unique solution to this equation is $C''=0$, $C=\cos^2\phi  $.
Then $C'=\sin^2\phi $, and
\begin{equation}\label{anggen}
 J=\cos^2\phi\,(L_5^L-L_5^R)+\sin^2\phi\,(R_8^L-R_8^R)
\end{equation}

The light-cone gauge fixing explicitly breaks part of the original
$\mathfrak{d}(2,1;\alpha )^2$ symmetry. By inspecting the symmetries
of the giant  magnons in $AdS_3\times S^3$ \cite{Lee:2008sk} and
drawing some intuition from the spin chain picture in the dual CFT,
David and Sahoo argued that the little group of the light-cone gauge
in $AdS_3\times S^3\times T^4$ is $PSU(1|1)\times PSU(1|1)$
\cite{David:2008yk}. The
$\mathfrak{psu}(1|1)\oplus\mathfrak{psu}(1|1)$ superalgebra admits a
three-parametric central extension \cite{Beisert:2007sk}, and all
three central charges appear in the symmetry algebra of the giant
magnon \cite{David:2008yk}.

Let us see how the residual symmetries arise in the coset
construction of $AdS_3\times S^3\times S^3\times S^1$. The global
symmetry transformations act on the coset representative
(\ref{bdecomposition}) from the left and in general will change the
background. Only those transformations that commute with $D$ and $J$
can be pulled through $g_B=\,{\rm e}\,^{\Xi }\,$ and will act on
$\,{\rm e}\,^{\mathbbm{X}}$. Even these transformation will be
non-linearly realized on $\mathbbm{X}$, with the exception of the
transformations from $\mathfrak{h}_0$, for which one can apply a
compensating right gauge  multiplication such that $\mathbbm{X}$
will transform in the adjoint. Thus the elements of $\mathfrak{h}_0$
which commute with $D$ and $J$ act on the transverse field
$\mathbbm{X}$ by conjugations and leave the action invariant. There
are three $U(1)$ charges that satisfy this condition:
\begin{equation}\label{u1s}
 q_1=S^L_0+S^R_0,\qquad q_2=L^L_5+L^R_5,\qquad q_3=R_8^L+R^R_8.
\end{equation}

However, this is not the end of the story. By analogy with
$AdS_5\times S^5$ and $AdS_4\times CP^3$, we may expect that the
supercharges that commute with $D-J$ will be also preserved,
although their algebra, as well as their action on the transverse
fields, can be deformed by the gauge fixing. The supercharges that
commute with $D-J$ are $Q^{L,R}_{\pm\pm\pm}$. They form two copies
of the $\mathfrak{psu}(1|1)$ algebra (which is just the 2d Clifford
algebra). We expect that the algebra gets centrally extended by the
mechanism described in \cite{Beisert:2005tm,Arutyunov:2006ak}.

The quadratic action for fluctuations around the BMN geodesic can be
read off from equation~(\ref{quadaction}). In particular, the mass-squared
operator for the BMN modes is
\begin{equation}\label{}
 \mathcal{M}^2=\mathop{\mathrm{ad}}{}^2(D+J).
\end{equation}
Given the tensor product structure of $D(2,1;\alpha )\times
D(2,1;\alpha )$, the mass spectrum can be expressed in terms of the
$U(1)$ charges (\ref{u1s}):
\begin{equation}\label{}
 M=|q_1+\cos^2\phi \, q_2+\sin^2\phi \,q_3|.
\end{equation}
The mass takes four possible values $0$, $1$, $\cos^2\phi $ or
$\sin^2\phi $, in agreement with the bosonic spectrum found in
\cite{Gukov:2004ym} and the full spectrum at the special value of
$\phi =\pi /4$  \cite{Lu:2002kw}. Each mass level contains four
states: two bosons and two fermions. This is the dimension  of the
bifundamental multiplet of $\mathfrak{psu}(1|1)\times
\mathfrak{psu}(1|1)$. The spectrum thus nicely fits into four
bifundamental multiplets of the unbroken $\mathfrak{psu}(1|1)\times
\mathfrak{psu}(1|1)$.

To write down the Lagrangian for the fluctuation modes, we can
choose the following parameterization of the coset element in
(\ref{bdecomposition}):
\begin{eqnarray}\label{xtransverse}
   \mathbbm{X}_1&=&\chi ^{a\alpha \dot{a}}\left(\tilde{Q}^L_{a\alpha \dot{a}}
   -i\tilde{Q}^R_{a\alpha \dot{a}}\right)
   \nonumber \\
 \mathbbm{X}_2&=&\frac{1}{\sqrt{2}}\,X\left(S_1^L+iS_2^L-S_1^R-iS_2^R\right)
  +\frac{1}{\sqrt{2}}\,\bar{X}\left(S_1^L-iS_2^L-S_1^R+iS_2^R\right)
 \nonumber \\ &&
  +\frac{\cos\phi }{\sqrt{2}}\,Y\left(L_3^L+iL_4^L-L_3^R-iL_4^R\right)
   +\frac{\cos\phi }{\sqrt{2}}\,\bar{Y}\left(L_3^L-iL_4^L-L_3^R+iL_4^R\right)
   \nonumber \\ &&
   +\frac{\sin\phi }{\sqrt{2}}\,Z\left(R_6^L+iR_7^L-R_6^R-iR_7^R\right)
   +\frac{\sin\phi }{\sqrt{2}}\,\bar{Z}\left(R_6^L-iR_7^L-R_6^R+iR_7^R\right)
   \nonumber \\ &&
   \sin\phi \,\cos\phi \,V\left(L_5^L-R_8^L-L_5^R+R_8^R\right)
   \nonumber \\
 \mathbbm{X}_3&=&\theta  ^{a\alpha \dot{a}}\left(\tilde{Q}^L_{a\alpha \dot{a}}
   +i\tilde{Q}^R_{a\alpha \dot{a}}\right),
\end{eqnarray}
where $\tilde{Q}_{a\alpha \dot{\alpha }}$ are defined in
(\ref{newbasis}). Plugging (\ref{cosetpar}), (\ref{dilgen}),
(\ref{anggen}), and (\ref{xtransverse}) into (\ref{quadaction}), and
adding the $S^1$ mode $U$, we get:
\begin{eqnarray}\label{}
 \mathcal{L}&=&\sqrt{-h}h^{\mathbf{a}\mathbf{b}}
 \left[
  -\frac{1}{2}\left(1+2X\bar{X}\right)\partial _\mathbf{a}t\,\partial _\mathbf{b}t
  +\frac{1}{2}\left(1-2\cos^4\phi \,Y\bar{Y}-2\sin^4\phi \,Z\bar{Z}
  \right)\partial _\mathbf{a}\varphi \,\partial _\mathbf{b}\varphi
  \right. \nonumber \\ && \left.
  +\partial _\mathbf{a}\bar{X}\,\partial _\mathbf{b}X
  +\partial _\mathbf{a}\bar{Y}\,\partial _\mathbf{b}Y
  +\partial _\mathbf{a}\bar{Z}\,\partial _\mathbf{b}Z
  +\frac{1}{2}\,\partial _\mathbf{a}V\,\partial _\mathbf{b}V
  +\frac{1}{2}\,\partial _\mathbf{a}U\,\partial _\mathbf{b}U
 \right]
 \nonumber \\ &&
 +\sum_{a\,\alpha\, \dot{\alpha }\,b\,\beta\, \dot{\beta }=\pm}^{}\varepsilon
 _{ab}\varepsilon _{\alpha \beta }\varepsilon _{\dot{\alpha }\dot{\beta }}
 \left[
  i\left(\sqrt{-h}h^{\mathbf{a}\mathbf{b}}+\varepsilon ^{\mathbf{a}\mathbf{b}}\right)
  \chi ^{a\alpha \dot{\alpha }}M_{\mathbf{a}}^{(b,\beta ,\dot{\beta })}\partial
  _\mathbf{b}\chi ^{b\beta \dot{\beta }}
  \right. \nonumber \\ && \left.
  +i\left(\sqrt{-h}h^{\mathbf{a}\mathbf{b}}-\varepsilon ^{\mathbf{a}\mathbf{b}}\right)
  \theta  ^{a\alpha \dot{\alpha }}M_{\mathbf{a}}^{(b,\beta ,\dot{\beta })}\partial
  _\mathbf{b}\theta  ^{b\beta \dot{\beta }}
  \right. \nonumber \\ && \left.
 -\left(\sqrt{-h}h^{\mathbf{a}\mathbf{b}}+\varepsilon^{\mathbf{a}\mathbf{b}}\right)
 M_{\mathbf{a}}^{(a,\alpha  ,\dot{\alpha })}M_{\mathbf{b}}^{(b,\beta ,\dot{\beta })}
 \chi ^{a\alpha \dot{\alpha }}\theta ^{b\beta \dot{\beta }}
 \right],
\end{eqnarray}
where
\begin{equation}\label{}
 M_{\mathbf{a}}^{(a,\alpha ,\dot{\alpha })}=a\partial _\mathbf{a}t+
 \left(\alpha \cos^2\phi +\dot{\alpha }\sin^2\phi \right)\partial
 _\mathbf{a}\varphi.
\end{equation}

At the quadratic level the light-cone gauge fixing amounts in
replacing the world-sheet metric by $\eta
_{\mathbf{a}\mathbf{b}}=\mathop{\mathrm{diag}}(+-)$, and the
light-cone coordinates by their background values: $\partial
_\mathbf{a}t=\delta _\mathbf{a}^0=\partial _\mathbf{a}\varphi $. The
action can be brought to a nice 2d form by introducing the notations
\begin{equation}\label{}
 X^1=\frac{V+iU}{\sqrt{2}}\,,\qquad X^2=Z,\qquad X^3=Y,\qquad X^4=X,
\end{equation}
and
\begin{equation}\label{}
 \psi _l^I=-4\alpha \dot{\alpha }m_I\theta ^{-\alpha \dot{\alpha }},
 \qquad
 \psi _r^I=-4\alpha \dot{\alpha }m_I\chi ^{-\alpha \dot{\alpha }},
 \qquad
 \bar{\psi }^I_l=\chi ^{+\alpha \dot{\alpha }},
 \qquad
 \bar{\psi }^I_r=\theta ^{+\alpha \dot{\alpha }},
\end{equation}
where $I=(2\alpha +\dot{\alpha }+5)/2$ and
\begin{equation}\label{}
 m_I=(0,\sin^2\phi ,\cos^2\phi ,1).
\end{equation}
In these notations, the light-cone action becomes:
\begin{equation}\label{}
 \mathcal{L}_{\rm l.c.}=\sum_{I=1}^{4}\left(
 \partial _\mathbf{a}\bar{X}^I\,\partial ^\mathbf{a}X^I
 -m_I^2\bar{X}^IX^I
 +i\bar{\psi }^I\rho ^\mathbf{a}\partial _\mathbf{a}\psi^I -m_I\bar{\psi
 }^I\psi ^I
 \right),
\end{equation}
where we have introduced the 2d Dirac spinors
\begin{equation}\label{}
 \psi ^I=
 \begin{pmatrix}
   \psi _l^I   \\
   \psi _r^I  \\
 \end{pmatrix},\qquad
 \bar{\psi }^I =\left(\bar{\psi }_l^I \,\,\,\bar{\psi }_r^I
 \right).
\end{equation}
For the 2d gamma matrices we take:
\begin{equation}\label{}
 \rho  ^\mathbf{a}=(\sigma ^1,i\sigma ^2).
\end{equation}

One can readily expand the action beyond the leading BMN order. For
instance, the cubic interaction terms for the bosonic fields are
\begin{equation}\label{}
 \mathcal{L}^{(3,b)}=2\sin\phi \,\cos\phi \,
 \left(\sin^2\phi \,\bar{Z}Z-\cos^2\phi \,\bar{Y}Y\right)
 \sqrt{-h}h^{\mathbf{a}\mathbf{b}}
 \left(\partial _\mathbf{a}V \,\partial _\mathbf{b}\varphi
 +\frac{1}{3}\,V\,\partial _\mathbf{a}\partial _\mathbf{b}\varphi
 \right),
\end{equation}
which gives after the gauge fixing:
\begin{equation}\label{}
 \mathcal{L}^{(3,b)}_{\rm l.c.}=2\sin\phi \,\cos\phi \,
 \left(\sin^2\phi \,\bar{Z}Z-\cos^2\phi \,\bar{Y}Y\right)\partial
 _0V.
\end{equation}
The interactions involving fermions are more complicated and
although it is straightforward to include them in the near-BMN
expansion, we will not do it here. In appendix~\ref{qrti} we present the
quartic interaction terms for bosons.

 The heaviest mode  ($X^4=X$) lies on the threshold of the
decay $4\rightarrow 2+3$, since the masses satisfy the sum rule
$m_2+m_3=\sin^2\phi +\cos^2\phi =1=m_4$. A similar sum rule holds in
$AdS_4\times CP^3$. There, the massive mode disappears from the
spectrum, once the quantum corrections are taken into account, by
mixing with the continuum of the two-particle states
\cite{Zarembo:2009au,Sundin:2009zu}. We expect that the same
mechanism is at work here and that the heaviest mode should not be
regarded as an elementary excitation.

The appearance of the massless BMN modes is quite unusual in the
$AdS$/CFT context. In the $AdS_5\times S^5$ and $AdS_4\times CP^3$
backgrounds, which in other respects are very similar to
$AdS_3\times S^3\times S^3\times S^1$, the BMN modes are all
massive. As we shall see, the massless modes hinder straightforward
application of integrability methods to $AdS_3\times S^3\times
S^3\times S^1$.  We will return later to the discussion of their
origin and possible implications for the $AdS_3/$CFT${}_2$ duality.

\section{The $AdS_3\times S^3\times T^4$ limit}
\label{sec6}

The coset action we have constructed in  section~\ref{sec2} for Type
IIB string theory on $AdS_3\times S^3\times S^3\times S^1$ has the
relative radii of the two $S^3$'s ($R_\pm$) and the radius of $S^1$
as free parameters (see equation~(\ref{supgraradii})). We can take,
say, $R_-\rightarrow\infty$ and decompactify the $S^3$. The
resulting theory should be Type IIB string theory on $AdS_3\times
S^3\times T^4$, after periodic identifications in the resulting
$\mathbbm{R}^3$. In this section we show explicitly how this
happens.

A coset action  for Type IIB string theory on $AdS_3\times S^3$ had
been written down
in~\cite{Pesando:1998wm,Rahmfeld:1998zn,Park:1998un,Metsaev:2000mv}
and basically follows the Metsaev-Tseytlin construction for
$AdS_5\times S^5$ \cite{Metsaev:1998it}. In the notation of our
paper, this action is based on the $\mathbbm{Z}_4$ automorphism
(\ref{omegad}) of the $\mathfrak{g}=\mathfrak{psu}(1,1|2)\oplus
\mathfrak{psu}(1,1|2)$ superalgebra, which puts the six-dimensional
Type IIB GS action on $AdS_3\times S^3$ into the general framework
of $\mathbbm{Z}_4$ cosets. The coset has 16 fermions, half of which
can be removed by fixing kappa-symmetry, leaving 8 physical degrees of
freedom: the correct number for a six-dimensional GS action.

We propose that the coset action on $AdS_3\times S^3$, when
supplemented with four  free bosons, in fact describes
ten-dimensional Type IIB GS strings on $AdS_3\times S^3\times T^4$
in a suitable (fully fixed) kappa-symmetry gauge. This may appear
puzzling at first, since, as we have just mentioned, the
six-dimensional coset
action~\cite{Rahmfeld:1998zn,Park:1998un,Metsaev:2000mv} has only 8
physical fermions: a factor of two short
of the 16 fermions required in ten dimensions. We are going to argue
that the extra $T^4$ factor in the action changes the number of
physical degrees of freedom in the coset sector. At first sight that
seems impossible, since one can always go to the conformal gauge,
where the $T^4$ completely decouples. Were this to be true, the coset
would lack half of the fermion degrees of freedom. However, the
decoupling of the $T^4$ is not complete, since the four bosons of
$T^4$ do interact with the coset fermions through the 2d metric
coupling or, in the conformal gauge, through the Virasoro
constraints. The key point here is that the metric transforms
non-trivially under kappa symmetry. The metric couplings of the
extra bosons violate the kappa symmetry of the action and, in
effect, keep all 16 fermions physical. Put differently, the addition
of the extra free bosons modifies the Virasoro constraints for the
model; a consequence of this modification is that kappa-symmetry of
the six-dimensional action is {\em not} a symmetry of the
ten-dimensional action. As a result the coset + $T^4$ model has more
fermions that just the coset\footnote{This is perhaps a known
mechanism, although we could not find it anywhere in the
literature.}.

We first discuss the $R_-\rightarrow\infty$ limit  for the
$\mathfrak{d}(2,1;\alpha)$ super-algebra. To reintroduce the
dependence on $R_-$ we re-scale $R_{\dot{m}}\rightarrow R_-
R_{\dot{m}}$. We should also take $\phi\rightarrow 0$ ({\em c.f.}
equations~(\ref{radiiphi})). The relevant (anti)-commutators then
reduce to
\begin{eqnarray}
[R_{\dot{m}} ,R_{\dot{n}} ]&=&0 \nonumber \\
 \{Q_{a\alpha \dot{\alpha }},Q_{b\beta \dot{\beta }}\}
 &=&i( \varepsilon \gamma ^\mu)_{ab}\,\varepsilon _{\alpha \beta }\,\varepsilon
 _{\dot{\alpha }\dot{\beta }}\,S_\mu
 -\varepsilon _{ab}\,(\varepsilon \gamma ^m)_{\alpha \beta }\,\varepsilon
 _{\dot{\alpha }\dot{\beta }}\,L_m\,.
\end{eqnarray}
In the limit we also find that the $AdS_3$ and $S^3$ radii are equal
$R_+=l$ ({\em c.f.} equation~(\ref{supgraradii})). In fact, it is
easy to convince oneself that the algebra is now
$\mathfrak{psu}(1,1|2)$ together with three commuting generators
$R_{\dot{m}}$. This is a well known property of
$\mathfrak{d}(2,1;\alpha)$ for $\alpha=1$ (see for
example~\cite{Frappat:1996pb}).

Our coset action  is based on a $\mathbbm{Z}_4$ automorphism of
$\mathfrak{d}(2,1;\alpha)\oplus\mathfrak{d}(2,1;\alpha)$ constructed
in section~\ref{sec2}. In the $R_-\rightarrow\infty$ limit this
automorphism reduces to the $\mathbbm{Z}_4$ automorphism for
$\mathfrak{psu}(1,1|2)\oplus \mathfrak{psu}(1,1|2)$ which is used to
construct the coset action on $AdS_3\times S^3$. Taking into account
the extra commuting generators $R_{\dot{m}}$, we then find that in
the $R_-\rightarrow\infty$ limit the Type IIB coset action on
$AdS_3\times S^3\times S^3\times S^1$ reduces to the coset action
for $AdS_3\times S^3$ together with four free bosons for $T^4$.
Since the original action for $AdS_3\times S^3\times S^3\times S^1$
has the interpretation of a Green-Schwarz action in the
kappa-gauge~(\ref{kappafix}), we conclude that the coset action for
$AdS_3\times S^3$ together with four free bosons for $T^4$ is a
Green-Schwarz action for $AdS_3\times S^3\times T^4$ in the kappa
gauge
\begin{equation}
\Gamma^{012345}\theta^I=\theta^I\,.
\end{equation}

\section{Integrability}
\label{sec7}

As we have shown in the previous sections, string theory on the
$AdS_3\times S^3\times S^3\times S^1$ background is described by an
integrable sigma-model. Its classical equations of motion can be
solved in a quite general form by the finite-gap integration method
\cite{Novikov:1984id}. The finite-gap method basically performs the
separation of variables, always possible in an integrable system. It
thus replaces the oscillator expansion in the flat space and serves
as a first step towards string quantization via Bethe ansatz. For
the $AdS_5\times S^5$
\cite{Kazakov:2004qf,Kazakov:2004nh,Beisert:2004ag,SchaferNameki:2004ik,Beisert:2005bm,Dorey:2006zj,Dorey:2006mx,Zarembo:2004hp}
and $AdS_4\times CP^3$ \cite{Gromov:2008bz} backgrounds the
finite-gap method yields a set of coupled integral equations, which
on the one hand parameterize possible classical solutions of the
sigma-model and on the other hand can be regarded as the classical
limit of the Bethe equations for the quantum spectrum of the string.
We first describe the general scheme of finite-gap integration, as
applied to the $\mathbbm{Z}_4$ cosets, and then specify the general
construction to the case at hand, the sigma model on $AdS_3\times
S^3\times S^3\times S^1$.

\subsection{The General Scheme}

The general construction mostly follows the derivation of the
finite-gap equations for $AdS_5\times S^5$ \cite{Beisert:2005bm},
but we will use more invariant group-theoretic language, and unlike
in all previous studies will not rely on an explicit supermatrix
representation of the Lax connection\footnote{Different matrix
representations give the same algebraic curve and the same set of
integral equations \cite{Beisert:2004ag}. The group-theoretic
construction makes this equivalence manifest.}.

As usual in integrable systems, instead of solving the non-linear
equations of motion one can study the linear problem for the Lax
operator (\ref{lax}). The fundamental solution of the linear problem
is the monodromy matrix  of the Lax connection\cite{Faddeev:1987ph}:
\begin{equation}\label{}
 \mathcal{M}({\tt x})={\rm
 P}\exp\oint_{C_{x_*x_*}}dx^\mathbf{a}L_\mathbf{a}(x;{\tt x}).
\end{equation}
The contour of integration $C_{x_*x_*}$ is a closed curve that links
the worldsheet\footnote{We discuss only the tree level of string
theory, and assume that the worldsheet has the topology of a
cylinder.}, but is otherwise arbitrary. The canonical choice is the
equal time section $x^0=x_*^0$. Because of the flatness condition
(\ref{flatness}) the monodromy matrix does not change under
continuous deformations of the contour.

The monodromy matrix is gauge-dependent and also depends on the base
point $x_*$. Under the gauge transformations the monodromy matrix
gets conjugated by an element of $H_0$: $\mathcal{M}\rightarrow
h^{-1}(x_*)\mathcal{M}h(x_*)$, $h(x_*)\in H_0$. If the base point is
shifted to $x_*'$, the monodromy matrix is conjugated by the
monodromy along the curve connecting $x_*$ and $x_*'$:
$\mathcal{M}\rightarrow U^{-1}\mathcal{M}U$, $U\equiv U(\Gamma
_{x_*x_*'})\in G$. Both transformation leave the monodromy matrix in
the same conjugacy class. Therefore, the conjugacy class of the
monodromy matrix is gauge invariant and independent of the base
point (time-independent).

Since the conjugacy class of $\mathcal{M}({\tt x})$ is
time-indepen\-dent for arbitrary spectral parameter $\x$, it
generates an infinite number of integrals of motion. The set of
conjugacy classes is isomorphic to the maximal torus of $G$ modulo
Weyl group, and by choosing a Cartan basis $H_l$, the monodromy
matrix can be locally represented as
$$
 \mathcal{M}({\tt x})=U^{-1}({\tt x})\exp\left(p_l({\tt x})H_l)
 \right)U({\tt x}).
$$
The quasi-momenta $p_l({\tt x})$ are the gauge-invariant generating
functions for the integrals of motion. They are defined up to
transformations from the Weyl group and shifts by integer multiples
of $2\pi $.

The monodromy matrix $\mathcal{M}({\tt x})$ is a meromorphic
function of the spectral parameter ${\tt x}$ whose only possible
singularities are located at ${\tt x}=\pm 1$. The nature of these
singularities will be discussed later. On the contrary, the
quasi-momenta $p({\tt x})$ in general are multi-valued functions of
${\tt x}$, defined up to Weyl transformations, and can have branch
points with the monodromy in the Weyl group. For simplicity, we only
consider the case when the monodromies are elementary Weyl
reflections (including generalized Weyl reflections specific to
supergroups \cite{Dobrev:1985qz,PenkovSerganova}). An arbitrary
element of the Weyl group is a product of Weyl reflections. In the
Bethe-ansatz language branch points with composite monodromies
correspond to stacks of Bethe roots
\cite{Beisert:2005di,Gromov:2007ky} describing composite quantum
states of several elementary excitations.

The Weyl reflection with respect to the $l$th root acts on the $l$th
quasi-momentum as $ p_l({\tt x})\rightarrow p_l({\tt
x})-A_{lm}p_m({\tt x})$, where $A_{lm}$ is the Cartan matrix of
$\mathfrak{g}$. We denote by $\{{\tt a}_{l,i}\}$ the set of branch
points of the quasi-momentum $p_l(\x)$ and by $\{C_{l,i}\}$ the set
of cuts that connect these branch points pairwise. The precise
nature of the singularity of $p(\x)$ at ${\tt a}_{l,i}$ is
determined by the monodromy
\begin{equation}\label{monodr}
 p_l({\tt x})\rightarrow p_l({\tt
x})-A_{lm}p_m({\tt x})+2\pi n_{l,i},
\end{equation}
and depends on whether the $l$th root of the superalgebra is bosonic
or fermionic. For the fermionic root, $A_{ll}=0$, and the
quasi-momentum  shifts by a (known) function which is analytic at
$\x={\tt a}_{l,i}$: $p_l\rightarrow p_l+\ldots $. Consequently, the
quasi-momentum $p_l(\x)$ has a logarithmic singularity. For the
bosonic root $A_{ll}=2$ and, in addition to the shift by an analytic
function, the quasi-momentum changes sign: $p_l\rightarrow
-p_l+\ldots $. This means that $p_l(\x)$ has a square root branch
point at $\x={\tt a}_{l,i}$.

In conclusion, the quasi-momenta are meromorphic functions on the
complex plane with cuts $C_{l,i}$, and can have logarithmic
(fermionic) or square root (bosonic) branch points. For the bosonic,
square-root cuts, the monodromy condition (\ref{monodr}) is
equivalent to an equation for the continuous part of the
quasi-momentum across the cut:
\begin{equation}\label{babybethe}
 A_{lm}/\!\!\!p_m({\tt x})=2\pi  n_{l,i},\qquad {\tt x}\in C_{l,i},
\end{equation}
where we define:
\begin{equation}\label{}
 /\!\!\!p_l({\tt x})=\frac{1}{2}\left(p_l({\tt x}+i0 )
 +p_l({\tt x}-i0 )\right).
\end{equation}
The same equation holds at the end-points of the fermionic cuts, in
which case $p_l(\x)$ drops out of the equation.

In addition to the branch cuts, $p_l$ has simple poles at ${\tt
x}=\pm 1$, where the Lax connection (\ref{lax}) itself has a
singularity:
\begin{equation}\label{}
 L_\mathbf{a}=\frac{1}{2\sqrt{-h}}\left(\sqrt{-h}h_{\mathbf{a}\mathbf{b}}\pm
 \varepsilon _{\mathbf{a}\mathbf{b}}\right)J_2^\mathbf{b}\,\frac{1}{{\tt x}\pm 1}
 +\ldots \qquad (\x\rightarrow \mp 1) .
\end{equation}
Hence,
\begin{equation}\label{}
 p_l({\tt x})=\frac{1}{2}\,\,\frac{\kappa _l\mp 2\pi m_l}{{\tt x}\pm 1}+\ldots
 \qquad  (\x\rightarrow \mp 1).
\end{equation}
Parameterization of the residues at $\x=1$ and $\x=-1$ by their sum
and difference is a matter of convenience.

The quasi-momenta carry all  the information about the conserved
quantities in the sigma-model. In particular, their asymptotic
behavior at infinity determines the global symmetry charges. Indeed,
\begin{equation}\label{}
 L_\mathbf{a}=g^{-1}\left(\partial _\mathbf{a}+\frac{1}{\tt x}\,\,
 \frac{2}{h}\,\varepsilon _{\mathbf{a}\mathbf{b}}k^\mathbf{b}\right)g+\ldots
 \qquad ({\tt x}\rightarrow \infty ),
\end{equation}
where $k^\mathbf{a}$ is the global symmetry current (\ref{globch}).
Thus,
\begin{equation}\label{}
 \mathcal{M}({\tt x})=1+\frac{2}{\tt x}\oint
 dx^\mathbf{a}\,\,\frac{1}{h}\,\varepsilon
 _{\mathbf{a}\mathbf{b}}k^{\mathbf{b}}+\ldots \qquad ({\tt x}\rightarrow \infty ),
\end{equation}
and
\begin{equation}\label{}
 p_l({\tt x})=-\frac{2}{\tt x}\,Q_l \qquad ({\tt x}\rightarrow \infty ).
\end{equation}
Further coefficients of the Taylor expansion constitute an infinite
set of conserved charges responsible for integrability of the model.
Using quasi-momenta, one can also build the canonical set of action
variables \cite{Dorey:2006mx,Vicedo:2008jy}.

The information contained in $p_l({\tt x})$ is actually redundant
because of the $\mathbbm{Z}_4$ symmetry. The symmetry acts on the
flat connection according to (\ref{z4action}), and it is not hard to
see that
\begin{equation}\label{}
 \Omega (L_\mathbf{a}({\tt x}))=L_\mathbf{a}(1/{\tt x}).
\end{equation}
The action of the $\Omega $ on the Lie algebra elements, such as the
Lax connection $L_\mathbf{a}(\x)$, can be lifted to the group action
with the help of the exponential map. Thus,
\begin{equation}\label{}
 \Omega (\mathcal{M}(\x))=\mathcal{M}(1/\x).
\end{equation}
Likewise, $\Omega $ acts on the the maximal torus, albeit the
$\mathbbm{Z}_4$ action is then defined up to the Weyl reflections.
Given the $\mathbbm{Z}_4$ action on the Cartan generators:
\begin{equation}\label{}
 \Omega (H_l)=H_mS_{ml},
\end{equation}
one can infer the transformation properties of the quasi-momenta
under the inversion in the spectral-parameter plane:
\begin{equation}\label{inversion}
 p_l(1/{\tt x})=S_{lm}p_m({\tt x}).
\end{equation}
In consequence, the knowledge of the quasi-momenta in the physical
region $|{\tt x}|>1$ is sufficient to reconstruct them everywhere in
the complex plane. In particular, all the branch points of the
quasi-momenta are invariant under inversion. If ${\tt a}_{l,i}$ is a
branch point of $p_l(\x)$, (possibly) other quasi-momentum will have
a branch point at $1/{\tt a}_{l,i}$.

A meromorphic function with the properties listed above can be
reconstructed from the discontinuities at its cuts, which for the
bosonic quasi-momenta\footnote{Calling the quasi-momenta bosonic or
fermionic is a slight abuse of terminology, because all $p_l({\tt
x})$ are even elements of the Grassmann algebra. We call the
quasi-momentum $p_l$ fermionic if the $l$th node of the Dynkin
diagram is fermionic.} we denote by $2\pi i\rho _l({\tt x})$. For
fermionic cuts, the monodromy condition (\ref{monodr}) completely
determines the discontinuity and there is no further freedom. The
quasi-momenta thus admit the spectral representation:
\begin{equation}\label{}
 p_l({\tt x})=-\frac{\kappa _l{\tt x}+2\pi m_l}{{\tt x}^2-1}
 +\int_{C_l}^{}d{\tt y}\,\frac{\rho _l({\tt y})}{{\tt x}-{\tt y}}
 +\int_{1/C_l}^{}d{\tt y}\,\frac{\tilde{\rho }_l({\tt y})}{{\tt x}-{\tt
 y}}\,,
\end{equation}
where $C_l$ denotes the collection of cuts in the physical domain
$|{\tt x}|>1$ on which $p _l({\tt x})$ has a discontinuity. Since
the quasi-momenta inside the unit circle can be reconstructed by
inversion, we chose to treat separately the cuts at $|\x|>1$ and
their images under $\x\rightarrow 1/\x$.

The inversion symmetry (\ref{inversion}) determines the densities
$\tilde{\rho }_l({\tt x})$ in terms of $\rho_k (1/{\tt x})$, and
imposes certain constraints on $\kappa _l$ and $m_l$. We find that
$m_l$'s must be integers\footnote{Under ${\tt x}\rightarrow 1/{\tt
x}$, the pole part of $p_l$ changes sign and shifts by $-2\pi m_l$,
provided that the equations below are satisfied. If $m_l$'s are
integers, the shift has no physical significance, since the
quasi-momentum is defined up to an integer multiple of $2\pi $.
Typically, the integers $m_l$ have the meaning of the winding
numbers \cite{Kazakov:2004nh}.} and that $\kappa _l$ and $m_l$
satisfy\footnote{Since $\Omega^2=(-1)^F $ and $H_l$ are
Grassmann-even, the matrix $S_{lm}$ squares to one and has
eigenvalues one or minus one.}
\begin{equation}\label{}
 S_{lk}\kappa _k=-\kappa _l,\qquad S_{lk}m_k=-m_l,
\end{equation}
Finally,
\begin{equation}\label{}
 p_l({\tt x})=-\frac{\kappa _l{\tt x}+2\pi m_l}{{\tt x}^2-1}
 +\int_{}^{}d{\tt y}\,
 \frac{\rho _l({\tt y})}{{\tt x}-{\tt y}}
 -S_{lm}\int_{}^{}\frac{d{\tt y}}{{\tt y}^2}\,\,\frac{\rho_m({\tt y}) }{{\tt x}-\frac{1}{\tt
 y}}\,.
\end{equation}
The integration contours lie entirely outside the unit circle. The
condition (\ref{babybethe}) becomes a set of integral equations for
the densities:
\begin{equation}\label{classbethe}
 A_{lm}-\!\!\!\!\!\!\!\!\;\int_{}^{}d{\tt y}\,\frac{\rho _m({\tt y})}{{\tt x}-{\tt y}}
 -A_{lk}S_{km}\int_{}^{}\frac{d{\tt y}}{{\tt y}^2}\,\,
 \frac{\rho_m({\tt y}) }{{\tt x}-\frac{1}{\tt
 y}}=A_{lk}\,\frac{\kappa _k{\tt x}+2\pi m_k}{{\tt
 x}^2-1}+2\pi n_{l,i},\qquad {\tt x}\in C_{l,i}.
\end{equation}
For bosonic nodes of the Dynkin diagram, these equations hold on the
cuts of the quasi-momenta. For the fermionic nodes, they hold at the
positions of singularities. Solutions of these equations describe
quasi-periodic solutions of the equations of motion in the
sigma-model. The conserved charges for a given  solution can be
computed by expanding the quasi-momenta at infinity. The classical
finite-gap equations have a direct quantum counterpart, the Bethe
equations for the quantum spectrum of the sigma-model.

\subsection{Classical Bethe equations for $D(2,1;\alpha )\times D(2,1;\alpha )$}

The symmetrized Cartan matrix of\footnote{In this section, we denote
quantities related to the left (right) $\mathfrak{d}(2,1;\alpha )$
by $+$($-$).} $D_+(2,1;\alpha )\times D_-(2,1;\alpha )$ is
\begin{equation}\label{Cartand21}
 A=
 \begin{pmatrix}
   4\sin^2\phi   & -2\sin^2\phi  & 0   \\
   -2\sin^2\phi   & 0 & -2\cos^2\phi   \\
   0 & -2\cos^2\phi  & 4\cos^2\phi
 \end{pmatrix}\otimes\mathbbm{1}.
\end{equation}
The second factor in the tensor product acts on the $\pm$ indices.
The $\mathbbm{Z}_4$ symmetry operator acts on the Cartan generators
simply by permutation of the two $D(2,1;\alpha )$ factors:
\begin{equation}\label{}
 S=\mathbbm{1}\otimes \sigma ^1.
\end{equation}
The vector $\kappa _l$ can be found by evaluating the monodromy
matrix on the vacuum solution (\ref{cosetpar}) and calculating the
residue of the quasi-momenta at ${\tt x}=\pm 1$:
\begin{equation}\label{}
 \kappa =2\pi \mathcal{E}
 \begin{pmatrix}
   0  \\
   1 \\
   0 \\
 \end{pmatrix}
 \otimes
 \begin{pmatrix}
     -1 \\
     1  \\
 \end{pmatrix}.
\end{equation}
Here, $\mathcal{E}=D/\sqrt{\lambda }$ is the ratio of the energy of
the string to its tension which we denote by\footnote{In the
$AdS_5/$CFT${}_4$ case, $\lambda $ is also the 't~Hooft coupling of
the dual super-Yang-Mills theory. The precise nature of the
parameter $\lambda $ in the CFT${}_2$ dual of the $AdS_3\times
S^3\times S^3\times S^1$ background is  not clear to us.}
$\sqrt{\lambda }/(2\pi )$.

\begin{figure}[t]
\centerline{\includegraphics[width=12cm]{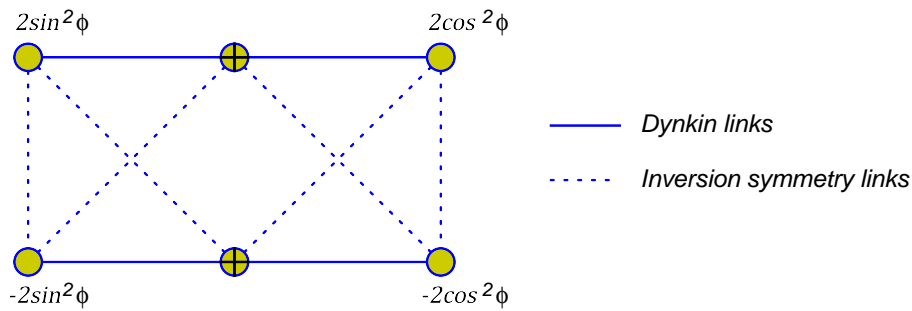}}
\caption{\label{dynkinclass}\small The Dynkin diagram of the
classical Bethe equations.}
\end{figure}

These data leads to the following set of classical Bethe equations:
\begin{eqnarray}\label{clbethe1}
 \pm 4\pi \sin^2\phi \,\frac{\mathcal{E}{\tt x}+m}
 {{\tt x}^2-1}+2\pi n_{1,i}^\pm&=&
 4\sin^2\phi -\!\!\!\!\!\!\!\!\,\int_{}^{}d{\tt y}\,\,\frac{\rho^\pm _1}{{\tt x}-{\tt y}}
 -2\sin^2\phi \int_{}^{}d{\tt y}\,\,\frac{\rho ^\pm_2}{{\tt x} - {\tt y}}
 \nonumber \\
 &&-4\sin^2\phi\int_{}^{}\frac{d{\tt y}}{{\tt y}^2}\,\,\frac{\rho^\mp
 _1}{{\tt x}-\frac{1}{{\tt y}}}
 +2\sin^2\phi \int_{}^{}\frac{d{\tt y}}{{\tt y}^2}\,\,\frac{\rho ^\mp_2}
 {{\tt x} - \frac{1}{{\tt y}}}
 \\ \label{clbethe2}
 2\pi n_{2,i}^\pm&=&
 -2\sin^2\phi \int_{}^{}d{\tt y}\,\,\frac{\rho^\pm
 _1}{{\tt x}-{\tt y}}
 -2\cos^2\phi \int_{}^{}d{\tt y}\,\,\frac{\rho ^\pm_3}{{\tt x} - {\tt y}}
 \nonumber \\
 &&+2\sin^2\phi\int_{}^{}\frac{d{\tt y}}{{\tt y}^2}\,\,\frac{\rho^\mp
 _1}{{\tt x}-\frac{1}{{\tt y}}}
 +2\cos^2\phi \int_{}^{}\frac{d{\tt y}}{{\tt y}^2}\,\,\frac{\rho ^\mp_3}
 {{\tt x} - \frac{1}{{\tt y}}}
 \\ \label{clbethe3}
 \pm 4\pi \cos^2\phi \,\frac{\mathcal{E}{\tt x}+m}
 {{\tt x}^2-1}+2\pi n_{3,i}^\pm&=&
 4\cos^2\phi -\!\!\!\!\!\!\!\!\,\int_{}^{}d{\tt y}\,\,\frac{\rho^\pm _3}{{\tt x}-{\tt y}}
 -2\cos^2\phi \int_{}^{}d{\tt y}\,\,\frac{\rho ^\pm_2}{{\tt x} - {\tt y}}
 \nonumber \\
 &&-4\cos^2\phi\int_{}^{}\frac{d{\tt y}}{{\tt y}^2}\,\,\frac{\rho^\mp
 _3}{{\tt x}-\frac{1}{{\tt y}}}
 +2\cos^2\phi \int_{}^{}\frac{d{\tt y}}{{\tt y}^2}\,\,\frac{\rho ^\mp_2}
 {{\tt x} - \frac{1}{{\tt y}}}\,.
\end{eqnarray}
The integer $m$ is the winding number of the string. The equations
can be summarized in a diagram shown in fig.~\ref{dynkinclass}. The
nodes of the diagram correspond to the densities $\rho _n^\pm$. The
left hand side of the equations is determined by the Dynkin labels,
and the right hand side by the links. The original Dynkin links from
the Cartan matrix (\ref{Cartand21}) determine coefficients in the
first term in (\ref{classbethe}). The second term, which is
associated with the inversion symmetry, produces additional links on
the Dynkin diagram, shown in fig.~\ref{dynkinclass} by broken lines.

The light-cone energy ($\mathcal{E}-\mathcal{J}=(D-J)/\sqrt{\lambda
}$) and the worldsheet momentum of the solution are given by
\begin{eqnarray}\label{}
 \mathcal{P}&=&-2\sum_{s=\pm}^{}s\left(
 \sin^2\phi \int_{}^{}\frac{d{\tt x}}{{\tt x}}\,\,\rho ^s_1({\tt x})
 +\cos^2\phi \int_{}^{}\frac{d{\tt x}}{{\tt x}}\,\,\rho ^s_3({\tt x})
 \right)
  \\
  \label{EminusJ}
 \mathcal{E}-\mathcal{J}
 &=&2\sum_{s=\pm}^{}\left(
 \sin^2\phi \int_{}^{}\frac{d{\tt x}}{{\tt x}^2}\,\,\rho ^s_1({\tt x})
 +\cos^2\phi \int_{}^{}\frac{d{\tt x}}{{\tt x}^2}\,\,\rho ^s_3({\tt x})
 \right).
\end{eqnarray}
Physical states should in addition satisfy the level-matching
condition $\mathcal{P}\in 2\pi \mathbbm{Z}$.

\subsection{BMN limit}

The trivial solution of the finite-gap equations, with zero
densities, describes the BMN vacuum.  The BMN modes correspond to
the vanishingly small cuts whose position is determined by the
no-force condition, the vanishing of the left-hand side the Bethe
equations. For instance, the left hand side of (\ref{clbethe1})
vanishes at
\begin{equation}\label{}
 \frac{1}{{\tt x}_n}=\frac{\mathcal{J}}{n}\left(\sin^2\phi
 -\sqrt{\sin^4\phi +\frac{n^2}{\mathcal{J}^2}}\right),
\end{equation}
where we have set the winding number $m$ to zero and also neglected
the difference between $\mathcal{E}$, which enters the classical
Bethe equations, and $\mathcal{J}$, which plays the role of the
length of the string in the light-cone gauge. This is justified for
small deviations from the BMN vacuum. The set of points $\{\x_n\}$
determines the locus at which short cuts with infinitesimal filling
fractions can emerge. The solution with infinitesimal cuts
corresponds to exciting a number of BMN modes. Their occupation
numbers $N_n$ are proportional to the filling fractions of the cuts:
\begin{equation}\label{}
 S_n=\int_{C_{1,n}}^{}d\x\,\rho (\x).
\end{equation}
The precise relationship between the occupation numbers are the
filling fractions is derived in \cite{Kazakov:2004qf}:
$$
 S_n=\frac{\pi N_n}{\sqrt{\lambda }}\left(1+\sqrt{1+\frac{n^2}{\mathcal{J}^2
 \sin^4\phi }}\right).
$$
The energy then is
$$
 \mathcal{E}-\mathcal{J}=\sum_{n}^{}\frac{2S_n\sin^2\phi }{{\tt x}_n^2}=
 \frac{2\pi}{\sqrt{\lambda }}\sum_{n}^{}N_n\left(\sqrt{\sin^4\phi +\frac{n^2}{\mathcal{J}^2}}-
 \sin^2\phi \right),
$$
Because the length of the string in the light-cone gauge is $2\pi
\mathcal{J}$, the combination $n/\mathcal{J}$ plays the role of the
worldsheet momentum. The spectrum of small fluctuations thus
describes particles with the dispersion relation
\begin{equation}\label{}
 \varepsilon (p)=\sqrt{p^2+\sin^4\phi }\,.
\end{equation}

Similarly, the densities $\rho ^\pm_3({\tt x})$ describe particles
with mass $\cos^2\phi $. The heavy modes are more tricky. They
correspond to stacks \cite{Beisert:2005di,Gromov:2007ky} that cross
from node $1$ to node $3$ through node $2$. The stack, roughly
speaking, is a set of overlapping densities on different nodes. In
this particular case it is a simultaneous solution of a pair of
equations
\begin{equation}\label{}
 \frac{2\sin^2\phi \,\mathcal{J}{\tt x}}{{\tt x}^2-1}=2\pi n,
 \qquad
 \frac{2\cos^2\phi \,\mathcal{J}{\tt x}}{{\tt x}^2-1}=2\pi m.
\end{equation}
The solution is only possible in the thermodynamic limit $n\sim
m\sim \mathcal{J}\rightarrow \infty $, since $n$ and $m$ must
satisfy $n/m=\tan^2\phi$, in which case the stack corresponds to a
particle of mass $1$. The stacks are also responsible for the
correct four-fold degeneracy at each mass level. For instance, the
bosonic members of the $PSU(1|1)\times PSU(1|1)$ multiplet of mass
$\sin^2\phi $ are the single-node solutions for the densities
$\rho_1^+$ and $\rho _1^-$. The fermions in the same multiplet are
the $1^+-2^+$ and $1^--2^-$ stacks.

We have correctly reproduced the massive part of the BMN spectrum.
However, the massless modes, that we also found in section~\ref{sec5},
are completely missing. As we explain below in section~\ref{sec8},
the finite-gap equations do not capture massless modes and describe
only those solutions of the sigma-model in which the massless modes
are not excited. Although this makes our analysis incomplete, we
will proceed with quantization of the classical Bethe equations
obtained above.

\subsection{Quantum Bethe equations}\label{QBEsec}

In the previously studied cases of $AdS_5\times S^5$ and
$AdS_4\times CP^3$ the relationship between classical and quantum
Bethe equations follows a regular pattern dictated by the structure
of the Dynkin diagram. At the quantum level, the densities $\rho
_l(\x)$ describe macroscopic distributions of the Bethe roots, the
solutions of the quantum Bethe equations. When the number of roots
is very large and the sums over the roots can be replaced by the
integrals over their densities, the quantum Bethe equations reduce
to the finite-gap integral equations for the densities. By way of
observation, one can notice that the quantum Bethe equations for
$AdS_5\times S^5$ and $AdS_4\times CP^3$ can be inverse engineered
from the finite gap equations by applying a simple set of regular
rules for each element in the Dynkin diagram. The key observation is
that only three distinct structures appear in the finite-gap
equations, namely: (i) the normal Dynkin links; (ii) the inversion
symmetry links between a momentum-carrying node and "wrong"
fermionic nodes (all possible links of type appear in the
equations); (iii) the inversions symmetry links that connect the
momentum-carrying nodes pairwise. It is quite remarkable that the
classical Bethe equations in our case also contain only these three
types of structural elements. We can thus apply the same set of
rules as in $AdS_5/$CFT${}_4$ and $AdS_4/$CFT${}_3$ to discretize
the classical Bethe equations derived above. There is one subtlety
though. The discretization is straightforward only when the elements
of the Cartan matrix are integers, which is not the case for the
$\mathfrak{d}(2,1;\alpha )$ algebra in general. The two exceptions
are $\phi =\pi /4$ (considered here) and $\phi =0$ (discussed
below), and we will restrict our attention to these two special
cases.

When $\phi =\pi /4$, the $\mathfrak{d}(2,1;\alpha )$ superalgebra
coincides with $\mathfrak{osp}(4|2)$ whose Cartan matrix has integer
entries. In quantum theory, the cuts of the classical spectral curve
get discretized and become the arrays of Bethe roots. The asymptotic
Bethe ansatz determines the positions of the roots in the spectral
plane,  ${\tt x}_{l,i}$, through a system of discrete functional
equations (the Bethe equations). The coupling constant of the
sigma-model (playing the role of $\hbar$) $2\pi /\sqrt{\lambda }$
does not appear in the equations explicitly and only enters through
the quantum parameters ${\tt x}^\pm$ defined by the Jukovsky map:
\begin{equation}\label{}
 {\tt x}^\pm+\frac{1}{{\tt x}^\pm}={\tt x }
 +\frac{1}{{\tt x}}\pm\frac{i}{2h(\lambda )}\,.
\end{equation}
The function $h(\lambda )$  cannot determined by integrability
alone, but at strong coupling should behave as
\begin{equation}\label{whatish}
 h(\lambda )\approx \frac{\sqrt{\lambda }}{2\pi }\qquad \left(\lambda \rightarrow \infty
 \right),
\end{equation}
in order to reproduce the correct dispersion relation $\varepsilon
(p)=\sqrt{p^2+1/4}$.

\begin{figure}[t]
\centerline{\includegraphics[width=12cm]{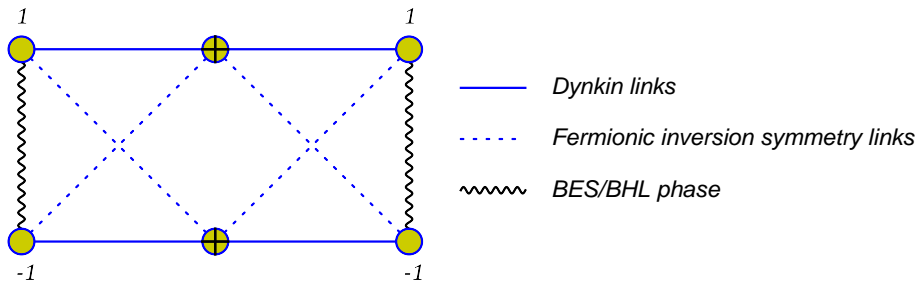}}
\caption{\label{qdynkin}\small The Dynkin diagram of the asymptotic
Bethe ansatz.}
\end{figure}

The Bethe equations are constructed according to the Dynkin diagram
in fig.~\ref{qdynkin}, which is obtained from the classical Dynkin
diagram~\ref{dynkinclass} by (i) assigning ${\tt x}-{\tt x^\pm}$
type interactions to the normal links; (ii) assigning the
$\x-1/\x^\pm$ type interactions to the anomalous fermionic links;
(iii) and finally associating the BES/BHL phase with the anomalous
bosonic links.

The conjectured set of Bethe equations thus reads\footnote{We denote
by $1$, $2$, and $3$ the upper nodes of $OSp_+(4|2)$ and by
$\bar{1}$, $\bar{2}$, and $\bar{3}$ the lower nodes of $OSp_-(4|2)$.
The nodes $1$ and $3$ ($\bar{1}$ and $\bar{3}$) carry momentum $1$
($-1$).}
\begin{eqnarray}\label{}
 \left(\frac{{\tt x}^+_{1,j}}{{\tt x}^-_{1,j}}\right)^L
 &=&\prod_{k\neq j}^{}\frac{{\tt x}^+_{1,j}-{\tt x}^-_{1,k}}{{\tt x}^-_{1,j}-{\tt x}^+_{1,k}}
 \,\,\frac{1-\frac{1}{{\tt x}^+_{1,j}{\tt x}^-_{1,k}}}{1-\frac{1}{{\tt x}^-_{1,j}{\tt x}^+_{1,k}}}
 \,\sigma ^2({\tt x}_{1,j},{\tt x}_{1,k})
 \nonumber \\
 && \times
 \prod_{k}^{}\frac{{\tt x}^-_{1,j}-{\tt x}_{2,k}}{{\tt x}^+_{1,j}-{\tt x}_{2,k}}
 \prod_{k}^{}\frac{1-\frac{1}{{\tt x}^-_{1,j}{\tt x}_{\bar{2},k}}}
 {1-\frac{1}{{\tt x}^+_{1,j}{\tt
 x}_{\bar{2},k}}}
 \prod_{k}^{}\sigma^{-2} ({\tt x}_{1,j},{\tt x}_{\bar{1},k})
 \nonumber \\
 1&=&
 \prod_{k}^{}\frac{{\tt x}_{2,j}-{\tt x}^+_{1,k}}{{\tt x}_{2,j}-{\tt x}^-_{1,k}}
 \prod_{k}^{}\frac{{\tt x}_{2,j}-{\tt x}^+_{3,k}}{{\tt x}_{2,j}-{\tt x}^-_{3,k}}
 \prod_{k}^{}\frac{1-\frac{1}{{\tt x}_{2,j}{\tt x}^+_{\bar{1},k}}}
 {1-\frac{1}{{\tt x}_{2,j}{\tt
 x}^-_{\bar{1},k}}}
 \prod_{k}^{}\frac{1-\frac{1}{{\tt x}_{2,j}{\tt x}^+_{\bar{3},k}}}
 {1-\frac{1}{{\tt x}_{2,j}{\tt
 x}^-_{\bar{3},k}}}
 \nonumber \\
 \left(\frac{{\tt x}^+_{3,j}}{{\tt x}^-_{3,j}}\right)^L
 &=&\prod_{k\neq j}^{}\frac{{\tt x}^+_{3,j}-{\tt x}^-_{3,k}}{{\tt x}^-_{3,j}-{\tt x}^+_{3,k}}
 \,\,\frac{1-\frac{1}{{\tt x}^+_{3,j}{\tt x}^-_{3,k}}}{1-\frac{1}{{\tt x}^-_{3,j}{\tt x}^+_{3,k}}}
 \,\sigma ^2({\tt x}_{3,j},{\tt x}_{3,k})
 \nonumber \\
 && \times
 \prod_{k}^{}\frac{{\tt x}^-_{3,j}-{\tt x}_{2,k}}{{\tt x}^+_{3,j}-{\tt x}_{2,k}}
 \prod_{k}^{}\frac{1-\frac{1}{{\tt x}^-_{3,j}{\tt x}_{\bar{2},k}}}
 {1-\frac{1}{{\tt x}^+_{3,j}{\tt
 x}_{\bar{2},k}}}
 \prod_{k}^{}\sigma^{-2} ({\tt x}_{3,j},{\tt x}_{\bar{3},k})
 \nonumber \\
 \left(\frac{{\tt x}^-_{\bar{1},j}}{{\tt x}^+_{\bar{1},j}}\right)^L
 &=&\prod_{k\neq j}^{}\frac{{\tt x}^+_{\bar{1},j}-{\tt x}^-_{\bar{1},k}}{{\tt x}^-_{\bar{1},j}-{\tt x}^+_{\bar{1},k}}
 \,\,\frac{1-\frac{1}{{\tt x}^+_{\bar{1},j}{\tt x}^-_{\bar{1},k}}}{1-\frac{1}{{\tt x}^-_{\bar{1},j}{\tt x}^+_{\bar{1},k}}}
 \,\sigma ^2({\tt x}_{\bar{1},j},{\tt x}_{\bar{1},k})
 \nonumber \\
 && \times
 \prod_{k}^{}\frac{{\tt x}^-_{\bar{1},j}-{\tt x}_{\bar{2},k}}{{\tt x}^+_{\bar{1},j}-{\tt x}_{\bar{2},k}}
 \prod_{k}^{}\frac{1-\frac{1}{{\tt x}^-_{\bar{1},j}{\tt x}_{2,k}}}
 {1-\frac{1}{{\tt x}^+_{\bar{1},j}{\tt
 x}_{2,k}}}
 \prod_{k}^{}\sigma^{-2} ({\tt x}_{\bar{1},j},{\tt x}_{1,k})
 \nonumber \\
 1&=&
 \prod_{k}^{}\frac{{\tt x}_{\bar{2},j}-{\tt x}^+_{\bar{1},k}}{{\tt x}_{\bar{2},j}-{\tt x}^-_{\bar{1},k}}
 \prod_{k}^{}\frac{{\tt x}_{\bar{2},j}-{\tt x}^+_{\bar{3},k}}{{\tt x}_{\bar{2},j}-{\tt x}^-_{\bar{3},k}}
 \prod_{k}^{}\frac{1-\frac{1}{{\tt x}_{\bar{2},j}{\tt x}^+_{1,k}}}
 {1-\frac{1}{{\tt x}_{\bar{2},j}{\tt
 x}^-_{1,k}}}
 \prod_{k}^{}\frac{1-\frac{1}{{\tt x}_{\bar{2},j}{\tt x}^+_{3,k}}}
 {1-\frac{1}{{\tt x}_{\bar{2},j}{\tt
 x}^-_{3,k}}}
 \nonumber \\
 \left(\frac{{\tt x}^-_{\bar{3},j}}{{\tt x}^+_{\bar{3},j}}\right)^L
 &=&\prod_{k\neq j}^{}\frac{{\tt x}^+_{\bar{3},j}-{\tt x}^-_{\bar{3},k}}{{\tt x}^-_{\bar{3},j}-{\tt x}^+_{\bar{3},k}}
 \,\,\frac{1-\frac{1}{{\tt x}^+_{\bar{3},j}{\tt x}^-_{\bar{3},k}}}{1-\frac{1}{{\tt x}^-_{\bar{3},j}{\tt x}^+_{\bar{3},k}}}
 \,\sigma ^2({\tt x}_{\bar{3},j},{\tt x}_{\bar{3},k})
 \nonumber \\
 && \times
 \prod_{k}^{}\frac{{\tt x}^-_{\bar{3},j}-{\tt x}_{\bar{2},k}}{{\tt x}^+_{\bar{3},j}-{\tt x}_{\bar{2},k}}
 \prod_{k}^{}\frac{1-\frac{1}{{\tt x}^-_{\bar{3},j}{\tt x}_{2,k}}}
 {1-\frac{1}{{\tt x}^+_{\bar{3},j}{\tt
 x}_{2,k}}}
 \prod_{k}^{}\sigma^{-2} ({\tt x}_{\bar{3},j},{\tt x}_{3,k})
\end{eqnarray}
The energy and momentum are given by
\begin{eqnarray}\label{}
 E&=&ih(\lambda )\sum_{l=1,3,\bar{1},\bar{3}}^{}\sum_{j}^{}
 \left(\frac{1}{{\tt x}_{l,j}^+}-\frac{1}{{\tt x}_{l,j}^-}\right)
 \nonumber \\
 \,{\rm
 e}\,^{\frac{2i\pi P}{\sqrt{\lambda }}}&=&\frac{\prod_{l=1,3}^{}}{\prod_{l=\bar{1},\bar{3}}^{}}\,
 \prod_{j}^{}\frac{{\tt x}^+_{l,j}}{{\tt x}^-_{l,j}}\equiv 1.
\end{eqnarray}
The elementary excitation associated to single Bethe roots then have
the dispersion relation
$$
 \varepsilon (p)=\sqrt{4h^2(\lambda )\sin^2\frac{\pi p}{\sqrt{\lambda
 }}+\frac{1}{4}},
$$
which reproduces the dispersion relation of the BMN modes provided
that the function $h(\lambda )$ behaves at strong coupling as
(\ref{whatish}).

The BES/BHL phase \cite{Beisert:2006ib,Beisert:2006ez} admits the
following integral representation \cite{Dorey:2007xn}:
\begin{equation}\label{}
 \sigma ({\tt x},{\tt y})=\exp\left[
 \sum_{r,s=\pm}^{}
 \frac{rs}{4\pi ^2}
 \oint
 \frac{dz\,dw}{\left({\tt x}^r-z\right)\left({\tt y}^s-w\right)}
 \,\,\ln\frac{\Gamma \left(1+ih(\lambda )\left(z+\frac{1}{z}-w-\frac{1}{w}\right)\right)}
 {\Gamma \left(1-ih(\lambda )\left(z+\frac{1}{z}-w-\frac{1}{w}\right)\right)}
 \right],
\end{equation}
where the integration contour is the unit circle: $|z|=1$, $|w|=1$.
From this representation one can readily infer the analytic
structure and the symmetry properties of $\sigma ({\tt x},{\tt y})$
\cite{Arutyunov:2009kf,Volin:2009uv}. At the lowest order in the
strong-coupling expansion, $h\rightarrow \infty $, the dressing
factor reduces to the AFS phase \cite{Arutyunov:2004vx}:
\begin{equation}\label{}
 \sigma ({\tt x},{\tt y})\approx
 \frac{1-\frac{1}{{\tt x}^-{\tt y}^+}}{1-\frac{1}{{\tt x}^+{\tt y}^-}}
 \left[
 \frac{\left(1-\frac{1}{{\tt x}^+{\tt y}^-}
 \right)
 \left(1-\frac{1}{{\tt x}^-{\tt y}^+}
 \right)}{\left(1-\frac{1}{{\tt x}^+{\tt y}^+}
 \right)\left(1-\frac{1}{{\tt x}^-{\tt y}^-}
 \right)}
 \right]^{ih\left({\tt x}+\frac{1}{{\tt x}}-{\tt y}-\frac{1}{{\tt y}}\right)}
 \qquad (h\rightarrow \infty ),
\end{equation}
and this is all one needs to reproduce the classical Bethe
equations, which indeed follow from the quantum equations above in
the $h\rightarrow \infty $ limit, upon the identification
\begin{equation}\label{}
 \rho _l^+({\tt x})=\frac{1}{h}\sum_{j}^{}\frac{{\tt x}_{l,j}^2}{{\tt x}_{l,j}^2-1}
 \,\delta ({\tt x}-{\tt x}_{l,j}),\qquad
 \rho _l^-({\tt x})=\frac{1}{h}\sum_{j}^{}\frac{{\tt x}_{\bar{l},j}^2}{{\tt
 x}_{\bar{l},j}^2-1}
 \,\delta ({\tt x}-{\tt x}_{\bar{l},j}).
\end{equation}

As we discussed above, the classical finite-gap equations describe
only massive modes of the string. The sole reason to quantize this
incomplete set of equations is their striking similarity to the
finite-gap equations in $AdS_5\times S^5$ and $AdS_4\times CP^3$. In
those cases, the resulting quantum Bethe equations describe the full
asymptotic spectrum at any coupling. For sure, this cannot be true
here, because the massless modes are missing already at the
classical level. Our best hope is that the equations above represent
a truncation of the hypothetical complete Bethe ansatz system that
also takes into account the massless modes of the string. In an
integrable system, in which action-angle variables separate, such
truncations are in many cases possible. For instance, the Bethe
equations for the $\mathfrak{su}(2)$ sector of $AdS_5/$CFT${}_4$
\cite{Arutyunov:2004vx} can be reconstructed from the finite-gap
equations on $S^3\times R^1$ \cite{Kazakov:2004qf}, which ignores
most part of the string modes in $AdS_5\times S^5$. At the moment we
cannot justify that ignoring the massless modes we get a consistent
truncation of the Bethe equations in $AdS_3/$CFT${}_2$. It is also
possible that the inclusion of massless modes modifies the equations
in a more essential way than just adding extra nodes into the Dynkin
diagram.

\subsection{$AdS_3\times S^3\times T^4$}

\begin{figure}[t]
\centerline{\includegraphics[width=12cm]{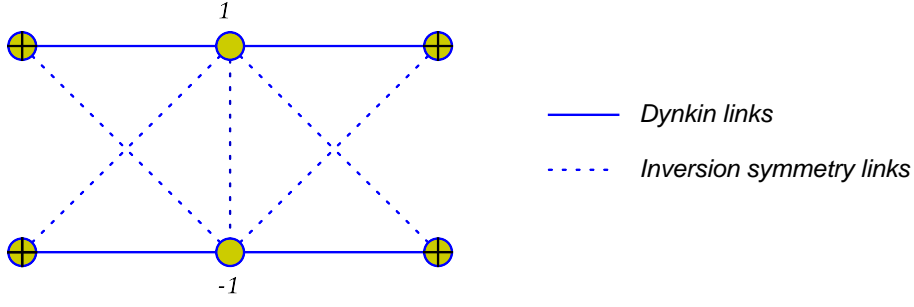}}
\caption{\label{dps}\small The Dynkin diagram for the
$PSU(1,1|2)\times PSU(1,1|2)/SU(1,1)\times SU(2)$ coset.}
\end{figure}

We can apply the same general formalism to the coset part of the
$AdS_3\times S^3\times T^4$ background, which is the sigma-model on
$PSU(1,1|2)\times PSU(1,1|2)/SU(1,1)\times SU(2)$. The Cartan matrix
of $PSU(1,1|2)\times PSU(1,1|2)$ is
\begin{equation}\label{Cartanpsu22}
 A=
 \begin{pmatrix}
   0   & -1  & 0   \\
   -1 & 2 & -1   \\
   0 & -1  & 0
 \end{pmatrix}\otimes\mathbbm{1}.
\end{equation}
The $\mathbbm{Z}_4$ symmetry operator again is the permutation:
\begin{equation}\label{}
 S=\mathbbm{1}\otimes \sigma ^1.
\end{equation}
From these data we infer the classical Bethe equations
(fig.~\ref{dps}):
\begin{eqnarray}\label{clbethepsu22_1}
 2\pi n_{1,i}^\pm&=&
 - \int_{}^{}d{\tt y}\,\,\frac{\rho^\pm _2}{{\tt x}-{\tt y}}
 +\int_{}^{}\frac{d{\tt y}}{{\tt y}^2}\,\,\frac{\rho^\mp
 _2}{{\tt x}-\frac{1}{{\tt y}}}
 \\ \label{clbethe2}
 \pm 4\pi \,\frac{\mathcal{E}{\tt x}+m}
 {{\tt x}^2-1}+2\pi n_{2,i}^\pm
 &=&
 - \int_{}^{}d{\tt y}\,\,\frac{\rho^\pm_1}{{\tt x}-{\tt y}}
 +2-\!\!\!\!\!\!\!\!\,\int_{}^{}d{\tt y}\,\,\frac{\rho^\pm _2}{{\tt x}-{\tt y}}
 - \int_{}^{}d{\tt y}\,\,\frac{\rho^\pm_3}{{\tt x}-{\tt y}}
 \nonumber \\ &&
 +\int_{}^{}\frac{d{\tt y}}{{\tt y}^2}\,\,\frac{\rho^\mp
 _1}{{\tt x}-\frac{1}{{\tt y}}}
 -2\int_{}^{}\frac{d{\tt y}}{{\tt y}^2}\,\,\frac{\rho^\mp
 _2}{{\tt x}-\frac{1}{{\tt y}}}
 +\int_{}^{}\frac{d{\tt y}}{{\tt y}^2}\,\,\frac{\rho^\mp
 _3}{{\tt x}-\frac{1}{{\tt y}}}
 \\
 2\pi n_{3,i}^\pm&=&
 - \int_{}^{}d{\tt y}\,\,\frac{\rho^\pm _2}{{\tt x}-{\tt y}}
 +\int_{}^{}\frac{d{\tt y}}{{\tt y}^2}\,\,\frac{\rho^\mp
 _2}{{\tt x}-\frac{1}{{\tt y}}}
\end{eqnarray}

\begin{figure}[t]
\centerline{\includegraphics[width=12cm]{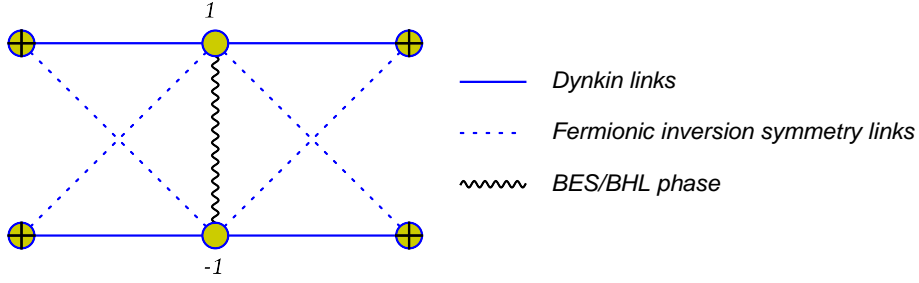}}
\caption{\label{dpsq}\small The Dynkin diagram for the quantum Bethe
equations.}
\end{figure}

The quantum Bethe equations can be reconstructed by applying the
same set of rules as before (fig.~\ref{dpsq}):
\begin{eqnarray}\label{t4bethe}
 1&=&\prod_{k}^{}\frac{\x_{1,j}-\x^+_{2,k}}{\x_{1,j}-\x^-_{2,k}}
 \prod_{k}^{}\frac{1-\frac{1}{\x_{1,j}\x^+_{\bar{2},k}}}
 {1-\frac{1}{\x_{1,j}\x^-_{\bar{2},k}}}
 \nonumber \\
 \left(\frac{\x^+_{2,j}}{\x^-_{2,j}}\right)^L&=&
 \prod_{k\neq
 j}^{}\frac{\x^+_{2,j}-\x^-_{2,k}}{\x^-_{2,j}-\x^+_{2,k}}\,\,
 \frac{1-\frac{1}{\x^+_{2,j}\x^-_{2,k}}}{1-\frac{1}{\x^-_{2,j}\x^+_{2,k}}}\,
 \sigma ^2(\x_{2,j},\x_{2,k})
 \prod_{k}^{}\frac{\x^-_{2,j}-\x_{1,k}}{\x^+_{2,j}-\x_{1,k}}
 \prod_{k}^{}\frac{\x^-_{2,j}-\x_{3,k}}{\x^+_{2,j}-\x_{3,k}}
 \nonumber \\ && \times
 \prod_{k}^{}\frac{1-\frac{1}{\x^-_{2,j}\x_{\bar{1},k}}}{1-\frac{1}{\x^+_{2,j}\x_{\bar{1},k}}}
 \prod_{k}^{}\frac{1-\frac{1}{\x^-_{2,j}\x_{\bar{3},k}}}{1-\frac{1}{\x^+_{2,j}\x_{\bar{3},k}}}
 \prod_{k}^{}\sigma ^{-2}(\x_{2,j},\x_{\bar{2},k})
 \nonumber \\
 1&=&\prod_{k}^{}\frac{\x_{3,j}-\x^+_{2,k}}{\x_{3,j}-\x^-_{2,k}}
 \prod_{k}^{}\frac{1-\frac{1}{\x_{3,j}\x^+_{\bar{2},k}}}
 {1-\frac{1}{\x_{3,j}\x^-_{\bar{2},k}}}
 \nonumber \\
 1&=&\prod_{k}^{}\frac{\x_{\bar{1},j}-\x^+_{\bar{2},k}}{\x_{\bar{1},j}-\x^-_{\bar{2},k}}
 \prod_{k}^{}\frac{1-\frac{1}{\x_{\bar{1},j}\x^+_{{2},k}}}
 {1-\frac{1}{\x_{\bar{1},j}\x^-_{{2},k}}}
 \nonumber \\
 \left(\frac{\x^+_{\bar{2},j}}{\x^-_{\bar{2},j}}\right)^L&=&
 \prod_{k\neq
 j}^{}\frac{\x^+_{\bar{2},j}-\x^-_{\bar{2},k}}{\x^-_{\bar{2},j}-\x^+_{\bar{2},k}}\,\,
 \frac{1-\frac{1}{\x^+_{\bar{2},j}\x^-_{\bar{2},k}}}{1-\frac{1}{\x^-_{\bar{2},j}\x^+_{\bar{2},k}}}\,
 \sigma ^2(\x_{\bar{2},j},\x_{\bar{2},k})
 \prod_{k}^{}\frac{\x^-_{\bar{2},j}-\x_{\bar{1},k}}{\x^+_{\bar{2},j}-\x_{\bar{2},k}}
 \prod_{k}^{}\frac{\x^-_{\bar{2},j}-\x_{\bar{3},k}}{\x^+_{\bar{2},j}-\x_{\bar{3},k}}
 \nonumber \\ && \times
 \prod_{k}^{}\frac{1-\frac{1}{\x^-_{\bar{2},j}\x_{{1},k}}}{1-\frac{1}{\x^+_{\bar{2},j}\x_{{1},k}}}
 \prod_{k}^{}\frac{1-\frac{1}{\x^-_{\bar{2},j}\x_{{3},k}}}{1-\frac{1}{\x^+_{\bar{2},j}\x_{{3},k}}}
 \prod_{k}^{}\sigma ^{-2}(\x_{\bar{2},j},\x_{{2},k})
 \nonumber \\
 1&=&\prod_{k}^{}\frac{\x_{\bar{3},j}-\x^+_{\bar{2},k}}{\x_{\bar{3},j}-\x^-_{\bar{2},k}}
 \prod_{k}^{}\frac{1-\frac{1}{\x_{\bar{3},j}\x^+_{{2},k}}}
 {1-\frac{1}{\x_{\bar{3},j}\x^-_{{2},k}}}\,.
\end{eqnarray}
Again it is not too hard to show that these equations describe
$4_B+4_F$ modes with mass $1$, and do not capture $4_B+4_F$ massless
modes. So the remarks at the end of sec.~\ref{QBEsec} apply to this
case as well.

\section{Outlook: Some comments on the massless modes}
\label{sec8}
\setcounter{equation}{0}

A new feature of the $AdS_3/$CFT${}_2$ dual theories,  as compared
with the $AdS_5/$CFT${}_4$ and $AdS_4/$CFT${}_3$ cases, is the
presence of massless modes. We have encountered these massless modes
as appearing either from the free boson(s) on $S^1$ or $T^4$, or
from an extra massless mode in the BMN limit of the $AdS_3\times
S^3\times S^3\times S^1$ theory; there are also fermionic massless
modes related to the bosonic modes by supersymmetry. The
integrability techniques we have used throughout this paper have
little to say about the massless degrees of freedom.

It is not difficult to see why the finite-gap methods fail to
capture the massless modes, just by looking at the bosonic string on
the $S^3\times S^3\times \mathbbm{R}^1$ subspace of $AdS_3\times
S^3\times S^3\times S^1$. The $\mathbbm{R}^1$ factor is the global
time direction in $AdS_3$. At the classical level, all three factors
are independent and related only by the Virasoro constraints. It is
known that the Lax connection, in effect, imposes the constraints
automatically \cite{Faddeev:1985qu}. In the case at hand the Lax
connection actually over-imposes the Virasoro by setting the energy
momentum tensors $T^{1,2}_{\pm\pm}$ for each of the three-spheres to
a constant independently, and thus kills two degrees of freedom
instead of one -- we want only the sum $T^1_{\pm\pm}+T^2_{\pm\pm}$
to be constrained. The massless excitation arises from the
combination of the longitudinal modes on the two spheres such that
their Virasoro-violating contributions mutually cancel. The
finite-gap method eliminates this mode from the very beginning.

By applying the finite-gap method to the bosonic $AdS_5\times S^5$
or $AdS_4\times CP^3$ backgrounds one imposes two constraints and
thus eliminates just the right amount of unphysical, longitudinal
modes of the string. In $AdS_3\times S^3\times S^3$, we eliminate
three degrees of freedom, one of which is actually physical.
Refs.~\cite{Gromov:2006dh,Gromov:2006cq,Gromov:2007fn} discuss how
longitudinal modes can be incorporated in the classical/quantum
Bethe ansatz. In the case of the $S^3$ sigma-model discussed there,
the longitudinal degree of freedom corresponds to the extra rapidity
variables in the Bethe equations. Perhaps our equations from
section~\ref{sec7} could also be augmented by on or more extra
rapidity node(s) in order to describe the missing massless degrees
of freedom.

In this section we collect together some first steps we have taken
towards understanding the zero-modes. We hope to return to these
issues in the future.

\subsection{Semi-classical lightlike geodesics}\label{otherlightg}

Throughout this section we ignore  the spacetime $S^1$, as it plays
no role in the classical analysis - its equation of motion is
decoupled from the others, apart from the Virasoro constraints, and
so can be set to zero for simplicity. We use complex embedding
co-ordinates $X_i$, $Y_i$ and $Z_i$ for $AdS_3$ and the two $S^3$'s
\begin{equation}
-X^1X_1+X^2X_2=-l^2\,,\qquad
Y^1Y_1+Y^2Y_2=R_+^2\,,\qquad
Z^1Z_1+Z^2Z_2=R_-^2\,,
\end{equation}
where $X^i\equiv X_i^*$, $Y^i\equiv Y_i^*$ and $Z^i\equiv Z_i^*$. The equations of motion are
\begin{eqnarray}
-\partial^2 X_i + \Lambda_X X_i&=&0\,,\qquad l^2\Lambda_X = -X^i\partial^2X_i\,,\\
-\partial^2 Y_i + \Lambda_Y Y_i&=&0\,,\qquad R_+^2\Lambda_Y = Y^i\partial^2Y_i\,,\\
-\partial^2 Z_i + \Lambda_Z Z_i&=&0\,,\qquad R_-^2\Lambda_Z = Z^i\partial^2Z_i\,.
\end{eqnarray}
The Virasoro constraints are
\begin{eqnarray}
-{\dot X^i}{\dot X_i}- X^i{}^\prime X_i{}^\prime+{\dot Y^i}{\dot Y_i}+ Y^i{}^\prime Y_i{}^\prime
+{\dot Z^i}{\dot Z_i}+ Z^i{}^\prime Z_i{}^\prime&=&0\,,\label{v1}\\
-{\dot X^i} X_i{}^\prime+{\dot Y^i} Y_i{}^\prime+{\dot Z^i} Z_i{}^\prime+\mbox{c.c.}&=&0\,.\label{v2}
\end{eqnarray}
The Cartan generators of the Noether charges are
\begin{eqnarray}
E&=&\frac{i}{4\pi\alpha'}\int_0^{2\pi}d\sigma\,\,( X_1{\dot X}^1-{\dot X}_1X^1)\,,\\
J&=&\frac{i}{4\pi\alpha'}\int_0^{2\pi}d\sigma\,\, (X_2{\dot X}^2-{\dot X}_2X^2)\,,\\
J_i^+&=&\frac{i}{4\pi\alpha'}\int_0^{2\pi}d\sigma\,\, (Y_i{\dot Y}^i-{\dot Y}_iY^i)\,,\\
J_i^-&=&\frac{i}{4\pi\alpha'}\int_0^{2\pi}d\sigma\,\, (Z_i{\dot
Z}^i-{\dot Z}_iZ^i)\,,
\end{eqnarray}
where above there is no summation on the index $i=1,2$. Using the global symmetry of
the background a point-particle world-line can always be rotated to
\begin{equation}
X_1=le^{i\kappa(t)}\,,\qquad Y_1=R_+e^{i\omega_1(t)}\,,\qquad Z_1=R_-e^{i\omega_2(t)}\,,
\label{geod1}
\end{equation}
for some, as yet undetermined, $\kappa, \omega_i$ functions  of the
world-line parameter $t$. Simple solutions of the equations of
motion\footnote{A particular case of such solutions was analyzed in
\cite{Sommovigo:2003kd}.} can be found for
\begin{equation}
\kappa(t)=\kappa t\,,\qquad\omega_i(t)=\omega_i t\,,\label{geod2}
\end{equation}
for constant $\kappa, \omega_i$. In this case the Lagrange multipliers are constant
\begin{equation}
\Lambda_X=\kappa^2\,,\qquad
\Lambda_Y=\omega_1^2\,,\qquad
\Lambda_Z=\omega_2^2\,,
\end{equation}
and the Virasoro constraints reduce to
\begin{equation}
\kappa^2=\omega_1^2+\omega_2^2\,,
\end{equation}
which can be solved in terms of an angle variable $\theta$
\begin{equation}
\omega_1=\kappa\cos\theta\,,\qquad
\omega_2=\kappa\sin\theta\,.
\label{lgsoln}
\end{equation}
The angle $\theta$ denotes  the relative angle between the great
circles on the two $S^3$ factors. In the coset language the  fact
that such a one-parameter family of geodesics exists corresponds to
the fact that (for $C''=0$) equation~(\ref{zeronorm}) is solved by
\begin{equation}
C=\cos\phi\cos\theta\,,\qquad
C'=\sin\phi\sin\theta\,.
\end{equation}
The Noether  charges of these solutions are
\begin{eqnarray}
E&=&\frac{l^2}{\alpha'}\kappa\,,\nonumber \\
J^+_1&=&\frac{R_+^2}{\alpha'}\omega_1=\kappa\frac{l^2}{\alpha'}\frac{\cos\theta}{\cos^2\phi}\,,\qquad
J^-_1=\frac{R_-^2}{\alpha'}\omega_2=\kappa\frac{l^2}{\alpha'}\frac{\sin\theta}{\sin^2\phi}\,.
\end{eqnarray}

These solutions correspond to a different choice of the light-like
geodesic, compared to the supersymmetric case studied in
section~\ref{sec5}. One picks a different linear combination of the
Noether charges to define the angular momentum:
\begin{equation}
{\cal J}\equiv j^+ \cos^2\phi J^+_1 + j^- \sin^2\phi J^-_1
\end{equation}
where the $\phi$ dependence is chosen for later  convenience. Since
${\cal J}$ has to be suitably normalised (in the measure given by
equation~(\ref{measure})), we must take
\begin{equation}
j_+^2\cos^2\phi+j_-^2\sin^2\phi=1\,,
\end{equation}
in other words we should choose them as
\begin{equation}
j_+=\frac{\cos\lambda}{\cos\phi}\,,\qquad
j_-=\frac{\sin\lambda}{\sin\phi}\,,\label{norm}
\end{equation}
for a free parameter $\lambda$. The choice of $j_+=j_-=1$, or
$\lambda=\phi$ corresponds to the BPS geodesic of
section~\ref{sec5}. Evaluating $E-{\cal J}$ on
solutions~(\ref{geod1})-(\ref{geod2}) we find
\begin{equation}
E-{\cal J}=\frac{l^2}{\alpha'}\left(\kappa-j^+\omega_1-j^-\omega_2\right)
=\kappa\frac{l^2}{\alpha'}\left(1-\cos(\lambda-\theta)\right)\,.
\end{equation}
The BPS groundstate corresponds to
\begin{equation}
\omega_1=\kappa\cos\lambda\cos\phi\,,
\qquad
\omega_1=\kappa\sin\lambda\sin\phi\,.
\end{equation}
In particular, for the choice $j_\pm=1$ made in section~\ref{sec5},
we find $\omega_1=\kappa\cos^2\phi$ and $\omega_2=\kappa\sin^2\phi$.
If we
define ${\cal J}^\perp$ as the operator 'orthogonal' to ${\cal J}$
in the coset metric
\begin{equation}
{\cal J}^\perp=\sin\lambda\cos\phi J_1^+-\cos\lambda\sin\phi J_1^-\,,
\end{equation}
then our light-like geodesic solutions carry the following ${\cal J}^\perp$ charges
\begin{equation}
\frac{\kappa l^2}{\alpha'}\left(\sin\lambda\cos\theta\sec\phi-\cos\lambda\sin\theta\csc\phi\right)\,.
\end{equation}
In particular, the BPS groundstate  has ${\cal J}^\perp=0$.
The existence of a continuous family of light-like geodesics is a
semi-classical manifestation of the massless modes we have
encountered; it is a new feature of the $AdS_3$ background.

\subsection{Large charge limit}

The finite gap equations do not capture the above semi-classical solutions. One way to see this is to consider the large charge limit of the string action~\cite{Kruczenski:2003gt,Kruczenski:2004kw,Hernandez:2004uw,Stefanski:2004cw,Kruczenski:2004cn,Hernandez:2004kr,Stefanski:2005tr,Stefanski:2007dp}. For simplicity, we restrict ourselves initially to a bosonic subsector of our action which is the analogue of the $SU(2)$ subsector originally considered in~\cite{Kruczenski:2003gt}. To take the large charge limit one redefines the global coordinates as follows
\begin{eqnarray}
X_1&=&le^{it(\tau\,\,\sigma)}\,,\qquad X_2=0\,.\\
Y_i&=&R_+e^{iv(\tau\,\,\sigma)}V_i\,,\\
Z_i&=&R_-e^{iw(\tau\,\,\sigma)}W_i\,.
\end{eqnarray}
After taking $t=\kappa\tau$ ($\kappa$ is the large-spin parameter here), redefining
\begin{eqnarray}
v&=&\cos^2\phi\,\,\kappa\tau+{\tilde v}(\tau\,,\,\sigma)\,,\\
w&=&\sin^2\phi\,\,\kappa\tau+{\tilde w}(\tau\,,\,\sigma)\,,
\end{eqnarray}
and rescaling $\tau\rightarrow\kappa\tau$, to leading order in $\kappa$, the Virasoro constraint~(\ref{v2}) reduces to
\begin{equation}
\partial_\sigma{\tilde v}+\partial_\sigma{\tilde w}+iV^i\partial_\sigma V_i+iW^i\partial_\sigma W_i=0\,,
\end{equation}
while the other Virasoro constraint~(\ref{v1}) determines $\partial_\tau({\tilde v}+{\tilde w})$ in terms of the other fields. The coordinates conjugate to the charges ${\cal J}$ and ${\cal J}^\perp$ are
\begin{eqnarray}
k&=&{\tilde v}+{\tilde w}\,,\\
k^\perp&=&(1+\tan\phi){\tilde v}+(1-\cot\phi){\tilde w}\,.
\end{eqnarray}
and we can eliminate the dependence of the Lagrangian on $k$ through the Virasoro constraints. The Lagrangian then reduces to
\begin{equation}
\frac{\alpha'}{l^2}{\cal L}_{\mbox{\scriptsize $AdS_3\times S^3\times S^3\times S^1$}}\rightarrow
2iV^i\partial_\tau V_i+2iW^i\partial_\tau W_i-\frac{\left|D_\sigma V_i\right|^2}{\cos^2\phi}
-\frac{\left|D_\sigma W_i\right|^2}{\sin^2\phi}-({\cal D}_\sigma k^\perp)^2\,,
\end{equation}
where
\begin{eqnarray}
{\cal D}_\sigma k^\perp&\equiv&\partial_\sigma k^\perp+i(1+\tan\phi) V^i\partial_\sigma V_i +i(1-\cot\phi) W^i\partial_\sigma W_i\,,\\
D_\sigma V_i&\equiv&\partial_\sigma V_i-V^j\partial_\sigma V_jV_i\,,\\
D_\sigma W_i&\equiv&\partial_\sigma W_i-W^j\partial_\sigma W_jW_i\,,
\end{eqnarray}
Notice that in this limit the field $k^\perp$ has no time derivative in the action. In the finite gap equations one simply ignores the ${\tilde w}$ dependent term above; the other terms reduce to standard $SU(2)/U(1)$ Landau-Lifshitz sigma models with a normalisation of the kinetic term that reflects the relative size of the two $CP^1$ factors.

We can also construct the Landau-Lifshitz sigma  model for the full
theory. To do this we need to identify the sub-group $H$ which
preserve the vacuum of the model. From the BMN analysis in
section~\ref{sec5} the stability sub-group in this case is
$H=(U(1)^3\ltimes PSU(1|1))^2$ where the bosonic generators are
$S_0^{L\,,\,R}\,,\,L_5^{L\,,\,R}\,,\,R_8^{L\,,\,R}$ and the
fermionic generators are $Q^{L\,,\,R}_{\pm\pm\pm}$. $H$ can be
thought of as two copies of the maximal central extension of
$PSU(1|1)^2$.~\footnote{This central extension has been discussed in
\cite{David:2008yk}.} The Landau-Lifshitz sigma model can then be
constructed on the coset
\begin{equation}
\frac{D(2,1;\alpha)^2}{U(1)^6\ltimes PSU(1|1)^2}\,,
\end{equation}
using the definition given in~\cite{Stefanski:2007dp} for a Landau-Lifshitz sigma model on a general coset $G/H$. This sigma model will have $6+6$ degrees of freedom as is evident from counting the super-dimension of the coset.~\footnote{Just as in the discussion above this sigma model will not describe the ${\tilde w}$ massless mode, nor the corresponding massless fermionic zero modes.}

\section{Discussion}

It would be very interesting to understand how the Bethe equations
we postulate arise in the dual CFT. The CFT dual for $M_4=T^4$ is
well understood (see for
example~\cite{Maldacena:1998re,Maldacena:1998bw,Seiberg:1999xz,Larsen:1999uk,Arutyunov:1997gi,Lunin:2001pw,Jevicki:1998bm}).
It is a deformation of the symmetric product orbifold ${\rm
Sym}^{Q_{1}Q_{5}}(M_4)$, where $Q_{1},Q_{5}$ are the numbers of
coincident D1- and D5-branes correspondingly, such that the latter
are wrapping $M_4$ and the former are transverse to $M_4$. These
CFTs possess "small" $\mathcal{N}=(4,4)$ super conformal symmetry
with four supercurrents\footnote{The dependence on just the product
$Q_{1}Q_{5}$ is due to U-duality which guarantee equivalence of
theories with the same product $Q_{1}Q_{5}$.}. The BMN limit of this
dual pair was investigated
in~\cite{Hikida:2002in,Lunin:2002fw,Gomis:2002qi,Gava:2002xb}. It
should be possible to construct a suitable spin-chain which could
presumably match our Bethe equations at weak coupling and perhaps
shed some light on the problem of massless modes. Indeed some first
steps in this direction have been taken in~\cite{David:2008yk}. By
contrast the CFT dual for $M_4=S^3\times S^1$ is probably one of the
most obscure amongst $AdS$/CFT
pairs~\cite{Cowdall:1998bu,Boonstra:1998yu,Gauntlett:1998kc,
Elitzur:1998mm,deBoer:1999rh,Giveon:2003ku,Gukov:2004ym}. The
symmetries dictate that this CFT has a "large" $\mathcal{N}=(4,4)$
super-conformal symmetry, which appears to be more difficult to
treat. As a result, to date a suitable dual CFT candidate has not
been identified. In fact, a number of potential duals do not appear
to satisfy the requirements of the duality~\cite{Gukov:2004ym}. In
this context, we hope that unraveling the integrability structures
of superstring theory on this background may be helpful for further
understanding of correct CFT dual for this background.

We have derived the classical Bethe equations for the supercoset
model on the $AdS_3\times S^3\times S^3\times S^1$ background. We
also made a guess for the quantum Bethe equations.  But both of
these equations miss the massless string modes. In addition, we only
see the rigid, $D(2,1;\alpha )\times D(2,1;\alpha )$ part of the
target-space Virasoro symmetry. We believe that these two problems
are related, and that understanding the target space symmetry
enhancement may also shed light on the massless modes. The
appearance of the target-space Virasoro algebra is understood at the
classical supergravity level \cite{Brown:1986nw} and at the quantum
level in the NSNS $AdS_3$ backgrounds
\cite{Giveon:1998ns,deBoer:1998pp}, including $AdS_3\times S^3\times
S^3\times S^1$ \cite{Elitzur:1998mm}. In an interesting recent
development, the target space Virasoro generators were constructed
for the string on $AdS_3\times S^3\times T^4$ with the RR flux
\cite{Ashok:2009jw}. Perhaps incorporating these results into the
integrability approach can provide the  missing information on the
massless degrees of freedom on the worldsheet.

{\bf Note added:} When we were preparing this paper for publication,
ref.~\cite{Pakman:2009mi} appeared on the ArXiv, in which the spin
chain for the symmetric orbifold CFT on $T^4$ was constructed. It
would be extremely interesting to see if the $\lambda \rightarrow 0$
limit of the Bethe equations (\ref{t4bethe}) is capable of capturing
a part of the spectrum of this spin chain.

\subsection*{Acknowledgments}
We would like to thank S.~Ashok, N.~Beisert, A.~Dabholkar,
S.~Elitzur, J. Gauntlett, N.~Gromov, V.~Kazakov, C.~Kristjansen,
J.~Minahan, O.~Ohlsson Sax, D.~Sorokin, J.~Troost and A.~Tseytlin
for interesting discussions.  B.S. is grateful to the Centro de
Ciencias de Benasque Pedro Pascual and the organisers of the String
Theory workshop held there where part of this work was undertaken.
A.B. is grateful to the Einstein center for support. B.S. is
supported by an ARF from EPSRC. The work of K.Z. was supported in
part by the BQR ENS, in part by the Swedish Research Council under
the contract 621-2007-4177, in part by the RFFI grant 09-02-00253,
and in part by the grant for support of scientific schools
NSH-3036.2008.2.

\appendix

\section{The $\mathfrak{d}(2,1;\alpha)$ superalgebra}\label{salgebra}

The bosonic subalgebra of $\mathfrak{d}(2,1;\alpha)$ consists of
three commuting $\mathfrak{sl}(2)$'s. The supercharges are in their
tri-spinor representation. We are interested in the real form of
$\mathfrak{d}(2,1;\alpha )$ in which one of the $sl(2)$'s is
non-compact and the other two are compact, so that the bosonic
subalgebra is
$\mathfrak{sl}(2,\mathbbm{R})\oplus\mathfrak{su}(2)\oplus\mathfrak{su}(2)$.
With this choice of the real form the parameter $\alpha $ must lie
between zero and one, and it is convenient to introduce the
trigonometric parameterization $\alpha =\cos^2\phi $. We denote the
$\mathfrak{sl}(2,\mathbbm{R})$, $\mathfrak{su}(2)_+$, and
$\mathfrak{su}(2)_-$ generators by $S_\mu$  ($\mu  =0,1,2$), $L_n$
($n=3,4,5$), and $R_{\dot{n}}$ ($\dot{n}=6,7,8$). The supercharges
are $Q_{a\alpha \dot{\alpha }}$. Their spinor indices $a$, $\alpha$,
and $\dot{\alpha }$ take values $+$ or $-$.

To describe the action of the $\mathfrak{sl}(2)$ generators on the
supercharges we introduce three sets of Pauli matrices:
\begin{equation}\label{3ddirac}
\gamma ^\mu =(i\sigma ^2,\sigma ^1,\sigma ^3),\qquad  \gamma
^n=(\sigma ^1,\sigma ^2,\sigma ^3),\qquad \gamma ^{\dot{n}}=(\sigma
^1,\sigma ^2,\sigma ^3).
\end{equation}
The (anti-)commutation relations of
$\mathfrak{d}(2,1;\cos^2\phi )$  then read:
\begin{eqnarray}\label{commutation_relations}
 {}[S_\mu ,S_\nu ]&=&\epsilon _{\mu \nu\lambda }S^\lambda \nonumber \\
 {}[L_m,L_n]&=&\epsilon _{mnp}L^p \nonumber \\
 {}[R_{\dot{m}} ,R_{\dot{n}} ]&=&\epsilon_{\dot{m}\dot{n}\dot{p}}
 R^{\dot{p}} \nonumber \\
 {}[S_\mu ,Q_{a\alpha \dot{\alpha }}]&=&
 -\frac{1}{2}\,Q_{b\alpha \dot{\alpha }}\gamma _{\mu \,a}^{b}\nonumber \\
 {}[L_m,Q_{a\alpha \dot{\alpha }}]&=&
 -\frac{i}{2}\,Q_{a\beta \dot{\alpha }}\gamma _{m\,\alpha }^{\beta }\nonumber \\
 {}[R_{\dot{m}},Q_{a\alpha \dot{\alpha }}]&=&
 -\frac{i}{2}\,
 Q_{a\alpha \dot{\beta }}\gamma _{\dot{m}\,\dot{\alpha }}^{\dot{\beta  }}\nonumber \\
 \{Q_{a\alpha \dot{\alpha }},Q_{b\beta \dot{\beta }}\}
 &=&i( \varepsilon \gamma ^\mu)_{ab}\,\varepsilon _{\alpha \beta }\,\varepsilon
 _{\dot{\alpha }\dot{\beta }}\,S_\mu
 -\cos^2\phi \,\varepsilon _{ab}\,(\varepsilon \gamma ^m)_{\alpha \beta }\,\varepsilon
 _{\dot{\alpha }\dot{\beta }}\,L_m
 \nonumber \\
 &&-\sin^2\phi \,\varepsilon _{ab}\,\varepsilon _{\alpha \beta }\,(\varepsilon
 \gamma ^{\dot{m}})_{\dot{\alpha }\dot{\beta }}\,R_{\dot{m}},
\end{eqnarray}
where $\epsilon _{012}=\epsilon _{345}=\epsilon _{678}=1$, and the
vector indices are raised and lowered by $\eta _{\mu \nu
}=\mathop{\mathrm{diag}} (-++)$, $\delta _{nm}$, and $\delta
_{\dot{n}\dot{m}}$.

The invariant bilinear form on $\mathfrak{d}(2,1;\cos^2\phi )$ is
given by
\begin{eqnarray}\label{str}
 \mathop{\mathrm{Str}}S_\mu S_\nu &=&\frac{1}{4}\,\eta _{\mu \nu }\nonumber \\
 \mathop{\mathrm{Str}}L_mL_n&=&\frac{1}{4\cos^2\phi }\,\delta _{mn}\nonumber \\
 \mathop{\mathrm{Str}}R_{\dot{m}}R_{\dot{n}}&=&\frac{1}{4\sin^2\phi }\,\delta
 _{\dot{m}\dot{n}}\nonumber \\
 \mathop{\mathrm{Str}}Q_{a\alpha \dot{\alpha }}Q_{b\beta\dot{\beta }}
 &=&\frac{i}{2}\,\varepsilon _{ab}\,\varepsilon _{\alpha \beta }\,\varepsilon _{\dot{\alpha }\dot{\beta
 }}\,,\label{measure}
\end{eqnarray}
and is consistent with the $(-+\ldots +)$ signature in the target
space.

We will also use another basis of supercharges:
\begin{equation}\label{newbasis}
 \tilde{Q}_{\pm\alpha \dot{\alpha }}=Q_{+\alpha \dot{\alpha }}\mp
 iQ_{-\alpha \dot{\alpha }}.
\end{equation}
The commutators with the $\mathfrak{sl}(2)$ generators then take the
form
\begin{equation}\label{}
[S_\mu ,\tilde{Q}_{a\alpha \dot{\alpha }}]=
 -\frac{i}{2}\,\tilde{Q}_{b\alpha \dot{\alpha }}\tilde{\gamma }_{\mu
 \,a}^{b},
\end{equation}
with the gamma-matrices rotated to a different basis: $\tilde{\gamma
}_\mu =(\sigma ^3,i\sigma ^2,-i\sigma ^1)$. The normalization also
changes:
\begin{equation}\label{}
 \mathop{\mathrm{Str}}\tilde{Q}_{a\alpha \dot{\alpha }}\tilde{Q}_{b\beta\dot{\beta }}
 =-\varepsilon _{ab}\,\varepsilon _{\alpha \beta }\,\varepsilon _{\dot{\alpha }\dot{\beta
 }}\,.
\end{equation}

\section{Background-field expansion for $\mathbbm{Z}_4$
cosets}\label{backexp}

In this appendix we expand the sigma-model action (\ref{action}) to
the second order in fluctuations around an arbitrary bosonic
background $g_B(x)$. This is necessary for comparison of the coset
model to the Green-Schwarz action, and also for fixing the
light-cone gauge in the BMN limit. The expansion can be done quite
generally, starting with the coset representative in the form
\begin{equation}\label{bdecomposition}
 g=g_B\,{\rm e}\,^{\mathbbm{X}},
\end{equation}
where
$\mathbbm{X}(x)=X^A(x)T_A\in\mathfrak{h}_1\oplus\mathfrak{h}_2\oplus\mathfrak{h}_3$
is the fluctuation field. For the background currents we introduce
the following notations:
\begin{eqnarray}\label{bcurrent}
 \left(g_B^{-1}\partial _{\bf a} g_B\right)_0&=&A_{\bf a} ,\nonumber \\
 \left(g_B^{-1}\partial _{\bf a} g_B\right)_2&=&K_{\bf a}.
\end{eqnarray}
According to the discussion in the main text, $A_{\bf
a}\in\mathfrak{h}_0 $ is the background gauge field. By $D_{\bf a} $
we will denote the corresponding covariant derivative:
\begin{equation}
 D_{\bf a} =\partial _{\bf a} +[A_{\bf a} ,\cdot ],
\end{equation}
and by $F_{\mathbf{a}\mathbf{b}}$ the field strength $F_{{\bf a}
{\bf b} }=\partial _{\bf a} A_{{\bf b} }-\partial _{\bf b} A_{\bf a}
+[A_{\bf a} ,A_{\bf b} ]$. The flatness of the current
$g_B^{-1}\partial _\mathbf{a}g_B$ implies that
\begin{eqnarray}\label{eqmo}
 && [K_{\bf a} ,K_{\bf b} ]+F_{{\bf a} {\bf b} }=0, \\
 && D_{\bf a} K_{\bf b} -D_{\bf b} K_{\bf a} =0.
\end{eqnarray}

The current (\ref{cur}) can be readily expanded in power series in
$\mathbbm{X}$:
\begin{equation}\label{curexp}
 J_{\bf a} =A_{\bf a} +K_{\bf a} +\frac{1-\,{\rm e}\,^{- \mathop{\mathrm{ad}}\mathbbm{X}}}
 {\mathop{\mathrm{ad}}\mathbbm{X}}\,\mathcal{D}_{\bf a} \mathbbm{X}
 =
 A_{\bf a} +K_{\bf a} + \mathcal{D}_{\bf a} \mathbbm{X}-\frac{ 1}{2}[\mathbbm{X},\mathcal{D}_{\bf a} \mathbbm{X}]
 +\ldots ,
\end{equation}
where the long derivative $\mathcal{D}_{\bf a} $ is defined by
\begin{equation}
 \mathcal{D}_{\bf a} =\partial _{\bf a} +[g_B^{-1}\partial _{\bf a} g_B,\cdot ]
 =D_{\bf a} +[K_{\bf a} ,\cdot ].
\end{equation}
Unlike the covariant derivative $D_{\bf a} $, which commutes with
the $\mathbbm{Z}_4$ grading, the long derivative $\mathcal{D}_{\bf
a} $ does not have definite $\mathbbm{Z}_4$ charge. Thus, $(D_{\bf
a} \mathbbm{X})_n=D_{\bf a} \mathbbm{X}_n$ for any $n$, also
$(\mathcal{D}_{\bf a} \mathbbm{X})_2=D_{\bf a} \mathbbm{X}_2$, but
$(\mathcal{D}_{\bf a} \mathbbm{X})_{1,3}=D_{\bf a}
\mathbbm{X}_{1,3}+[K_{\bf a} ,\mathbbm{X}_{3,1}]$.

We can now expand the action (\ref{action}) in powers of
$\mathbbm{X}$. To simplify the result, one should use the identity:
\begin{equation}
 \varepsilon ^{\mathbf{a}\mathbf{b}}\mathop{\mathrm{Str}}D_\mathbf{a}
 \mathbbm{X}_1D_\mathbf{b} \mathbbm{X}_3
 =-\varepsilon ^{\mathbf{a}\mathbf{b}}\mathop{\mathrm{Str}}
 [K_\mathbf{a},\mathbbm{X}_1][K_\mathbf{b},\mathbbm{X}_3]
 +{\rm total~derivative},
\end{equation}
which follows from (\ref{eqmo}). Then, to the second order in
fluctuations,
\begin{eqnarray}\label{quadaction}
 S&=&\int_{}^{}d^2x\,\mathop{\mathrm{Str}}
 \left\{\sqrt{-h}h^{{\bf a} {\bf b} }K_\mathbf{a}K_\mathbf{b}
 -2\mathbbm{X}_2\nabla_\mathbf{a}K^\mathbf{a}
 \right.\nonumber \\
 &&\left. +\sqrt{-h}h^{{\bf a} {\bf b} }
 \left(
 D_{\bf a} \mathbbm{X}_2D_{\bf b} \mathbbm{X}_2-[K_{\bf a} ,\mathbbm{X}_2][K_{\bf b} ,\mathbbm{X}_2]
  \right)\right. \nonumber \\ &&\left.
 +\left(\sqrt{-h}h^{\mathbf{a}\mathbf{b}}+\varepsilon ^{\mathbf{a}\mathbf{b}}\right)
  \mathbbm{X}_1[K_{{\bf a} },D_\mathbf{b}\mathbbm{X}_1]
 +\left(\sqrt{-h}h^{\mathbf{a}\mathbf{b}}-\varepsilon ^{\mathbf{a}\mathbf{b}}\right)
 \mathbbm{X}_3[K_{{\bf a} },D_\mathbf{b}\mathbbm{X}_3]
 \right. \nonumber \\ &&\left.
 -\left(\sqrt{-h}h^{\mathbf{a}\mathbf{b}}+\varepsilon ^{\mathbf{a}\mathbf{b}}\right)
 [K_{{\bf a} },\mathbbm{X}_1][K_{{\bf b} },\mathbbm{X}_3]
 -\left(\sqrt{-h}h^{\mathbf{a}\mathbf{b}}-\varepsilon ^{\mathbf{a}\mathbf{b}}\right)
 [K_{{\bf a} },\mathbbm{X}_3][K_{{\bf b} },\mathbbm{X}_1]
 \right\}.
 \nonumber \\
\end{eqnarray}
The second term contains the covariantized derivative
$\nabla_\mathbf{a}K^\mathbf{b}=D_\mathbf{a}K^\mathbf{b}+\Gamma
^\mathbf{b}_{\mathbf{a}\mathbf{c}}K^\mathbf{c}$ and vanishes
on-shell, when $K^\mathbf{a}$ satisfies the equations of motion. Let
us stress that we have not used the equations of motion for
$K^\mathbf{a}$ in deriving (\ref{quadaction}). The equations
(\ref{eqmo}) that we used are identities valid for any
$K_\mathbf{a}$ and $A_\mathbf{a}$ of the form (\ref{bcurrent}).

\section{Gamma matrices}\label{lesnouvelles}

We pick the following representation for the 10d Dirac matrices:
\begin{eqnarray}\label{gamma1}
\Gamma^\mu&=&\sigma^1\otimes \sigma ^2\otimes \gamma ^\mu\otimes
\mathbbm{1} \otimes \mathbbm{1}\,,\qquad \mu=0,1,2 \\
\label{gamma2} \Gamma^{n}&=& \sigma^1\otimes
\sigma^1\otimes \mathbbm{1} \otimes \gamma ^{n} \otimes \mathbbm{1}\,,\qquad n=3,4,5\\
\label{gamma3}
\Gamma^{\dot{n}}&=&\sigma^1\otimes \sigma^3 \otimes \mathbbm{1} \otimes \mathbbm{1}
\otimes \gamma ^{\dot{n}}\,,\qquad \dot{n}=6,7,8\\
\Gamma^9&=&-\sigma^2\otimes \mathbbm{1} \otimes \mathbbm{1} \otimes
\mathbbm{1} \otimes \mathbbm{1},
\end{eqnarray}
where the 3d gamma-matrices $\gamma ^i$ are taken from
(\ref{3ddirac}).

In this basis,
\begin{eqnarray}\label{}
\Gamma ^{012}&=&\sigma ^1\otimes\sigma ^2\otimes
\mathbbm{1}\otimes\mathbbm{1}\otimes\mathbbm{1}\\
\Gamma ^{345}&=&i\sigma ^1\otimes\sigma ^1\otimes
\mathbbm{1}\otimes\mathbbm{1}\otimes\mathbbm{1}\\
\Gamma ^{678}&=&i\sigma ^1\otimes\sigma ^3\otimes
\mathbbm{1}\otimes\mathbbm{1}\otimes\mathbbm{1}\\
\Gamma ^{012345}&=&\mathbbm{1}\otimes\sigma ^3\otimes
\mathbbm{1}\otimes\mathbbm{1}\otimes\mathbbm{1}\\
\Gamma ^{012678}&=&-\mathbbm{1}\otimes\sigma ^1\otimes
\mathbbm{1}\otimes\mathbbm{1}\otimes\mathbbm{1}\\
\Gamma &=&\sigma ^3\otimes\mathbbm{1}\otimes
\mathbbm{1}\otimes\mathbbm{1}\otimes\mathbbm{1}
\end{eqnarray}
and
\begin{eqnarray}\label{whatnot}
 \Gamma ^{\mu \nu }&=&-\varepsilon ^{\mu \nu \lambda }
 \mathbbm{1}\otimes\mathbbm{1}\otimes\gamma
 _\lambda\otimes\mathbbm{1}\otimes\mathbbm{1}\nonumber \\
 \Gamma ^{mn}&=&i\varepsilon ^{mnp}
 \mathbbm{1}\otimes\mathbbm{1}\otimes\mathbbm{1}\otimes\gamma _{p}
 \otimes\mathbbm{1}\nonumber \\
 \Gamma ^{\dot{m}\dot{n}}&=&
 i\varepsilon ^{\dot{m}\dot{n}\dot{p}}\mathbbm{1}\otimes\mathbbm{1}\otimes\mathbbm{1}\otimes
 \mathbbm{1}\otimes\gamma _{\dot{p}}.
\end{eqnarray}
 The charge conjugation matrix is
\begin{equation}\label{chargeconjmatr}
 C=i\sigma ^2\otimes\sigma ^2\otimes\sigma ^2\otimes\sigma ^2\otimes\sigma
 ^2.
\end{equation}

The kappa-symmetry projectors introduced in section~\ref{sec3},
eq.~(\ref{kappaprojector}), are of the form
\begin{equation}\label{kapexp}
 K^\pm=\mathbbm{1}\otimes\,\frac{1}{2}\left(1\pm\cos \phi\, \sigma ^3
 \mp\sin \phi \,\sigma ^1\right)\otimes\mathbbm{1}\otimes\mathbbm{1}\otimes\mathbbm{1}
 =\mathbbm{1}\otimes
 \left|\pm\right\rangle_\phi {\vphantom{\left|+\right\rangle}}_\phi\! \left\langle\pm\right|
 \otimes\mathbbm{1}\otimes\mathbbm{1}\otimes\mathbbm{1},
\end{equation}
where
\begin{equation}\label{spinors}
 \left|+\right\rangle_\phi =
 \begin{pmatrix}
    \cos\frac{\phi }{2}  \\
     -\sin\frac{\phi }{2}   \\
 \end{pmatrix},
 \qquad \left|-\right\rangle_\phi =
 \begin{pmatrix}
   \sin\frac{\phi }{2}    \\
   \cos\frac{\phi }{2}   \\
 \end{pmatrix}.
\end{equation}
In these notations,
\begin{equation}\label{chiralprojector}
 \frac{1}{2}\left(1\pm\Gamma\right)=
 \left|\pm\right\rangle
 _0 {\vphantom{\left|+\right\rangle}}_0\!\left\langle \pm\right|
 \otimes\mathbbm{1}\otimes\mathbbm{1}
 \otimes\mathbbm{1}\otimes\mathbbm{1}.
\end{equation}
The following identities are useful in the calculations in
section~\ref{sec4}:
\begin{equation}\label{spinident}
 \left\langle\mp\right|\sigma ^1\left|\pm\right\rangle=
     \cos\phi,\qquad
 \left\langle\mp\right|\sigma ^2\left|\pm\right\rangle=
 \pm i,\qquad
     \left\langle\mp\right|\sigma ^3\left|\pm\right\rangle=
     \sin\phi.
\end{equation}

\section{Quartic terms in the near-BMN Lagrangian}\label{qrti}

The quartic near-BMN Lagrangian is already rather complicated. We
list here its bosonic part before the gauge fixing:
\begin{eqnarray}\label{}
 \mathcal{L}^{(4,b)}&=&
 -\frac{2}{3}\left(\bar{X}X\right)^2\left(\partial
 _\mathbf{a}t\right)^2
 +\frac{1}{3}\left[2\cos^6\phi \,\left(\bar{Y}Y\right)^2
 +\sin^2\phi \,\cos^6\phi \,V^2\bar{Y}Y
 \right. \nonumber \\ && \left.
 +2\sin^6\phi \,\left(\bar{Z}Z\right)^2
 +\cos^2\phi \,\sin^6\phi \,V^2\bar{Z}Z\right]\left(\partial _\mathbf{a}\varphi
 \right)^2 \vphantom{\frac{1}{3}}
 \nonumber \\ &&
 -\frac{1}{3}\,\cos^2\phi \,\sin^2\phi \,\left(\bar{Y}Y+\bar{Z}Z\right)
 \left(\partial _\mathbf{a}V\right)^2
 +\frac{1}{3}\,\bar{X}X\partial _\mathbf{a}\bar{X}\partial
 ^\mathbf{a}X
 \nonumber \\ &&
 -\frac{1}{3}\,\cos^2\phi \,\left(\sin^2\phi \,V^2+\bar{Y}Y\right)\partial
 _\mathbf{a}\bar{Y}\partial ^\mathbf{a}Y
 -\frac{1}{3}\,\sin^2\phi \,\left(\cos^2\phi \,V^2+\bar{Z}Z\right)\partial
 _\mathbf{a}\bar{Z}\partial ^\mathbf{a}Z
 \nonumber \\ &&
 -\frac{1}{6}\left[\left(\partial _\mathbf{a}\bar{X}\right)^2X^2+\bar{X}^2\left(
 \partial _\mathbf{a}X\right)^2\right]
 +\frac{1}{6}\,\cos^2\phi \,\left[
 \left(\partial _\mathbf{a}\bar{Y}\right)^2Y^2+\bar{Y}^2
 \left(\partial _\mathbf{a}Y\right)^2\right]
 \nonumber \\ &&
 +\frac{1}{6}\,\sin^2\phi \,\left[
 \left(\partial _\mathbf{a}\bar{Z}\right)^2Z^2
 +\bar{Z}^2\left(\partial _\mathbf{a}Z\right)^2
 \right]
 \nonumber \\ &&
 +\frac{1}{3}\,\cos^2\phi \,\sin^2\phi \,V\partial _\mathbf{a}V
 \left(
 \partial ^\mathbf{a}\bar{Y}Y+\bar{Y}\partial ^\mathbf{a}Y
 +\partial ^\mathbf{a}\bar{Z}Z+\bar{Z}\partial ^\mathbf{a}Z
 \right).
\end{eqnarray}

\bibliographystyle{nb}
\bibliography{refs}

\begin{thebibliography}{100}
\ifx\href\asklfhas\newcommand{\href}[2]{#2}\fi
\raggedright
\small
\parskip 0pt

\bibitem{Strominger:1996sh}
A.~Strominger and C.~Vafa,
\textit{``{Microscopic Origin of the Bekenstein-Hawking Entropy}''},
\textsf{Phys.~Lett.~B379,~99~(1996)},
\href{http://arXiv.org/abs/hep-th/9601029}{\texttt{hep-th/9601029}}.
%
\bibitem{Maldacena:1998re}
J.~M.~Maldacena,
\textit{``The large N limit of superconformal field theories and
  supergravity''},
\textsf{Adv.~Theor.~Math.~Phys.~2,~231~(1998)},
\href{http://arXiv.org/abs/hep-th/9711200}{\texttt{hep-th/9711200}}.
%
\bibitem{Giveon:1998ns}
A.~Giveon, D.~Kutasov and N.~Seiberg,
\textit{``{Comments on string theory on $AdS_3$}''},
\textsf{Adv.~Theor.~Math.~Phys.~2,~733~(1998)},
\href{http://arXiv.org/abs/hep-th/9806194}{\texttt{hep-th/9806194}}.
%
\bibitem{Elitzur:1998mm}
S.~Elitzur, O.~Feinerman, A.~Giveon and D.~Tsabar,
\textit{``{String theory on $AdS_3\times S^3\times S^3\times S^1$}''},
\textsf{Phys.~Lett.~B449,~180~(1999)},
\href{http://arXiv.org/abs/hep-th/9811245}{\texttt{hep-th/9811245}}.
%
\bibitem{Maldacena:2000hw}
J.~M.~Maldacena and H.~Ooguri,
\textit{``{Strings in AdS(3) and SL(2,R) WZW model. I}''},
\textsf{J.~Math.~Phys.~42,~2929~(2001)},
\href{http://arXiv.org/abs/hep-th/0001053}{\texttt{hep-th/0001053}}.
%
\bibitem{Maldacena:2000kv}
J.~M.~Maldacena, H.~Ooguri and J.~Son,
\textit{``{Strings in AdS(3) and the SL(2,R) WZW model. II: Euclidean black
  hole}''},
\textsf{J.~Math.~Phys.~42,~2961~(2001)},
\href{http://arXiv.org/abs/hep-th/0005183}{\texttt{hep-th/0005183}}.
%
\bibitem{Maldacena:2001km}
J.~M.~Maldacena and H.~Ooguri,
\textit{``{Strings in AdS(3) and the SL(2,R) WZW model. III: Correlation
  functions}''},
\textsf{Phys.~Rev.~D65,~106006~(2002)},
\href{http://arXiv.org/abs/hep-th/0111180}{\texttt{hep-th/0111180}}.
%
\bibitem{Berkovits:1999im}
N.~Berkovits, C.~Vafa and E.~Witten,
\textit{``{Conformal field theory of AdS background with Ramond-Ramond
  flux}''},
\textsf{JHEP~9903,~018~(1999)},
\href{http://arXiv.org/abs/hep-th/9902098}{\texttt{hep-th/9902098}}.
%
\bibitem{Rahmfeld:1998zn}
J.~Rahmfeld and A.~Rajaraman,
\textit{``{The GS string action on $AdS_3\times S^3$ with Ramond-Ramond
  charge}''},
\textsf{Phys.~Rev.~D60,~064014~(1999)},
\href{http://arXiv.org/abs/hep-th/9809164}{\texttt{hep-th/9809164}}.
%
\bibitem{Park:1998un}
J.~Park and S.-J.~Rey,
\textit{``{Green-Schwarz superstring on $AdS_3\times S^3$}''},
\textsf{JHEP~9901,~001~(1999)},
\href{http://arXiv.org/abs/hep-th/9812062}{\texttt{hep-th/9812062}}.
%
\bibitem{Metsaev:2000mv}
R.~R.~Metsaev and A.~A.~Tseytlin,
\textit{``{Superparticle and superstring in $AdS_3\times S^3$ Ramond-Ramond
  background in light-cone gauge}''},
\textsf{J.~Math.~Phys.~42,~2987~(2001)},
\href{http://arXiv.org/abs/hep-th/0011191}{\texttt{hep-th/0011191}}.
%
\bibitem{Chen:2005uj}
B.~Chen, Y.-L.~He, P.~Zhang and X.-C.~Song,
\textit{``{Flat currents of the Green-Schwarz superstrings in $AdS_5\times S^1$
  and $AdS_3\times S^3$ backgrounds}''},
\textsf{Phys.~Rev.~D71,~086007~(2005)},
\href{http://arXiv.org/abs/hep-th/0503089}{\texttt{hep-th/0503089}}.
%
\bibitem{Adam:2007ws}
I.~Adam, A.~Dekel, L.~Mazzucato and Y.~Oz,
\textit{``{Integrability of type II superstrings on Ramond-Ramond backgrounds
  in various dimensions}''},
\textsf{JHEP~0706,~085~(2007)},
\href{http://arXiv.org/abs/hep-th/0702083}{\texttt{hep-th/0702083}}.
%
\bibitem{David:2008yk}
J.~R.~David and B.~Sahoo,
\textit{``{Giant magnons in the D1-D5 system}''},
\textsf{JHEP~0807,~033~(2008)},
\href{http://arXiv.org/abs/0804.3267}{\texttt{0804.3267}}.
%
\bibitem{Bena:2003wd}
I.~Bena, J.~Polchinski and R.~Roiban,
\textit{``Hidden symmetries of the {$AdS_5\times S^5$} superstring''},
\textsf{Phys.~Rev.~D69,~046002~(2004)},
\href{http://arXiv.org/abs/hep-th/0305116}{\texttt{hep-th/0305116}}.
%
\bibitem{Kazakov:2004qf}
V.~A.~Kazakov, A.~Marshakov, J.~A.~Minahan and K.~Zarembo,
\textit{``Classical/quantum integrability in AdS/CFT''},
\textsf{JHEP~0405,~024~(2004)},
\href{http://arXiv.org/abs/hep-th/0402207}{\texttt{hep-th/0402207}}.
%
\bibitem{Beisert:2005bm}
N.~Beisert, V.~A.~Kazakov, K.~Sakai and K.~Zarembo,
\textit{``{The algebraic curve of classical superstrings on $AdS_5\times
  S^5$}''},
\textsf{Commun.~Math.~Phys.~263,~659~(2006)},
\href{http://arXiv.org/abs/hep-th/0502226}{\texttt{hep-th/0502226}}.
%
\bibitem{Beisert:2005fw}
N.~Beisert and M.~Staudacher,
\textit{``Long-range $PSU(2,2|4)$ Bethe ansaetze for gauge theory and
  strings''},
\textsf{Nucl.~Phys.~B727,~1~(2005)},
\href{http://arXiv.org/abs/hep-th/0504190}{\texttt{hep-th/0504190}}.
%
\bibitem{Staudacher:2004tk}
M.~Staudacher,
\textit{``The factorized S-matrix of CFT/AdS''},
\textsf{JHEP~0505,~054~(2005)},
\href{http://arXiv.org/abs/hep-th/0412188}{\texttt{hep-th/0412188}}.
%
\bibitem{Beisert:2005tm}
N.~Beisert,
\textit{``{The su(2|2) dynamic S-matrix}''},
\textsf{Adv.~Theor.~Math.~Phys.~12,~945~(2008)},
\href{http://arXiv.org/abs/hep-th/0511082}{\texttt{hep-th/0511082}}.
%
\bibitem{Gromov:2008bz}
N.~Gromov and P.~Vieira,
\textit{``{The AdS4/CFT3 algebraic curve}''},
\textsf{JHEP~0902,~040~(2009)},
\href{http://arXiv.org/abs/0807.0437}{\texttt{0807.0437}}.
%
\bibitem{Gromov:2008qe}
N.~Gromov and P.~Vieira,
\textit{``{The all loop AdS4/CFT3 Bethe ansatz}''},
\textsf{JHEP~0901,~016~(2009)},
\href{http://arXiv.org/abs/0807.0777}{\texttt{0807.0777}}.
%
\bibitem{Janik:2006dc}
R.~A.~Janik,
\textit{``The $AdS_5 \times S^5$ superstring worldsheet S-matrix and crossing
  symmetry''},
\textsf{Phys.~Rev.~D73,~086006~(2006)},
\href{http://arXiv.org/abs/hep-th/0603038}{\texttt{hep-th/0603038}}.
%
\bibitem{Arutyunov:2004vx}
G.~Arutyunov, S.~Frolov and M.~Staudacher,
\textit{``Bethe ansatz for quantum strings''},
\textsf{JHEP~0410,~016~(2004)},
\href{http://arXiv.org/abs/hep-th/0406256}{\texttt{hep-th/0406256}}.
%
\bibitem{Gromov:2009tv}
N.~Gromov, V.~Kazakov and P.~Vieira,
\textit{``{Integrability for the Full Spectrum of Planar AdS/CFT}''},
\href{http://arXiv.org/abs/0901.3753}{\texttt{0901.3753}}.
%
\bibitem{Bombardelli:2009ns}
D.~Bombardelli, D.~Fioravanti and R.~Tateo,
\textit{``{Thermodynamic Bethe Ansatz for planar AdS/CFT: a proposal}''},
\textsf{J.~Phys.~A42,~375401~(2009)},
\href{http://arXiv.org/abs/0902.3930}{\texttt{0902.3930}}.
%
\bibitem{Gromov:2009bc}
N.~Gromov, V.~Kazakov, A.~Kozak and P.~Vieira,
\textit{``{Integrability for the Full Spectrum of Planar AdS/CFT II}''},
\href{http://arXiv.org/abs/0902.4458}{\texttt{0902.4458}}.
%
\bibitem{Arutyunov:2009ur}
G.~Arutyunov and S.~Frolov,
\textit{``{Thermodynamic Bethe Ansatz for the $AdS_5 \times S^5$ Mirror
  Model}''},
\textsf{JHEP~0905,~068~(2009)},
\href{http://arXiv.org/abs/0903.0141}{\texttt{0903.0141}}.
%
\bibitem{Gromov:2009tq}
N.~Gromov,
\textit{``{Y-system and Quasi-Classical Strings}''},
\href{http://arXiv.org/abs/0910.3608}{\texttt{0910.3608}}.
%
\bibitem{Metsaev:1998it}
R.~R.~Metsaev and A.~A.~Tseytlin,
\textit{``Type IIB superstring action in $AdS_5 \times S^5$ background''},
\textsf{Nucl.~Phys.~B533,~109~(1998)},
\href{http://arXiv.org/abs/hep-th/9805028}{\texttt{hep-th/9805028}}.
%
\bibitem{Arutyunov:2008if}
G.~Arutyunov and S.~Frolov,
\textit{``{Superstrings on $AdS_4 \times CP^3$ as a Coset Sigma-model}''},
\textsf{JHEP~0809,~129~(2008)},
\href{http://arXiv.org/abs/0806.4940}{\texttt{0806.4940}}.
%
\bibitem{Stefanski:2008ik}
j.~Stefa\'nski,~B.,
\textit{``{Green-Schwarz action for Type IIA strings on $AdS_4\times CP^3$}''},
\textsf{Nucl.~Phys.~B808,~80~(2009)},
\href{http://arXiv.org/abs/0806.4948}{\texttt{0806.4948}}.
%
\bibitem{Gomis:2008jt}
J.~Gomis, D.~Sorokin and L.~Wulff,
\textit{``{The complete $AdS_4\times CP^3$ superspace for the type IIA
  superstring and D-branes}''},
\textsf{JHEP~0903,~015~(2009)},
\href{http://arXiv.org/abs/0811.1566}{\texttt{0811.1566}}.
%
\bibitem{Berkovits:1999zq}
N.~Berkovits, M.~Bershadsky, T.~Hauer, S.~Zhukov and B.~Zwiebach,
\textit{``{Superstring theory on $AdS_2\times S^2$ as a coset
  supermanifold}''},
\textsf{Nucl.~Phys.~B567,~61~(2000)},
\href{http://arXiv.org/abs/hep-th/9907200}{\texttt{hep-th/9907200}}.
%
\bibitem{Pesando:1998wm}
I.~Pesando,
\textit{``{The GS type IIB superstring action on AdS(3) x S(3) x T**4}''},
\textsf{JHEP~9902,~007~(1999)},
\href{http://arXiv.org/abs/hep-th/9809145}{\texttt{hep-th/9809145}}.
%
\bibitem{Cowdall:1998bu}
P.~M.~Cowdall and P.~K.~Townsend,
\textit{``{Gauged supergravity vacua from intersecting branes}''},
\textsf{Phys.~Lett.~B429,~281~(1998)},
\href{http://arXiv.org/abs/hep-th/9801165}{\texttt{hep-th/9801165}}.
%
\bibitem{Boonstra:1998yu}
H.~J.~Boonstra, B.~Peeters and K.~Skenderis,
\textit{``{Brane intersections, anti-de Sitter spacetimes and dual
  superconformal theories}''},
\textsf{Nucl.~Phys.~B533,~127~(1998)},
\href{http://arXiv.org/abs/hep-th/9803231}{\texttt{hep-th/9803231}}.
%
\bibitem{Gauntlett:1998kc}
J.~P.~Gauntlett, R.~C.~Myers and P.~K.~Townsend,
\textit{``{Supersymmetry of rotating branes}''},
\textsf{Phys.~Rev.~D59,~025001~(1999)},
\href{http://arXiv.org/abs/hep-th/9809065}{\texttt{hep-th/9809065}}.
%
\bibitem{deBoer:1999rh}
J.~de~Boer, A.~Pasquinucci and K.~Skenderis,
\textit{``{AdS/CFT dualities involving large 2d N = 4 superconformal
  symmetry}''},
\textsf{Adv.~Theor.~Math.~Phys.~3,~577~(1999)},
\href{http://arXiv.org/abs/hep-th/9904073}{\texttt{hep-th/9904073}}.
%
\bibitem{Gukov:2004ym}
S.~Gukov, E.~Martinec, G.~W.~Moore and A.~Strominger,
\textit{``{The search for a holographic dual to AdS(3) x S**3 x S**3 x
  S**1}''},
\textsf{Adv.~Theor.~Math.~Phys.~9,~435~(2005)},
\href{http://arXiv.org/abs/hep-th/0403090}{\texttt{hep-th/0403090}}.
%
\bibitem{Giveon:2003ku}
A.~Giveon and A.~Pakman,
\textit{``{More on superstrings in AdS(3) x N}''},
\textsf{JHEP~0303,~056~(2003)},
\href{http://arXiv.org/abs/hep-th/0302217}{\texttt{hep-th/0302217}}.
%
\bibitem{Kac:1977qb}
V.~G.~Kac,
\textit{``{A Sketch of Lie Superalgebra Theory}''},
\textsf{Commun.~Math.~Phys.~53,~31~(1977)}.
%
\bibitem{Frappat:1996pb}
L.~Frappat, P.~Sorba and A.~Sciarrino,
\textit{``{Dictionary on Lie superalgebras}''},
\href{http://arXiv.org/abs/hep-th/9607161}{\texttt{hep-th/9607161}}.
%
\bibitem{Lu:2002kw}
H.~Lu and J.~F.~Vazquez-Poritz,
\textit{``{Penrose limits of non-standard brane intersections}''},
\textsf{Class.~Quant.~Grav.~19,~4059~(2002)},
\href{http://arXiv.org/abs/hep-th/0204001}{\texttt{hep-th/0204001}}.
%
\bibitem{Green:1983wt}
M.~B.~Green and J.~H.~Schwarz,
\textit{``{Covariant Description of Superstrings}''},
\textsf{Phys.~Lett.~B136,~367~(1984)}.
%
\bibitem{Henneaux:1984mh}
M.~Henneaux and L.~Mezincescu,
\textit{``{A Sigma Model Interpretation of Green-Schwarz Covariant Superstring
  Action}''},
\textsf{Phys.~Lett.~B152,~340~(1985)}.
%
\bibitem{McArthur:1999dy}
I.~N.~McArthur,
\textit{``{Kappa-symmetry of Green-Schwarz actions in coset superspaces}''},
\textsf{Nucl.~Phys.~B573,~811~(2000)},
\href{http://arXiv.org/abs/hep-th/9908045}{\texttt{hep-th/9908045}}.
%
\bibitem{Arutyunov:2009}
G.~Arutyunov and S.~Frolov,
\textit{``{Foundations of the AdS(5) x S**5 superstring: I.}''},
\textsf{J.~Phys.~A:~Math.~Theor~42,~1~(2009)},
\href{http://arXiv.org/abs/0901.4937}{\texttt{0901.4937}}.
%
\bibitem{Roiban:2000yy}
R.~Roiban and W.~Siegel,
\textit{``{Superstrings on $AdS_5\times S^5$ supertwistor space}''},
\textsf{JHEP~0011,~024~(2000)},
\href{http://arXiv.org/abs/hep-th/0010104}{\texttt{hep-th/0010104}}.
%
\bibitem{Cagnazzo:2009zh}
A.~Cagnazzo, D.~Sorokin and L.~Wulff,
\textit{``{String instanton in $AdS_4\times CP^3$}''},
\href{http://arXiv.org/abs/0911.5228}{\texttt{0911.5228}}.
%
\bibitem{Cvetic:1999zs}
M.~Cvetic, H.~Lu, C.~N.~Pope and K.~S.~Stelle,
\textit{``{T-Duality in the Green-Schwarz Formalism, and the Massless/Massive
  IIA Duality Map}''},
\textsf{Nucl.~Phys.~B573,~149~(2000)},
\href{http://arXiv.org/abs/hep-th/9907202}{\texttt{hep-th/9907202}}.
%
\bibitem{Berenstein:2002jq}
D.~E.~Berenstein, J.~M.~Maldacena and H.~S.~Nastase,
\textit{``{Strings in flat space and pp waves from N = 4 super Yang Mills}''},
\textsf{JHEP~0204,~013~(2002)},
\href{http://arXiv.org/abs/hep-th/0202021}{\texttt{hep-th/0202021}}.
%
\bibitem{Gubser:2002tv}
S.~S.~Gubser, I.~R.~Klebanov and A.~M.~Polyakov,
\textit{``{A semi-classical limit of the gauge/string correspondence}''},
\textsf{Nucl.~Phys.~B636,~99~(2002)},
\href{http://arXiv.org/abs/hep-th/0204051}{\texttt{hep-th/0204051}}.
%
\bibitem{Parnachev:2002kk}
A.~Parnachev and A.~V.~Ryzhov,
\textit{``{Strings in the near plane wave background and AdS/CFT}''},
\textsf{JHEP~0210,~066~(2002)},
\href{http://arXiv.org/abs/hep-th/0208010}{\texttt{hep-th/0208010}}.
%
\bibitem{Callan:2003xr}
C.~G.~Callan,~Jr. et~al.,
\textit{``{Quantizing string theory in $AdS_5\times S^5$: Beyond the pp-
  wave}''},
\textsf{Nucl.~Phys.~B673,~3~(2003)},
\href{http://arXiv.org/abs/hep-th/0307032}{\texttt{hep-th/0307032}}.
%
\bibitem{Callan:2004uv}
C.~G.~Callan,~Jr., T.~McLoughlin and I.~Swanson,
\textit{``{Holography beyond the Penrose limit}''},
\textsf{Nucl.~Phys.~B694,~115~(2004)},
\href{http://arXiv.org/abs/hep-th/0404007}{\texttt{hep-th/0404007}}.
%
\bibitem{Lee:2008sk}
B.-H.~Lee, R.~R.~Nayak, K.~L.~Panigrahi and C.~Park,
\textit{``{On the giant magnon and spike solutions for strings on AdS$_3\times$
  S$^3$}''},
\textsf{JHEP~0806,~065~(2008)},
\href{http://arXiv.org/abs/0804.2923}{\texttt{0804.2923}}.
%
\bibitem{Beisert:2007sk}
N.~Beisert and B.~I.~Zwiebel,
\textit{``{On Symmetry Enhancement in the psu(1,1|2) Sector of N=4 SYM}''},
\textsf{JHEP~0710,~031~(2007)},
\href{http://arXiv.org/abs/0707.1031}{\texttt{0707.1031}}.
%
\bibitem{Arutyunov:2006ak}
G.~Arutyunov, S.~Frolov, J.~Plefka and M.~Zamaklar,
\textit{``The off-shell symmetry algebra of the light-cone $AdS_5 \times S^5$
  superstring''},
\href{http://arXiv.org/abs/hep-th/0609157}{\texttt{hep-th/0609157}}.
%
\bibitem{Zarembo:2009au}
K.~Zarembo,
\textit{``{Worldsheet spectrum in $AdS_4/CFT_3$ correspondence}''},
\href{http://arXiv.org/abs/0903.1747}{\texttt{0903.1747}}.
%
\bibitem{Sundin:2009zu}
P.~Sundin,
\textit{``{On the worldsheet theory of the type IIA $AdS_4\times CP^3$
  superstring}''},
\href{http://arXiv.org/abs/0909.0697}{\texttt{0909.0697}}.
%
\bibitem{Novikov:1984id}
S.~Novikov, S.~V.~Manakov, L.~P.~Pitaevsky and V.~E.~Zakharov,
\textit{``Theory of solitons. the inverse scattering method''},
Consultants Bureau (1984),
New York, USA,
276p,
Contemporary Soviet Mathematics.
%
\bibitem{Kazakov:2004nh}
V.~A.~Kazakov and K.~Zarembo,
\textit{``Classical/quantum integrability in non-compact sector of AdS/CFT''},
\textsf{JHEP~0410,~060~(2004)},
\href{http://arXiv.org/abs/hep-th/0410105}{\texttt{hep-th/0410105}}.
%
\bibitem{Beisert:2004ag}
N.~Beisert, V.~A.~Kazakov and K.~Sakai,
\textit{``{Algebraic curve for the SO(6) sector of AdS/CFT}''},
\textsf{Commun.~Math.~Phys.~263,~611~(2006)},
\href{http://arXiv.org/abs/hep-th/0410253}{\texttt{hep-th/0410253}}.
%
\bibitem{SchaferNameki:2004ik}
S.~Schafer-Nameki,
\textit{``{The algebraic curve of 1-loop planar N = 4 SYM}''},
\textsf{Nucl.~Phys.~B714,~3~(2005)},
\href{http://arXiv.org/abs/hep-th/0412254}{\texttt{hep-th/0412254}}.
%
\bibitem{Dorey:2006zj}
N.~Dorey and B.~Vicedo,
\textit{``{On the dynamics of finite-gap solutions in classical string
  theory}''},
\textsf{JHEP~0607,~014~(2006)},
\href{http://arXiv.org/abs/hep-th/0601194}{\texttt{hep-th/0601194}}.
%
\bibitem{Dorey:2006mx}
N.~Dorey and B.~Vicedo,
\textit{``{A symplectic structure for string theory on integrable
  backgrounds}''},
\textsf{JHEP~0703,~045~(2007)},
\href{http://arXiv.org/abs/hep-th/0606287}{\texttt{hep-th/0606287}}.
%
\bibitem{Zarembo:2004hp}
K.~Zarembo,
\textit{``{Semiclassical Bethe ansatz and AdS/CFT}''},
\textsf{Comptes~Rendus~Physique~5,~1081~(2004)},
\href{http://arXiv.org/abs/hep-th/0411191}{\texttt{hep-th/0411191}}.
%
\bibitem{Faddeev:1987ph}
L.~D.~Faddeev and L.~A.~Takhtajan,
\textit{``Hamiltonian methods in the theory of solitons''},
Springer (1987),
Berlin, Germany,
592p,
Springer Series In Soviet Mathematics.
%
\bibitem{Dobrev:1985qz}
V.~K.~Dobrev and V.~B.~Petkova,
\textit{``{Group Theoretical Approach To Extended Conformal Supersymmetry:
  Function Space Realizations And Invariant Differential Operators}''},
\textsf{Fortschr.~Phys.~35,~537~(1987)}.
%
\bibitem{PenkovSerganova}
I.~Penkov and V.~Serganova,
\textit{``{Representations of classical Lie superalgebras of type I}''},
\textsf{Indag.~Math.~N.S.3(4),~419~(1992)}.
%
\bibitem{Beisert:2005di}
N.~Beisert, V.~A.~Kazakov, K.~Sakai and K.~Zarembo,
\textit{``{Complete spectrum of long operators in N = 4 SYM at one loop}''},
\textsf{JHEP~0507,~030~(2005)},
\href{http://arXiv.org/abs/hep-th/0503200}{\texttt{hep-th/0503200}}.
%
\bibitem{Gromov:2007ky}
N.~Gromov and P.~Vieira,
\textit{``{Complete 1-loop test of AdS/CFT}''},
\textsf{JHEP~0804,~046~(2008)},
\href{http://arXiv.org/abs/0709.3487}{\texttt{0709.3487}}.
%
\bibitem{Vicedo:2008jy}
B.~Vicedo,
\textit{``{Semiclassical Quantisation of Finite-Gap Strings}''},
\textsf{JHEP~0806,~086~(2008)},
\href{http://arXiv.org/abs/0803.1605}{\texttt{0803.1605}}.
%
\bibitem{Beisert:2006ib}
N.~Beisert, R.~Hernandez and E.~Lopez,
\textit{``A crossing-symmetric phase for $AdS_5 \times S^5$ strings''},
\textsf{JHEP~0611,~070~(2006)},
\href{http://arXiv.org/abs/hep-th/0609044}{\texttt{hep-th/0609044}}.
%
\bibitem{Beisert:2006ez}
N.~Beisert, B.~Eden and M.~Staudacher,
\textit{``{Transcendentality and crossing}''},
\textsf{J.~Stat.~Mech.~0701,~P021~(2007)},
\href{http://arXiv.org/abs/hep-th/0610251}{\texttt{hep-th/0610251}}.
%
\bibitem{Dorey:2007xn}
N.~Dorey, D.~M.~Hofman and J.~M.~Maldacena,
\textit{``{On the singularities of the magnon S-matrix}''},
\textsf{Phys.~Rev.~D76,~025011~(2007)},
\href{http://arXiv.org/abs/hep-th/0703104}{\texttt{hep-th/0703104}}.
%
\bibitem{Arutyunov:2009kf}
G.~Arutyunov and S.~Frolov,
\textit{``{The Dressing Factor and Crossing Equations}''},
\textsf{J.~Phys.~A42,~425401~(2009)},
\href{http://arXiv.org/abs/0904.4575}{\texttt{0904.4575}}.
%
\bibitem{Volin:2009uv}
D.~Volin,
\textit{``{Minimal solution of the AdS/CFT crossing equation}''},
\textsf{J.~Phys.~A42,~372001~(2009)},
\href{http://arXiv.org/abs/0904.4929}{\texttt{0904.4929}}.
%
\bibitem{Faddeev:1985qu}
L.~D.~Faddeev and N.~Y.~Reshetikhin,
\textit{``{Integrability of the principal chiral field model in
  (1+1)-dimension}''},
\textsf{Ann.~Phys.~167,~227~(1986)}.
%
\bibitem{Gromov:2006dh}
N.~Gromov, V.~Kazakov, K.~Sakai and P.~Vieira,
\textit{``{Strings as multi-particle states of quantum sigma- models}''},
\textsf{Nucl.~Phys.~B764,~15~(2007)},
\href{http://arXiv.org/abs/hep-th/0603043}{\texttt{hep-th/0603043}}.
%
\bibitem{Gromov:2006cq}
N.~Gromov and V.~Kazakov,
\textit{``{Asymptotic Bethe ansatz from string sigma model on $S^3\times R$
  }''},
\textsf{Nucl.~Phys.~B780,~143~(2007)},
\href{http://arXiv.org/abs/hep-th/0605026}{\texttt{hep-th/0605026}}.
%
\bibitem{Gromov:2007fn}
N.~Gromov, V.~Kazakov and P.~Vieira,
\textit{``{Classical limit of quantum sigma-models from Bethe ansatz}''},
\textsf{PoS~SOLVAY,~005~(2006)},
\href{http://arXiv.org/abs/hep-th/0703137}{\texttt{hep-th/0703137}}.
%
\bibitem{Sommovigo:2003kd}
L.~Sommovigo,
\textit{``{Penrose limit of $AdS_3\times S^3\times S^3\times S^1$ and its
  associated sigma-model}''},
\textsf{JHEP~0307,~035~(2003)},
\href{http://arXiv.org/abs/hep-th/0305151}{\texttt{hep-th/0305151}}.
%
\bibitem{Kruczenski:2003gt}
M.~Kruczenski,
\textit{``{Spin chains and string theory}''},
\textsf{Phys.~Rev.~Lett.~93,~161602~(2004)},
\href{http://arXiv.org/abs/hep-th/0311203}{\texttt{hep-th/0311203}}.
%
\bibitem{Kruczenski:2004kw}
M.~Kruczenski, A.~V.~Ryzhov and A.~A.~Tseytlin,
\textit{``{Large spin limit of AdS(5) x S**5 string theory and low energy
  expansion of ferromagnetic spin chains}''},
\textsf{Nucl.~Phys.~B692,~3~(2004)},
\href{http://arXiv.org/abs/hep-th/0403120}{\texttt{hep-th/0403120}}.
%
\bibitem{Hernandez:2004uw}
R.~Hernandez and E.~Lopez,
\textit{``{The SU(3) spin chain sigma model and string theory}''},
\textsf{JHEP~0404,~052~(2004)},
\href{http://arXiv.org/abs/hep-th/0403139}{\texttt{hep-th/0403139}}.
%
\bibitem{Stefanski:2004cw}
B.~Stefa\'nski,~Jr. and A.~A.~Tseytlin,
\textit{``{Large spin limits of AdS/CFT and generalized Landau- Lifshitz
  equations}''},
\textsf{JHEP~0405,~042~(2004)},
\href{http://arXiv.org/abs/hep-th/0404133}{\texttt{hep-th/0404133}}.
%
\bibitem{Kruczenski:2004cn}
M.~Kruczenski and A.~A.~Tseytlin,
\textit{``{Semiclassical relativistic strings in $S^5$ and long coherent
  operators in N = 4 SYM theory}''},
\textsf{JHEP~0409,~038~(2004)},
\href{http://arXiv.org/abs/hep-th/0406189}{\texttt{hep-th/0406189}}.
%
\bibitem{Hernandez:2004kr}
R.~Hernandez and E.~Lopez,
\textit{``{Spin chain sigma models with fermions}''},
\textsf{JHEP~0411,~079~(2004)},
\href{http://arXiv.org/abs/hep-th/0410022}{\texttt{hep-th/0410022}}.
%
\bibitem{Stefanski:2005tr}
B.~Stefa\'nski,~Jr. and A.~A.~Tseytlin,
\textit{``{Super spin chain coherent state actions and AdS(5) x S**5
  superstring}''},
\textsf{Nucl.~Phys.~B718,~83~(2005)},
\href{http://arXiv.org/abs/hep-th/0503185}{\texttt{hep-th/0503185}}.
%
\bibitem{Stefanski:2007dp}
B.~Stefa\'nski,~Jr.,
\textit{``{Landau-Lifshitz sigma-models, fermions and the AdS/CFT
  correspondence}''},
\textsf{JHEP~0707,~009~(2007)},
\href{http://arXiv.org/abs/0704.1460}{\texttt{0704.1460}}.
%
\bibitem{Maldacena:1998bw}
J.~M.~Maldacena and A.~Strominger,
\textit{``{AdS(3) black holes and a stringy exclusion principle}''},
\textsf{JHEP~9812,~005~(1998)},
\href{http://arXiv.org/abs/hep-th/9804085}{\texttt{hep-th/9804085}}.
%
\bibitem{Seiberg:1999xz}
N.~Seiberg and E.~Witten,
\textit{``{The D1/D5 system and singular CFT}''},
\textsf{JHEP~9904,~017~(1999)},
\href{http://arXiv.org/abs/hep-th/9903224}{\texttt{hep-th/9903224}}.
%
\bibitem{Larsen:1999uk}
F.~Larsen and E.~J.~Martinec,
\textit{``{U(1) charges and moduli in the D1-D5 system}''},
\textsf{JHEP~9906,~019~(1999)},
\href{http://arXiv.org/abs/hep-th/9905064}{\texttt{hep-th/9905064}}.
%
\bibitem{Arutyunov:1997gi}
G.~E.~Arutyunov and S.~A.~Frolov,
\textit{``{Four graviton scattering amplitude from S(N) R**8 supersymmetric
  orbifold sigma model}''},
\textsf{Nucl.~Phys.~B524,~159~(1998)},
\href{http://arXiv.org/abs/hep-th/9712061}{\texttt{hep-th/9712061}}.
%
\bibitem{Lunin:2001pw}
O.~Lunin and S.~D.~Mathur,
\textit{``{Three-point functions for M(N)/S(N) orbifolds with N = 4
  supersymmetry}''},
\textsf{Commun.~Math.~Phys.~227,~385~(2002)},
\href{http://arXiv.org/abs/hep-th/0103169}{\texttt{hep-th/0103169}}.
%
\bibitem{Jevicki:1998bm}
A.~Jevicki, M.~Mihailescu and S.~Ramgoolam,
\textit{``{Gravity from CFT on S**N(X): Symmetries and interactions}''},
\textsf{Nucl.~Phys.~B577,~47~(2000)},
\href{http://arXiv.org/abs/hep-th/9907144}{\texttt{hep-th/9907144}}.
%
\bibitem{Hikida:2002in}
Y.~Hikida and Y.~Sugawara,
\textit{``{Superstrings on PP-wave backgrounds and symmetric orbifolds}''},
\textsf{JHEP~0206,~037~(2002)},
\href{http://arXiv.org/abs/hep-th/0205200}{\texttt{hep-th/0205200}}.
%
\bibitem{Lunin:2002fw}
O.~Lunin and S.~D.~Mathur,
\textit{``{Rotating deformations of AdS(3) x S(3), the orbifold CFT and strings
  in the pp-wave limit}''},
\textsf{Nucl.~Phys.~B642,~91~(2002)},
\href{http://arXiv.org/abs/hep-th/0206107}{\texttt{hep-th/0206107}}.
%
\bibitem{Gomis:2002qi}
J.~Gomis, L.~Motl and A.~Strominger,
\textit{``{pp-wave / CFT(2) duality}''},
\textsf{JHEP~0211,~016~(2002)},
\href{http://arXiv.org/abs/hep-th/0206166}{\texttt{hep-th/0206166}}.
%
\bibitem{Gava:2002xb}
E.~Gava and K.~S.~Narain,
\textit{``{Proving the pp-wave / CFT(2) duality}''},
\textsf{JHEP~0212,~023~(2002)},
\href{http://arXiv.org/abs/hep-th/0208081}{\texttt{hep-th/0208081}}.
%
\bibitem{Brown:1986nw}
J.~D.~Brown and M.~Henneaux,
\textit{``{Central Charges in the Canonical Realization of Asymptotic
  Symmetries: An Example from Three-Dimensional Gravity}''},
\textsf{Commun.~Math.~Phys.~104,~207~(1986)}.
%
\bibitem{deBoer:1998pp}
J.~de~Boer, H.~Ooguri, H.~Robins and J.~Tannenhauser,
\textit{``{String theory on $AdS_3$}''},
\textsf{JHEP~9812,~026~(1998)},
\href{http://arXiv.org/abs/hep-th/9812046}{\texttt{hep-th/9812046}}.
%
\bibitem{Ashok:2009jw}
S.~K.~Ashok, R.~Benichou and J.~Troost,
\textit{``{Asymptotic Symmetries of String Theory on $AdS_3\times S^3$ with
  Ramond-Ramond Fluxes}''},
\textsf{JHEP~0910,~051~(2009)},
\href{http://arXiv.org/abs/0907.1242}{\texttt{0907.1242}}.
%
\bibitem{Pakman:2009mi}
A.~Pakman, L.~Rastelli and S.~S.~Razamat,
\textit{``{A Spin Chain for the Symmetric Product $CFT_2$}''},
\href{http://arXiv.org/abs/0912.0959}{\texttt{0912.0959}}.
%
\end{thebibliography}

\end{document}